\newif\ifShowKeys
\ShowKeystrue

\documentclass[11pt]{article}
\usepackage[T1]{fontenc}

\usepackage{listings}
\lstnewenvironment{arkady}  
{\lstset{language=C,frame=trbl,basicstyle = \footnotesize \ttfamily , breaklines = true,showstringspaces=false}}{}

\usepackage[usenames,dvipsnames]{xcolor}
\usepackage[setpagesize=false,pagebackref=false, 
linktocpage, bookmarksopen=true, colorlinks=true, 
linkcolor=Maroon,citecolor=Maroon,urlcolor=Maroon]{hyperref}

\usepackage[parsep]{collref}				

\usepackage{amsmath, amssymb,amsthm}
\usepackage{stackrel}
\numberwithin{equation}{section}
\usepackage{bm,environ,mathrsfs,array,arydshln}
\usepackage{booktabs,float,slashed}
\usepackage{appendix}
\usepackage[mathcal]{euscript}
\usepackage{tensor} 						
\usepackage{mathabx}
\usepackage[vcentermath]{youngtab}

\usepackage{graphicx,epsfig,epic}
\usepackage{tikz} 
\usetikzlibrary{arrows,decorations.pathreplacing,decorations.markings,snakes}
\usetikzlibrary{cd}

\usepackage{datetime}

\usepackage{framed}						
\definecolor{shadecolor}{rgb}{0.9996078, 0.984314, 0.960784}
\definecolor{TFTitleColor}{RGB}{1,1,1}
\definecolor{TFFrameColor}{RGB}{249	218	181}		
\definecolor{TFFrameColor}{RGB}{230 230 230 }

\usepackage{comment}					

\allowdisplaybreaks

\definecolor{myred}{RGB}{233, 33, 45}

\newcommand{\bs}{\begin{shaded}}
\newcommand{\es}{\end{shaded}\noindent}

\def\ba#1\ea{\begin{align}#1\end{align}}		
\newcommand{\be}{\begin{equation}}
\newcommand{\ee}{\end{equation}}
\newcommand{\bea}{\begin{equation} \begin{aligned}} 
\newcommand{\eea}{\end{aligned} \end{equation}}
\newcommand{\mc}{\mathcal }
\newcommand{\wh}{\widehat}

\newcommand{\la}{\label}
\newcommand{\eps}{\varepsilon}

\newcommand{\lp}{\notag \\ & }

\DeclareMathOperator{\Tr}{\text{Tr}}
\DeclareMathOperator{\vol}{vol}

\newcommand{\cf}{cf.}
\newcommand{\ie}{i.e.}
\newcommand{\eg}{e.g.}

\renewcommand{\l}{\lambda}

\newcommand{\T}{{\mathsf T}}

\newcommand{\ul}{\underline}
\newcommand{\vp}{\varphi}

\newcommand{\sfH}{\mathsf{H}}
\newcommand{\we}{\wedge}

\newcommand{\LL}{\mathrm{L}}

\newcommand{\TermF}{\textrm{{\rm \bf h}}}
\newcommand{\TermFF}{\textrm{{\rm \bf  hh}}}
\newcommand{\TermMix}{\textrm{\bf  hT }}

\newcommand{\vk}{\varkappa}
\newcommand{\kk}{\textrm{k}}

\def \ci {\cite}
\def \bb {{\rm b}} 
\def \foot {\footnote}

\def \b {\beta} 

\def \adsst {AdS$_3\times S^3$ }

\def\ov{\over}
\newcommand{\rf}[1]{(\ref{#1})}
\def \la {\label}
\def \del{ \partial}
\def \ed {
\small 
\bibliography{BT-biblio}
 \bibliographystyle{JHEP-v2.9}
\end{document}}
\def \iffa {\iffalse}

\newcommand{\uh}[1]{ {\underline{ \hat{ #1} }} }
\newcommand{\cd}{ { \mathcal{\nabla} } }
\newcommand{\XX}{\natural}
\newcommand{\ads}{ { \text{AdS} } }
\DeclareMathOperator{\tr}{tr}

\newcommand{\WW}{ \mathrm{Weyl}^2 }

\def \td {\tilde}

\begin{document}
\begin{titlepage}


\vspace*{3mm}
\begin{center}

{\Large\bf   Semiclassical quantization of M5 brane probes}\vskip 4pt
{\Large\bf   wrapped on  $\ads_3 \times S^3$  and defect anomalies } 

{\small 

\vspace*{6mm}

{\large M. Beccaria$^{a}$, \ \ L. Casarin$^{b,c}$, \ \ A.A. Tseytlin$^{d,}$\footnote{\ Also at the Institute for Theoretical and Mathematical Physics (ITMP) of MSU
   and Lebedev Institute.}} 
\vspace*{3mm}

${}^a$ Universit\`a del Salento, Dipartimento di Matematica e Fisica \textit{Ennio De Giorgi}\\ 
		and INFN - sezione di Lecce, Via Arnesano, I-73100 Lecce, Italy
		
\vskip 0.1cm
${}^{b}$ Institut f\"ur Theoretische Physik Leibniz Universit\"at Hannover \\
 Appelstra\ss{}e 2, 30167 Hannover, Germany
 \vskip 0.1cm
 
 ${}^{c}$ Max-Planck-Institut f\"ur Gravitationsphysik (Albert-Einstein-Institut) \\
  Am M\"uhlenberg 1, DE-14476 Potsdam, Germany
\vskip 0.1cm
${}^{d}$ Blackett Laboratory, Imperial College London SW7 2AZ, U.K.
\vskip 0.2cm {\small 
 \texttt{matteo.beccaria@le.infn.it, lorenzo.casarin@itp.uni-hannover.de, tseytlin@ic.ac.uk}}
\vspace*{0.2cm}
}
\end{center}
 \begin{abstract}  
\noindent
We consider two supersymmetric M5 brane probe solutions  in $\ads_7 \times S^4$
 and   one in $\ads_4 \times S^7$  that all have the   $\ads_3 \times S^3$  world-volume geometry. 
The values of the  classical  action of the first  two  M5 probes   (with  $S^3$ in $\ads_7$  or  in $S^4$) 
 are related to the leading  $N^2$ parts in the anomaly b-coefficient in the  (2,0) theory corresponding to a
  spherical surface  defect  in symmetric or antisymmetric  $SU(N)$ representations. 
   We  present a detailed computation of the  corresponding one-loop  M5 brane partition functions 
  finding  that they  vanish (in a particular  regularization). This implies the vanishing of the order $N^0$   part
   in the  b-anomaly coefficients,    in agreement with    earlier predictions for  their  exact   values. 
  It remains, however,  a puzzle of how  to reproduce the non-vanishing  order $N$  terms in these coefficients 
  within   the semiclassical M5-brane probe  setup. 
\end{abstract}
\vskip 0.5cm
	{	}
\end{titlepage}
{\small 
\tableofcontents
}
\def \Te {\Theta}
\def \vp {\varphi}  \def \dd  {{\rm d}}
\def \cc  {{\rm c}} 
\def \La {\Lambda}
\def \half {{1\ov 2}}
\def  \Ia {\textbf{Ia}}
\def  \Ib {\textbf{Ib}}
\def \II {\textbf{II}} 
 \def \bb {{\rm b}}\def \OO {{\cal O}}\def \four {{\textstyle {1\ov 4}}}\def \ka {\kappa} \def \ha {{\textstyle {1\ov 2}}}
\def \edd {\end{document}}
\def \ff {{\rm f}} 
\def \L {{\cal L}}
\def \f {{\rm h}} \def \bg {g} 
\def \bDelta  {{\bm \Delta}}
\def \z {\zeta} \def \no {\notag} 
\def \TT {\vartheta} 
\def \U {{\cal U}}
\def \edo {\end{document}}
\def \hb {b}
 \def \Gh {\widehat {\Gamma}}
\def \te {\textstyle}
\def \TF {{\rm F}}
\def \bre {e}
\def \WW {{\rm W}}
\def \Ss {{\mc R}}
 \def \bD  {\bm \Delta}
\def \bDelta {\bm \Delta}

 \setcounter{footnote}{0}
 \section{Introduction} 
 
 The defect anomaly coefficients in 6d  (2,0)  theory  (see, \eg, \cite{Estes:2018tnu,Jensen:2018rxu,Chalabi:2020iie,Chalabi:2021jud,Capuozzo:2023fll}  
 and refs. there)  can be studied via  AdS/CFT   correspondence 
 by  considering BPS  M-brane probes   in $\ads_{7}\times S^{4}$  \ci{Lunin:2007ab,DHoker:2008wvd}
 and semiclassically  quantizing them 
 \ci{Drukker:2020swu,Drukker:2023jxp,Drukker:2023bip,Jiang:2024wzs}.\foot{Examples of 
   one-loop  computations for  M-branes  in AdS  backgrounds  were  discussed  also  in \ci{Giombi:2023vzu,Beccaria:2023ujc,Beccaria:2023sph,Beccaria:2023cuo}.}
 
  Ref.   \ci{Drukker:2020swu}   considered an M2 brane probe wrapped on $\ads_3\subset \ads_7$ in  $\ads_7 \times S^4$  
background    that intersected the boundary over $S^2$. The effective  dimensionless  M2  tension 
 in this case is ${\T}_2={2\ov \pi} N$  
  where $N$ is the number of M5 branes forming the $\ads_7 \times S^4$  
  background (or rank of the  (2,0)   boundary  CFT).  A semiclassical  expansion of the M2 brane free energy $F$
  then determines the large $N$ expansion  of the ``central charge'' or  b-anomaly coefficient 
  of the $S^2$ defect in the (2,0) theory.  The resulting  classical and one-loop  M2 brane contributions  were found to be    \ci{Drukker:2020swu}
  \foot{To recall,  in  the presence of 2d defects in a    CFT  defined on a curved space 
 its  stress-tensor trace anomaly can be written as the sum  of the  ambient  space  contribution 
 and the following  additional term localized on the defect 
 \cite{Graham:1999pm,Henningson:1999xi,Asnin:2008ak,Schwimmer:2008yh,Chalabi:2020iie}:
$ T^{\mu}_{\mu}\big|_{\rm defect} = -\frac{1}{24\pi}\big({\rm b}\,\wh R +{\rm d_{1}} {\Pi}^{\mu}_{ij}\, {\Pi}^{ij}_{\mu}-{\rm d}_{2}\, W\indices{_{ij}^{ij}}\big)$. Here  $\wh R$ is the Ricci scalar for  the induced metric on the defect, 
   ${\Pi}^{\mu}_{ij}$  is  the traceless second fundamental form
  of the defect and $W_{ijk\ell}$ is the pull-back of the Weyl tensor. 
  We follow  \ci{Jensen:2018rxu}  and denote the ``central charge'' coefficient as b. 
  Following \cite{Jensen:2013lxa,Estes:2014hka,Gentle:2015jma} one 
   may  compute holographically the entanglement entropy (EE) of a spherical region centred on the 
 2d defect or of a semi-circle centred on the 2d boundary. 
  After subtracting 
  the EE of the ambient CFT the  coefficient of the  logarithmic in the UV cutoff 
   term  may be denoted as  $\frac{1}{3}{ b}$. For  a CFT$_{d}$
 this  ``central charge'' $b$  obeys \cite{Kobayashi:2018lil,Jensen:2018rxu}
$ {b} = {\rm b}-\frac{d-3}{d-1}{\rm d}_{2}.$}  
  \be \la{01}
\bb  = {12 }  N -  9 + \OO(N^{-1}) \,. \ee
  This  turns out  to be  consistent   with the expression for b-anomaly   found    from the entanglement entropy  computation for the 
``bubbling''  M5-M2 geometry \ci{DHoker:2008lup}.  The general expression for  b-anomaly corresponding to 
   a   $1\ov  2$-BPS  surface defect 
operator in (2,0)  theory   in a  $SU(N)$ representation  with the  Young tableau  with a large number  of boxes is 
  \cite{Estes:2018tnu,Jensen:2018rxu,Chalabi:2020iie}
\be \la{02}
\bb= 24(\rho, \lambda) +3 (\lambda, \lambda)   \,. \ee
Here $\rho$ is the Weyl vector of  $SU(N)$   and $\lambda$ is the  highest weight of the  $SU(N)$ 
representation.\foot{To be 
 precise, the status of \rf{02} as an exact in $N$ expression as found in  \cite{Estes:2018tnu}
 could   still  be   viewed as conjecture. In \ci{Chalabi:2020iie} 
  a  similar expression for the  d$_2$   anomaly  coefficient (see \rf{1.8},\rf{H1}) 
 was derived as an exact result from a  superconformal index  computation  (it also follows from the 5d Wilson Loop 
 localization computation as in  \ci{Mori:2014tca}, see Appendix \ref{WL} below). 
 Given that $ \bb$  and d$_2$   appear on an equal footing in the spherical entanglement
  entropy \ci{Jensen:2018rxu,Rodgers:2018mvq}
one may expect  
 that the expression for $\bb$  should   also  be exact. 
 Indeed, in \ci{Wang:2020xkc} the same  expression \rf{02} was found   using  't Hooft anomaly  considerations. 
}


If we formally   assume  that \rf{02} is  valid not  just for large representations but also for the ones
 with finite number of boxes  then  for a single M2 brane  corresponding to the
  surface operator in the fundamental representation (with $(\rho, \lambda)= {N-1\ov 2} , \  (\lambda, \lambda) = {N-1\ov N}$)
  one   finds 
\be \la{03}
\bb  = 12N  ( 1 +\four {N^{-1}} )  (1-N^{-1})  =   12 N - 9  - {3N^{-1}}    \,. \ee
The  first two terms  here match the classical and one-loop terms in 
\rf{01}  while the $N^{-1}$  term should  correspond to the 2-loop M2 brane correction. 

In the   case of $k$-symmetric and $k$-antisymmetric representations \rf{02} gives  (\cf\ (\ref{H.5}, \ref{H.6}))
\ba
\la{5}
&{\rm b}_{(k)} = 12kN\, \big(1+\four {k}N^{-1}\big)\big(1-N^{-1} \big)= 12k N -   3k (4-k)  - {3k^2N^{-1}}   \,,       \\  
&{\rm b}_{[k]} = 12kN\,  \big(1+\four N^{-1}\big)\big(1- k N^{-1} \big) = 12k N -   3k ( 4k-1)  - {3k^2N^{-1}}  \,,  \la{6}
\ea
which of course  reduce to \rf{03} for $k=1$.  The case $k>1$   should correspond to a system of multiple M2  brane  probes 
which it is not clear how to quantize directly. However,   for large $k\sim N \gg 1$  one may expect that such M2 brane configuration 
 should ``blow up'' 
into  a   single M5 brane (wrapped on $S^3 \subset \ads_{7}$ in the case of $k$-symmetric representation 
and on $S^3 \subset  S^4 $ in the case of $k$-antisymmetric one) 
with a non-zero world-volume 3-form flux representing the M2 brane charge $k$. 

The two corresponding classical M5  brane  probe solutions in $\ads_7 \times S^4$  that have $\ads_3 \times S^3$
 world volume geometry  were found  in \ci{Lunin:2007ab} (see  also \ci{Chen:2007ir,Mori:2014tca}).
 We will refer to them  as \Ia\  and \Ib\  probes below.
  These solutions should apply in    the limit when 
\be
\la{110}
N, k \gg 1\,, \qquad\qquad  \kappa^2 \equiv  \frac{k}{2N} = \text{fixed}\,.
\ee 
 $\kappa$ plays the role of   a  free  parameter of  an  M5 brane   solution  related to its location in $ \ads_{7 }\times S^4$
and also to the value of the world-volume 3-form  field $H_3$.
Expressing \rf{5},\rf{6} in terms of $N$ and $\ka$  we get 
\ba
\la{15}
&{\rm b}_{(k)}= 24 N (N-1) \,   \ka^2 (1 + \ha  \ka^2)  =  24N^2  \ka^2 (1 + \ha  \ka^2)   - 24 N \ka^2 (1 + \ha \ka^2 )   \,,       \\  
&{\rm b}_{[k]}= 24N(N+ \four)\,   \ka^2 (1  - 2 \ka^2)     =  24N^2 \ka^2 (1  - 2 \ka^2)   + 6 N \ka^2 ( 1 - 2 \ka^2 )  
   \,.  \la{150}
\ea
By analogy with the M2 brane probe case in \rf{01} one may conjecture that \rf{15},\rf{150} 
may be reproduced by  semiclassically quantizing the  corresponding M5 brane probe. 

The effective  dimensionless M5  brane tension 
here is  ${\T}_5= {2\ov \pi^3}  N^2 = {1\ov 2\pi} ({\T_2})^2$ (see \rf{27})
and the leading $N^2$ terms in \rf{15} and \rf{150} are indeed  reproduced by the  values of the 
classical  M5 brane action  for the two corresponding solutions. 

However, the subleading terms in \rf{15},\rf{150}   do not  appear to have a natural  interpretation 
within the semiclassical M5 brane expansion, \ie\ the expansion in powers of $({\rm T}_5)^{-1} \sim N^{-2}$ for fixed $\ka$. 
The order $N$ terms in \rf{15},\rf{150}   look  as if they are     coming, in fact,  from a classical M2 brane  action
or ``$\ha$-loop'' order of M5 brane perturbation theory.\foot{Surprisingly, 
the $\ka$-dependence of the leading (order $N^2$)  and the subleading (order $N$)  terms in \rf{15},\rf{150} 
 happens to be   the same. One   could then  conjecture that these  expressions  correspond to the classical M5 brane 
 contribution  but with ``renormalized''   M5 tension.  It is not clear, however, how to justify this possibility given, in particular, that 
 this ``renormalization''   happens  to be different in the two cases  in \rf{15} and \rf{150} (\cf\ also \rf{1.12} and \rf{120}  below).}

Regardless the resolution of the puzzle of the order $N$ terms, the    expressions \rf{15},\rf{150}  do not contain 
order $N^0$   terms implying that one-loop M5 brane corrections to the b-coefficient  should be zero. 
Our  aim    here will be to  demonstrate  this by  
directly computing    the one-loop  corrections to the  free energy of the  corresponding two M5 brane  probe 
solutions    in $\ads_7 \times S^4$  found in \ci{Lunin:2007ab}. 
 We will also   consider a similar M5 brane solution  in   $\ads_4 \times S^7$ 
  having again   the   $\ads_3 \times S^3$   world volume
 (this solution was found in \ci{Lunin:2007ab} and also in \ci{Arean:2007nh} 
 and the study of  bosonic fluctuations around it was initiated in \cite{Fiol:2010wf,Fiol:2010un}).

 This   will require  a non-trivial extension of the  earlier  computations of   one-loop partition functions 
  of M5 branes  wrapped on $S^1 \times S^5$ in a twisted version of  $\ads_4 \times S^7$  in \ci{Beccaria:2023cuo}
    and   on   $\ads_5 \times S^1$  in $\ads_7 \times S^4$  in  \ci{Jiang:2024wzs}
    to the cases with  a non-zero  $H_3$  world-volume  field. 
    The  presence of the $H_3$  background   introduces a  complication due to the self-duality constraint on the 
    world-volume 3-form field requiring to use the detailed   structure  of the M5 brane action 
    \ci{Pasti:1997gx,Bandos:1997ui,Aganagic:1997zq,Howe:1997fb,Bandos:1997gm,Kallosh:2005yu,Ko:2013dka} (see also 
    \ci{Mkrtchyan:2019opf,Avetisyan:2022zza,Bansal:2023pnr}).

    \iffa  
  Another case studied in the same reference was that of a co-dimension 2 four dimensional defect 
in the  6d $(2,0)$ theory by means of a $\frac{1}{2}$-BPS 
 M5  brane probe in $\ads_7 \times S^4$  
 wrapped on $S^1 \subset \ads_5$  and $S^1\subset S^4$ with 
  induced geometry   $\ads_5 \times S^1$ \ci{DHoker:2008wvd}, \cf\ also \ci{Lunin:2007ab}. In this case 
 the M5 brane  effective   action   was  proportional to $\vol(\ads_5) = \pi^2 \log (r \Lambda_{\rm IR} )$  and computed 
the  $a$-anomaly of  an $S^4$ defect in $A_{N-1}$  $(2,0)$  theory, \cf\  \ci{Chalabi:2021jud}.
    \fi

  \subsection{Review}

To put the  discussion in a broader context, 
 let  us review   
some  facts about supersymmetric     M-brane probes in $\ads_{7}\times S^{4}$ and $\ads_{4}\times S^{7}$
and their relation to defect anomalies. 
     

Supergravity in 11d admits two special  maximally supersymmetric solutions \cite{Freund:1980xh} -- 
$\ads_{7}\times S^{4}$ (near-horizon limit of a stack of
 M5 branes)  and $\ads_{4}\times S^{7}$ (near-horizon limit of a stack of M2 branes).
The dual  6d and 3d CFT's  have total of 32 supersymmetries. 
For  $\ads_{7}\times S^{4}$ the  bosonic  isometries are $SO(2,6)\times SO(5)\subset OSp(8^{*}|4)$, while
for  $\ads_{4}\times S^{7}$ they   are $SO(2,3)\times SO(8)\subset OSp(8|4, \mathbb{R})$.

 M-brane probe configurations in these backgrounds that  preserve 16 supersymmetries  are listed 
 in Table \ref{tab:probe-supersymmetry} below (see \cite{Lunin:2007ab,DHoker:2008wvd}).\footnote{The case {\bf Ia} was discussed  in \cite{Lunin:2007ab} but was not  mentioned explicitly  in Table 4 of \cite{DHoker:2008wvd}.}
\begin{table}[H]
\be
\def\arraystretch{1.3}
\begin{array}{ccccccc}
\toprule
\textsc{ } & \text{Background} & \text{Probe} &  \text{World-volume}   &   & \text{One-loop correction} \\
\midrule
{\textbf{Ia}} & {\ads_{7}\times S^{4}} & {M5} & {\ads_{3}\times S^{3}} & {S^{3}\subset \ads_{7}}    &\text{here} \\
\textbf{Ib} & \ads_{7}\times S^{4} & M5 & \ads_{3}\times S^{3} & S^{3}\subset S^{4} &   \text{here} \\
\textbf{I'} & \ads_{7}\times S^{4} & M5 & \ads_{5}\times S^{1} &  & \cite{Jiang:2024wzs} \\
\textbf{ I''} & \ads_{7}\times S^{4} & M2 & \ads_{3}&  & \cite{Drukker:2020swu}  \\
\midrule
\textbf{II} & \ads_{4}\times S^{7} & M5 & \ads_{3}\times S^{3} &  & \text{here}  \\
\textbf{  II'} & \ads_{4}\times S^{7} & M2 & \ads_{2}\times S^{1} &  & \cite{Giombi:2023vzu} \\
\bottomrule
\end{array}\notag
\ee
\caption{Brane probes  in $\ads_{7}\times S^{4}$ and $\ads_{4}\times S^{7}$ preserving 16 supersymmetries. 
}
\la{tab:probe-supersymmetry}
\end{table}
We introduced the labels ({\textbf{Ia}}, \Ib, etc.)    for the different probes   that will be used below. 
We also included a column with references to the  computations of the one-loop  corrections to the corresponding  M-brane partition functions. 

Our focus  will be on cases \Ia, \Ib,  \II\   that all have the $\ads_{3}\times S^{3}$   world-volume geometry.
Their  bosonic  isometry  is  $SO(2,2)\times SO(4)\times SO(4)$ 
which is a  part of the  subalgebra of $OSp(8^{*}|4)$ in the cases {\bf Ia, Ib},   and of $OSp(8|4, \mathbb{R})$ in the case {\bf II}.
The  corresponding   supergroups   are given   in Table \ref{tab:preserved-susy}
(which is adapted from \cite{DHoker:2008wvd}).\footnote{In the case {\bf Ia}  the brane is wrapped on $\ads_{3}\times S^{3}\subset \ads_{7}$    thus   having 
$SO(2,2)\times SO(4)$ symmetry,   and    is  also localized at a point in $S^{4}$  leading to the  extra $SO(4)$  factor. 
In the case {\bf Ib}  the additional $SO(4)$ symmetry   comes  from the 3-sphere part of  
$ds_{\ads_{7}}^{2}= L^{2}(du^{2}+\cosh^{2}u\, ds^{2}_{\ads_{3}}+\sinh^{2}u\, ds^{2}_{S^{3}})$, while 
in the  case {\bf II}  it  comes  from the second $S'^{3}$ in 
$ds_{S^{7}}^{2} = L^{2}(d\theta^{2}+\cos^{2}\theta\, ds^{2}_{S^{3}}+\sin^{2}\theta\, ds^{2}_{S'^{3}})$.}
\begin{table}[H]
\be
\def\arraystretch{1.3}
\begin{array}{ccc}
\toprule
\textsc{ } & \text{Background} & \text{Supergroup} \\
\midrule
{\textbf{Ia}} & {\ads_{7}\times S^{4}} &  OSp(4^{*}|2)\times OSp(4^{*}|2)\subset OSp(8^{*}|4)  \\
\textbf{Ib} & \ads_{7}\times S^{4} &  OSp(4^{*}|2)\times OSp(4^{*}|2)\subset OSp(8^{*}|4)  \\
\midrule
\textbf{II} & \ads_{4}\times S^{7} & OSp(4|2,\mathbb R)\times OSp(4|2,\mathbb R)\subset OSp(8|4,\mathbb R) \\
\bottomrule
\end{array}\notag
\ee
\caption{Supersymmetry algebras preserved by the M5 brane probes with world-volume $\ads_{3}\times S^{3}$.}
\la{tab:preserved-susy}
\end{table}
In the cases {\Ia}  and \Ib\   the  boundary of the $\ads_{3}$  part of 
M5 probe represents a 2-dimensional defect   in the dual 6d $(2,0)$ CFT.
In general, one  may distinguish  the  two cases: 

 (i)  the standard 
global $\ads_{7}$  with  the boundary  $S^6$  and thus  with 
the boundary of $\ads_{3}$  being  $S^2$; 

 (ii) ``thermal''  $\ads_{7}$  with the   boundary  $S^1_\b \times S^5$  and   thus  with 
the boundary of the corresponding ``thermal''  $\ads_{3}$  being  $S^1_\b \times S^1$  ($\b$ is the length of the circle).

While here we  will be mostly interested in  the first  case  when $S^2$ 
 corresponds to the spherical Wilson surface defect operator in 6d CFT  (see \ci{Drukker:2020swu}  and refs. there),
 let us add some comments on 
the second (ii) case.
 There  the dual (2,0) CFT   will be  defined on $S^1_\b \times S^5$ 
and thus may be  connected   to the 5d SYM theory  (with $\b$   related  to the YM coupling constant). Then the $S^1_\b \times S^1$  
 defect  may be interpreted in terms of the  $S^1$  Wilson loop (WL) in the 5d SYM.
 
In the case  {\bf I'}  of Table~\ref{tab:probe-supersymmetry}, \ie\  of a single 
M2 brane  probe    in  $\ads_7 \times S^4$,   the  corresponding  WL   should be in the 
 fundamental representation \ci{Beccaria:2023sph}. 
When the  defect  is taken in a  large-rank representation of the gauge group $SU(N)$,  one may conjecture that 
its dual   description  
may be in terms of a single   M5  brane probe    carrying  M2  brane charge and wrapped on $\ads_3$ and also  on  $S^3\subset \ads_7$ for the symmetric representation or $S^3 \subset S^4$  for  the antisymmetric representation  \ci{Mori:2014tca}. 
The  connection to the 
  5d SYM then suggests to compare  
  the  localization prediction   \ci{Mori:2014tca}   for the corresponding 
  $S^1$  WL expectation value  following from the  Chern-Simons  matrix model 
 (\cf\  \ci{Minahan:2013jwa})  to the   M5 brane probe partition function.

 The  large $\b$  limit  of the second (ii) case   is related  also to the ${\rm d}_{2}$ anomaly  coefficient of a surface  defect in the  (2,0) theory.
 \iffa 
 \foot{To recall,  in  the presence of 2d defects in a    CFT  defined on a curved space 
 its  stress-tensor trace anomaly can be written as the sum  of the  ambient  space  contribution 
 and the following  additional term localized on the defect 
 \cite{Graham:1999pm,Henningson:1999xi,Asnin:2008ak,Schwimmer:2008yh,Chalabi:2020iie}:
$ T^{\mu}_{\mu}\big|_{\rm defect} = -\frac{1}{24\pi}\big({\rm b}\,\wh R +{\rm d_{1}} {\Pi}^{\mu}_{ij}\, {\Pi}^{ij}_{\mu}-{\rm d}_{2}\, W\indices{_{ij}^{ij}}\big)$. Here  $\wh R$ is the Ricci scalar for  the induced metric on the defect, 
   ${\Pi}^{\mu}_{ij}$  is  the traceless second fundamental form
  of the defect and $W_{ijk\ell}$ is the pull-back of the Weyl tensor. 
  We follow  \ci{Jensen:2018rxu}  and denote the ``central charge'' coefficient as b. 
  Following \cite{Jensen:2013lxa,Estes:2014hka,Gentle:2015jma} one 
   may  compute holographically the entanglement entropy (EE) of a spherical region centred on the 
 2d defect or of a semi-circle centred on the 2d boundary. 
  After subtracting 
  the EE of the ambient CFT the  coefficient of the  logarithmic in the UV cutoff 
   term  may be denoted as  $\frac{1}{3}{ b}$. For  a CFT$_{d}$
 this  ``central charge'' $b$  obeys \cite{Kobayashi:2018lil,Jensen:2018rxu}
$ {b} = {\rm b}-\frac{d-3}{d-1}{\rm d}_{2}.$}
\fi
  In the present context  of the M5 brane probes  with  the geometry $\ads_{3}\times S^{3}$  
 one may expect that   the  anomaly coefficients b and ${\rm d}_{2}$ 
 can be extracted from the corresponding  free energy $F=-\log Z$  \cite{Chalabi:2020iie}
as follows. 

 If the boundary of $\ads_{3}$ is $S^{2}$, its  regularized volume is $\vol(\ads_{3}) = -2\pi \log (r\Lambda_{\rm IR})$,
 where $r$ is the  radius and $\Lambda_{\rm IR}$ an  IR cutoff (the latter  we shall  not explicitly indicate in what follows). 
 Then  
\ba
\la{1.2}
\partial \ads_{3}=S^{2}:\qquad\qquad \qquad\qquad    F = -\tfrac{1}{3}\text{b}\log r \,.
\ea
If  the boundary of $\ads_{3}$ is $ S^{1}_{\beta}\times S^1$, 
then $\vol(\ads_{3}) = -\ha \pi \b$  and ${\rm d}_{2}$ is proportional to the Casimir energy 
 that  corresponds to  the  large $\b$ asymptotics of  $F$ 
 \be
 \la{1.3}
 \partial \ads_{3}=S^{1}_\b \times S^{1}:\qquad\qquad \qquad \qquad 
 F_{\beta\gg 1} = -\tfrac{1}{12}\,\text{d}_{2}\,\beta\,.
 \ee
The  exact expressions   for ${\rm d}_{2}$  for symmetric and antisymmetric representations 
are similar to the ones  for the b-coefficient given in \rf{5},\rf{6}
\ci{Estes:2018tnu,Jensen:2018rxu,Chalabi:2020iie}  (in general, the analog of \rf{02} is 
${\rm d_2} = 24(\rho, \lambda) +6 (\lambda, \lambda)$,   see also  \rf{H.5}, \rf{H.6})
\ba
\la{1.8}
{\rm d}_{2\, (k)} =12N\, k\, \big(1+\tfrac{1}{2} k N^{-1}\big)\big(1-{N}^{-1}\big)\,,
\qquad \quad 
{\rm d}_{2\, [k]} =12N\, k\, \big(1+\tfrac{1}{2}{N^{-1}}\big) \big(1- {k}{N}^{-1}\big)\,.
\ea
The  same  expressions  were found  in \ci{Mori:2014tca} by a saddle point analysis 
 of the localization   matrix model that computes 
the corresponding  5d SYM   WL expectation value ($F = -\log\langle W\rangle$).  As we will explain 
below in Appendix \ref{WL}, 
the reason why the  saddle point approximation is  enough to get 
the exact value  of  d$_2$  is  because the  subleading at large  $N$ and $\beta$ corrections to the $N,\beta\gg 1$ limit
 turn out  to be  exponentially suppressed. 
 
The semiclassical M5  probe  expansion corresponds to  taking $N$ and $k$ large with $\ka$   defined in \rf{110}   being fixed. 
In this case we  can express \rf{1.8} in the same form as in \rf{15},\rf{150}
\ba
{\rm d}_{2\,(k)} &=24N(N-1) \,  \kappa^2(1+\kappa^2)\  =24N^2 \kappa^2(1+\kappa^2)\,-24N\kappa^2(1+\kappa^2)    \,, \la{1.12}  \\
{\rm d}_{2\,[k]} &= 24N(N + \ha) \,  \kappa^2(1-2\kappa^2)  = 24N^2 \kappa^2(1-2\kappa^2)\,+12N\kappa^2(1-2\kappa^2)   \,. \la{120}
\ea
The leading $N^2$ terms in these expressions 
 are indeed reproduced by the classical actions of an M5 brane probe   in the corresponding cases
\Ia\  and \Ib\  with  $ \partial \ads_{3}=S^{1}_\b \times S^{1}$  \ci{Mori:2014tca}. 
However,  like in the b-anomaly case in  \rf{15},\rf{150}, 
  the semiclassical M5 brane   interpretation of the order $N$   corrections in \rf{1.12},\rf{120}  
remains an open problem. 

In general,  to find the coefficient   d$_{2}$   defined  as in  \rf{1.3} from the semiclassical  M-brane probe perspective 
beyond the leading large $N$ order 
one  may actually   need  to replace the $\ads_7 \times S^4$ 
 background   by its ``twisted'' version (see \cite{Beccaria:2023sph}   and refs. there). 
Indeed,   to get   the  $\b$-dependent  WL expectation value  on the M-theory side
one should start   with an M-brane  in the  product of  the ``thermal'' $\ads_{7,\b}$   and  ``twisted'' $\tilde S^4$ (with one angle $z\to z + i \tau$, where
$\tau \in (0, \beta))$ is the ``time'' coordinate of $\ads_{7,\b}$). This deformation  does  not change
the value of  classical M-brane probe  or   the leading term in $F$ in  the semiclassical expansion (\ref{110}), but 
may alter the   subleading  corrections. 

 This  can  be seen  already  in the $k=1$ case of the fundamental representation, when 
 the WL  expectation value should be reproduced by a   single M2  brane probe wrapped on $\ads_{3,\b}$ \cite{Beccaria:2023sph}.
  According to  \rf{1.8},   in this case we should get  ${\rm d}_{2} =12N -  6 - 6  N^{-1}$.
  The leading term here is the same as in b  in \rf{03}   but the  subleading term is different. 
   This  implies  that the subleading correction  in  ${\rm d}_{2}$ cannot be reproduced by
    the one-loop   M2 brane probe computation in $\ads_7 \times S^4$  as in 
   \ci{Drukker:2020swu} with the only modification  being that  the  $\ads_3$  wrapped by M2
  has  now the  boundary $S^1_{\b \gg 1} \times S^1$ instead of $S^2$.\foot{Since for large $\b$  the space 
    $S^1_{\b \gg 1} \times S^1$  is the same as $\mathbb R\times S^1$   which is conformal to $S^2$ one would  get  
    the one-loop  contribution to the  free energy on $\ads_3$ (which is  a homogeneous space)   
    in  a universal form    $F^{(1)}=  a \vol(\ads_3)$. One would  then 
     find the same ratio of the  leading  (order $N$) and subleading  (order $ N^{0}$) terms in both 
   b and ${\rm d}_{2}$,  but this  is not  the case.}
   In fact, it  was shown in \cite{Beccaria:2023sph} that quantizing M2 brane in 
    the  ``twisted''  $\ads_{7,\b}\times \tilde S^{4}$ background 
  one indeed reproduces the localization result $ \langle W \rangle_{\beta\gg 1} = \exp[(N-\frac{1}{2}+\cdots)\beta]$. 
  Hence,   one finds (\cf\ \rf{1.3}) \ 
  ${\rm d}_{2} = {12}{\beta}^{-1} \log\langle W \rangle_{\beta\gg 1} = 12N-6+\cdots$, which is 
   in agreement with (\ref{1.8}).

\subsection{Summary} 

As already stated above, our  aim below   will be to
 compute the   one-loop   corrections  to the  free energies of the $\ads_{3}\times S^{3}$
M5  brane probes 
in the  three  cases  {\Ia},  \Ib\ and \II\  with 
the boundary of $\ads_3$  assumed to be  $S^2$.\foot{Note that the  general 
expressions for the quadratic  fluctuation  actions
and the resulting structure of the  one-loop  partition functions derived below 
   will apply   also to the case of $\ads_3$ with  $S^1_\b \times S^1$ boundary.}
 
 We will   start with the  M5 brane action   and expand  it to quadratic order 
in fluctuations near a given   classical solution. 
The presence of a non-zero  $H_3$ background will  introduce a non-trivial  complication   due to  a mixing between 
the fluctuations of the $B_2$  potential   and one of the scalar  coordinates. 

Expanding  all the 6d fluctuation fields 
 in modes on $S^3$ (labelled by level $\ell$) we  will   ``diagonalize'' that mixing   and 
 then organize the towers of fluctuations 
 into a   collection of  massive short  supermultiplets on $\ads_{3}$  thus   maintaining the underlying supersymmetry. 
We  will   then  evaluate the resulting one-loop $\ads_{3}$
  determinants  using the  standard relations for the case when  the $\ads_{3}$ boundary is $S^2$. 
Summing  all the  contributions together,  we will find that there are many non-trivial   cancellations
 with the 
final results   for the 
 one-loop  corrections to the free energies   given  simply  by
\be\la{18}
F^{(1)}_{\bf Ia} = F^{(1)}_{\bf Ib}  = -\tfrac{3}{2\pi} \vol(\ads_3) \sum_{\ell=1}^\infty \ell \,.
\ee
 Note  that despite  the  theory on M5 brane  probe being a 6d  one in a non-trivial background,
   there are no logarithmic 
divergences.\foot{We will   verify this  independently by showing that 
 the corresponding  Seeley coefficient vanishes. The same was found 
 also  in the  case of the M5 brane  probe   with the $S^1 \times S^5$ \ci{Beccaria:2023cuo} 
and the $S^1 \times \ads_5$ \ci{Jiang:2024wzs} 
 geometries  (with $H_3=0$)  which are conformally flat and have zero 6d Euler density.
   Note that logarithmic  divergences are automatically absent at one-loop   level in the case 
of the 3d theory on the M2 brane  (as illustrated, \eg, by the  examples  considered   in \ci{Giombi:2023vzu,Beccaria:2023sph}).
} 
This is of course a necessary requirement for being able to compare   M5 brane free energy with the one 
that determines the defect anomaly.

The sum over $\ell$ in \Ia\  and \Ib\  cases 
  is   quadratically divergent and  thus  the final result   depends on a  choice of   regularization.
  Similar    divergent sums requiring a specific regularization appeared in related  contexts, see,  \eg, \ci{Mansfield:2003gs,ArabiArdehali:2013vyp,Beccaria:2014xda} (\cf\ also a  discussion in  \ci{Bobev:2023dwx}).
 One regularization procedure used in the past  that  led to  consistent results is to introduce 
  a  sharp cutoff $\ell \leq \Lambda$  and drop all power divergent terms  (see a discussion below \rf{4.12}). 
  Adopting it here  we conclude that the coefficient in \rf{18}   should be   set to zero, so that 
  \be 
  \la{189}
  F^{(1)}_{\bf Ia} = F^{(1)}_{\bf Ib}  =0\,. \ee
    This  conclusion is   then consistent   with  the  fact that  the 
   exact   expressions for the ``central charge'' coefficients \rf{15} and \rf{150} 
corresponding to the cases \Ia\  and \Ib\ 
do not contain  order $N^0$  term.

 
In the case {\bf II} in Table 1 where the M5 brane 
 is embedded into $\ads_{4}\times S^{7}$ the classical action takes the form (\cf\ \rf{110},\rf{15},\rf{1.2})
\be
\la{1.9}
F^{(0)}_{\bf II} =\tfrac{1}{4\pi} \big(N-\tfrac{1}{2}\kk^{2}\big) \vol(\ads_{3})=
\tfrac{1}{4\pi} N  \big(1-\vk^{2}\big) \vol(\ads_{3})
\,, \qquad \ \ \ \ \te  \vk^2= {\kk^2\ov 2 N} \,.
\ee
Here $\kk$  is  an integer  parameter  of  $H_3$  that 
   has  an    interpretation of the  M2  brane charge carried by M5 brane  \ci{Lunin:2007ab}  
   with $\vk^2$ being the fixed semiclassical parameter  (here the 5-brane tension is proportional to $ N$). 
In contrast to the cases 
\Ia\  and \Ib,  the limit  of $\kk=1$    should not have a description in terms of a single M2  brane 
embedded into $\ads_{4}\times S^{7}$  as the   intersection of  M2 branes (of a probe one  with $N$ copies at the boundary)
 over a 2-surface is not a  BPS one. 
 The   computation of the M5 brane one-loop correction to \rf{1.9}  is similar to the  cases \Ia\  and \Ib\  and gives 
\be 
F^{(1)}_{\bf II} = 0 
\,. \la{20}
 \ee
Here the  vanishing of the one-loop  contribution   happens  before the summation over $\ell$, \ie\ 
at each  $S^3$ 
level $\ell$ independently:
 the contributions of states in the $\ads_3$ supermultiplet  with fixed $\ell$   cancel each other. 
 

\

The  plan of the rest of this  paper is  as follows. 
In section 2  we will present the three  M5 brane classical solutions  corresponding to the probes \Ia,\, \Ib\  and \II\  in Table 1
that have \adsst  world volume geometry  with $\del \ads_3=S^2$. We will compute the corresponding  values of the M5 brane action reproducing the leading large $N$ terms in the corresponding defect anomaly coefficients in  \rf{5} (\Ia)   and \rf{6} (\Ib) 
and also obtaining \rf{1.9}. 

In section 3 we will  study the  quadratic fluctuations of the bosonic   fields in  the M5 brane action 
near the three  probe solutions (the fermionic fluctuation operators will be found in Appendix \ref{fer}).  
We will derive the general expressions for the corresponding  fluctuation determinants
that appear in the one-loop  M5 brane partition function. 
We will then expand the 6d fluctuation fields in modes on $S^3$ 
 and present the corresponding mass  and scaling dimension spectra  of  the KK towers of fields on $\ads_3$. 

In section 4 we  first organize these $\ads_3$   fields into  supermultipets corresponding to the supergroups 
which represent  the required symmetry in each of the three cases (details of these will  be discussed in Appendix \ref{mult}). 
We will then  compute the  corresponding free  energies deriving the expressions in \rf{18} and \rf{20}. 

Some open   questions will be mentioned  in section \ref{op}.
In Appendix \ref{apz} we will work out  the explicit form  of the  one-loop  partition function of the 
gauge-invariant  rank 2 antisymmetric tensor defined on \adsst space with generic radii. 
In Appendix~\ref{s34} we will  comment on  a  close analytic continuation  relation between the 
fluctuation spectra in the \Ib\ and \II\ cases. 

In Appendix \ref{apdi} we will  discuss the  structure  of UV  divergences 
of the one-loop free energies,   explaining, in particular, why the logarithmic   divergences 
 are absent separately  in the  contributions of each of the 6d fluctuation fields. 
In Appendix~\ref{pfo}  we will recall  some  facts about   spectra of  $p$-form Laplacians on  $S^d$ 
and their decompositions into transverse and longitudinal parts. 
In Appendix \ref{cas}  we will  discuss  the values of the Casimir energies and the expressions for 
``thermal'' single particle partition functions associated with the  $\ads_3$  fluctuation field supermultiplets 
presented in Table \ref{muld}. 

Finally, in Appendix \ref{WL} we will  discuss  the large $N$  expansion of  the  5d SYM Wilson  loop expectation value 
in the symmetric or antisymmetric
representation of  $SU(N)$  represented by the  localization matrix model integral. We will 
 demonstrate that in  the  large $\beta$ limit  the 
saddle-point result of \ci{Mori:2014tca}   that  matches
the  expressions for $\dd_2$ in \rf{1.8} is, in fact,  exact up to exponential corrections.

 \section{Classical  solutions and actions for  M5 brane  probes  \la{s2} }
 

The bosonic part of the  PST  action for an M5 brane   in 11d supergravity background is given by  \cite{Pasti:1997gx,Bandos:1997ui,Ko:2013dka} (see also 
\ci{Aganagic:1997zq,Howe:1997fb}) 
\bea
\la{2.1}
S &=\ T_{5}\Big[\int d^{6}\xi\ \Big(-\sqrt{-|G_{ij}+\hat {\sfH}_{ij}|}+\frac{\sqrt{-|G|}}{4(\partial a)^2 }\partial_{i}a\, {\star \sfH}^{ijk}{\sfH}_{jk\ell}\, \partial^{\ell}a\Big)  +\int\big(\tfrac{1}{2}H_{3}\wedge C_{3}+C_{6}\big)\Big]\,,
\eea
where $T_{5} = \frac{1}{(2\pi)^{5}\ell_{P}^{6}}$   and 
 $G_{ij}$, $C_{3}$ and $C_{6}$ are pull-backs of the supergravity background  fields  to the 
  world volume  ($i,j,\, \ell,...=0,1,\dots, 5$)
  \be\la{21}
G_{ij}=G_{MN}(X(\xi))\,\partial_{i}X^{M}\partial_{j}X^{N}, \qquad
C_{ijk} = C_{MNL}(X(\xi))\,\partial_{i}X^{M}\partial_{j}X^{N}\partial_{k}X^{L} \,. 
\ee
$X^M(\xi)$ and $H_3= d B_2$ are the dynamical  world-volume fields  while  the scalar $a(\xi)$   will be gauge-fixed   as  
\be
a(\xi) = \xi^{1} \ \la{211}.
\ee
We use the  definitions $(\partial a)^2 = G^{ij} \del_i a \del_j a$  and\footnote{${\star \sfH}$ in \rf{2.1} 
 is 6d dual 3-form; we will use    $\star$ to denote also 11d dual forms. 
  The 6d  antisymmetric tensor   with raised  indices is  assumed to be 
   numerical with $\eps^{012345}=+1$, while 
   $\eps_{i_{1}\cdots i_{6}} $ is given by $G\, \eps^{i_{1}\cdots i_{6}}$,  where $G\equiv |G|=\det G_{ij}$.
   In components,  we thus have
   \(
   6 \sqrt{-|G|}\,  (\star \sfH)^{ijk}= \varepsilon^{ijkmnr}\sfH_{mnr}
   \).
   } 
\be
\la{2.3}
{\sfH}_{ijk} = H_{ijk}-C_{ijk}\,, \qquad\qquad \ \  \hat{\sfH}^{ij} = \frac{1}{6\sqrt{-|G|}}\frac{1}{\sqrt{-(\del a)^2 }}\ \eps^{ijk\ell mn}\partial_{k}a\  {\sfH}_{\ell m n }\,.\ee
The 11d field $C_6$ is defined   in terms of  the dual of $F_4$ 
\be 
dC_{6}=\star F_{4}-\tfrac{1}{2}C_{3}\wedge F_{4}\,,\qquad \qquad \ \ F_{4}=dC_{3}\,.\la{24}
\ee

\subsection{
M5  branes wrapping  $\ads_{3}\times S^{3}\subset \ads_{7}\times S^{4}$
\la{s21}}

We will  parametrize the $\ads_{7}\times S^{4}$ background as\foot{We will use the notation $\vol_M$  for  
the  volume form of  a space $M$ and $\vol(M)$   for its integral.}
\ba
\la{2.5}
ds_{11}^{2}&= L^{2}(du^{2}+\cosh^{2}u\, ds^{2}_{\ads_{3}}+\sinh^{2}u\, ds^{2}_{S^{3}})+\tfrac{1}{4}L^{2}ds^{2}_{S^{4}}\,,\\
\la{2.6}
F_{4} &= dC_{3} = \tfrac{3}{8}L^{3}\vol_{S^{4}}\,, \qquad \qquad L^{3} = 8\pi N\ell_{P}^{3} \,. 
\ea
The corresponding dimensionless tensions of M2 and M5 brane probes in this background are then 
\be\la{27}
 \T_2 = L^3 T_{2}=\tfrac{L^3}{(2\pi)^{2}\ell_{P}^{3}} = \tfrac{2}{\pi }N\,,\qquad\qquad 
 \T_{5} =L^6 T_5= \tfrac{L^6}{(2\pi)^{5}\ell_{P}^{6}}=   \tfrac{1}{2\pi}(\T_{2})^{2} = \tfrac{2}{\pi^{3} }N^{2}\,.
\ee
We may   assume   the  boundary of $\ads_3$ to be $S^{2}$   so that (in Euclidean coordinates)
  \be ds^{2}_{\ads_{3}} = dw^{2}+\sinh^{2}w\, ds^{2}_{S^{2}}\,. \la{28}  \ee 
  We may  also  
    consider the  ``thermal''   case  of $\ads_{3,\b} \subset \ads_{7,\beta}$ 
    with the  boundary  $S^{1}_\b\times S$  when  
\be
 ds^{2}_{\ads_{3}} = dw^{2}+\cosh^{2}w\, d\tau^{2}+\sinh^{2}w\, d\phi^{2}\,, \qquad\qquad 
\tau\in(0,\beta), \ \ \phi\in(0,2\pi)\,. \la{29}
\ee
In the limit $\b\to \infty$ when the boundary of \rf{29}  becomes $\mathbb  R\times S^1$  the metrics \rf{29}  and \rf{28} are related   by a coordinate transformation. The   Minkowski  signature  version  of \rf{29} with the boundary $\mathbb  R\times S^1$ is 
\ba\la{30}
ds^{2}_{\ads_{3}} &= -\cosh^{2}w\, dt^{2}+dw^{2}+\sinh^{2}w\, d\phi^{2}.
\ea
The metric of  $S^{4}$ in \rf{2.5}   can be represented as  
\ba
\la{2.10}
ds^{2}_{S^{4}} &= d\theta^{2}+\sin^{2}\theta\, ds^{2}_{S'^{3}}\,, \qquad\qquad  \theta\in(0,\pi)\,.
\ea
We will consider the  M5 brane probes that  wrap $\ads_{3}\times S^{3}$ with  $S^{3}\subset \ads_7$   (``probe \Ia'') 
 or  with  $S^{3}=S'^3 \subset S^{4}$  (``probe \Ib'').
 
{
In this subsection  we shall  label the coordinates in \rf{2.5},\rf{30},\rf{2.10}  as 
\be
\def\arraystretch{1.3}
\begin{array}{ccccc}
\toprule
\ads_{3} (t,w,\phi ) &  S^{3} & \   u\  & \theta \ &   S'^{3}  \\
 0,1,2 & 3,4,5 &  \  6 \ & 7\ &  8, 9, \natural\\
\bottomrule
\end{array}\la{lab1}
\ee
}
\subsubsection{Probe \Ia \la{s211}} 

The  solution for the M5 brane wrapped on $\ads_{3}\times S^{3} \subset \ads_7 $ was found in \cite{Lunin:2007ab}. 
As follows from \rf{2.6},  we have  $F_{4}\wedge C_{3}=0$ and then \rf{24} gives\foot{\, $e^A$ is the canonically normalized   basis  of 
1-forms for the metric \rf{2.5}.}
\ba
\la{2.11}
dC_{6} &= \star F_{4} = \tfrac{3}{8}L^{3}(\tfrac{2}{L})^{4} \star\big(e^{7}\we e^{8}\we e^{9}\we e^{\natural}\big)
= 6L^{6}\cosh^{3}u\, \sinh^{3}u\  du\wedge \vol_{\ads_{3}}\wedge \vol_{S^{3}} \,, \\
 \la{2.12}
C_{6} &= L^{6}(\sinh^{6}u+\tfrac{3}{2}\sinh^{4}u)\, \vol_{\ads_{3}}\wedge \vol_{S^{3}}\,,
\ea
where we fixed the integration constant  so that $C_6(u=0) =0$.

The  relevant BPS M5 brane  solution  is obtained  by  identifying the 
 coordinates of  $\ads_{3}\times S^{3}$ (labelled by 0,1,2,3,4,5 above) with  the  world-volume coordinates  $\xi^i$ 
  and  also assuming that  it is  localized at one point in $S^{4}$, \ie\   at $\theta=0$. 
Then the  remaining $\ads_{7}$ coordinate $u$   should be  fixed to a constant value $u_0$ and  $H_3$
 should be chosen to be  proportional to 
the volume form of $S^{3}$
\be
\la{2.13}
u=u_{0}\,, \qquad \sinh u_{0} = \kappa\,, \qquad \theta =0 \,, \ \ 
 \qquad H_{3}=\kappa^{2}L^{3}\vol_{S^{3}}\,, \qquad  \kappa^{2}=\tfrac{k}{2N}  \,. 
\ee
Here $\kappa$ is a free parameter.  It is related to  $k\in\mathbb Z$  which 
  is the M2  brane charge that is carried by the  M5 brane  due to 
 the $H_3$ flux  through $S^3$ being non-zero \cite{Camino:2001at,Lunin:2007ab}
\be \la{213}
T_{2}\int_{S^{3}}H_{3} = 2\pi k  \,, \qquad {\rm \ie}  \ \ \ \ \ \ \    2 \pi \ka^2 \T_2 =k \,.
\ee
Here we used  \rf{27} and that $\vol(S^{3})= \int \vol_{S^{3}} =2\pi^{2}$. 
The resulting induced \adsst  world-volume  metric and  the 6-form  in \rf{2.12}  are given by 
\ba
\la{2.15}
ds^{2} &= L^{2}\Big[(1+\kappa^{2})ds^{2}_{\ads_{3}}+\kappa^{2}ds^{2}_{S^{3}}\Big]\,, \\
C_{6} &=  \kappa^{4}(\kappa^{2}+\tfrac{3}{2})\, L^{6}\,  \vol_{\ads_{3}}\wedge \vol_{S^{3}}.\la{215}
\ea
As the M5 brane is localised in $S^4$, the pull-back of  $C_3$ in \rf{2.6}  is  zero
so  that in \rf{2.3} 
\be \la{2155}
\sfH_{3} = H_{3}=  \kappa^{2}L^{3}\vol_{S^{3}} \,. \ee 
As a result, the ${\star \sfH} {\sfH}$ term in the action \rf{2.1}   vanishes. 

We shall use the notation $g_A$ and $g_S$ for the  unit-radius metrics on $\ads_3$  and $S^3$, so that 
\rf{2.15} implies 
\be\la{2.18}
\sqrt{-|G|}= L^{6}\kappa^{3}(1+\kappa^{2})^{3/2}\sqrt{-g_{A}}\sqrt{g_{S}} \,. 
\ee
The components of the tensor $\hat\sfH^{ij} $ defined in \rf{2.3} then are (using the gauge condition \rf{211})
\ba
\la{2.19}
\hat\sfH^{ij} &= \frac{1}{6\sqrt{-|G|}}\sqrt{-G_{11}}\eps^{ij1\ell mn} {\sfH}_{\ell m n } 
= \frac{i}{L^{2}\kappa(1+\kappa^{2})\,  \sqrt{-g_{A}}    }\eps^{ij1} \,, 
\\
\la{2.20}
\hat\sfH_{02} &= -\hat\sfH_{20} =  G_{00}G_{22}\hat\sfH^{02}  
=iL^{2}\tfrac{1+\kappa^{2}}{\kappa}\,  \sqrt{-g_{A}}  \,. 
\ea
Thus
\ba
&G_{ij}+\hat\sfH_{ij} =\begin{pmatrix}
-L^{2}(1+\kappa^{2})\cosh^{2}w & 0 & iL^{2}\frac{1+\kappa^{2}}{\kappa}\sinh w\cosh w & 0 \\
0 & L^{2}(1+\kappa^{2}) & 0 & 0 \\
-iL^{2}\frac{1+\kappa^{2}}{\kappa}\sinh w\cosh w & 0 & L^{2}(1+\kappa^{2})\sinh^{2}w  & 0 \\
0 & 0 & 0 & L^{2}\kappa^{2}g_{S}
\end{pmatrix}\,,\notag  \\
\la{2.21}
&\sqrt{-|G_{ij}+\hat\sfH_{ij}|} =  L^{6}\kappa^{2}(1+\kappa^{2})^{2}\sqrt{-g_{A}}\sqrt{g_{S}} \,, \qquad \qquad  
 \sqrt{-g_{A}} = \sinh w\cosh w\,. 
\ea
Using \rf{2.21} and \rf{215}  we get from \rf{2.1} the following value for the classical M5 brane action\foot{Let us note that 
 the value of the classical action for the M5  wrapped on AdS$_3$  with $S^2$ boundary   and on 
$S^3$ (probe \Ia)  or on $S'^3$  (probe \Ib) 
 was not explicitly computed in  \ci{Lunin:2007ab}. The  observation  that  it matches 
 the  large $N$ value of  the $\bb$    defect anomaly  coefficient  \ci{Estes:2018tnu,Jensen:2018rxu} 
  for the  $k$-symmetric or $k$-antisymmetric representations   effectively  follows  from 
     the entanglement entropy computation in  \ci{Rodgers:2018mvq}.} 
\be
\la{2.24}
S=-T_{5}L^{6}\kappa^{2}(1+\tfrac{1}{2}\kappa^{2})\int \vol_{\ads_{3}}\we\vol_{S^{3}}.  
\ee
Continuing to the  Euclidean signature  ($S\to -S_E$) 
 and  
assuming that the  $\ads_3$   has metric \rf{28}  with $S^2$ as its boundary we get 
for the tree-level  contribution to the M5 brane free  energy 
\be\la{2.241} 
F^{(0)}_{\bf Ia} = S_{\rm E} =  \T_{5} \kappa^{2}(1+\tfrac{1}{2}\kappa^{2}) \vol(\ads_{3})\vol(S^{3}) 
= - 8 N^2 \ka^2 (1+\tfrac{1}{2}\kappa^{2})  \log r \,, 
\ee
where we used \rf{27}  and  that $\vol(\ads_{3})= -2\pi \log r$. 
Comparing this with \rf{1.2}   we conclude that  the corresponding leading large $\T_5$  or large $N$, fixed $\ka$ 
 contribution 
to the b-anomaly coefficient   is  
\be \la{2.25} 
\bb^{(0)}_{\bf Ia}= 24 N^2 \kappa^{2}(1+\tfrac{1}{2}\kappa^{2}) = 12 k N\big(1+\tfrac{1}{4} k {N}^{-1}\big)\,, 
\ee
which reproduces the leading term in \rf{15}. 

Naively,  one could expect to get  the same  value of the action  (\ref{2.24})
in the case when $\ads_3$ has $S^1_\b \times S^1$ boundary,   now with 
$\vol(\ads_{3}) = -\frac{1}{2}\pi \beta$ (see, \eg,  (A.3) in  \cite{Beccaria:2023sph}). 
It turns out, however, that  there is a subtlety -- one is to use a different gauge choice for $C_6$ 
 \cite{Mori:2014tca,Rodgers:2018mvq}.  
This then   gives \rf{2.24} with $1+\tfrac{1}{2}\kappa^{2}\to 1+\kappa^{2}$  or\foot{Ref. 
 \cite{Mori:2014tca} used  different coordinates  in which 
$
ds^{2}_{\ads_{7}} = L^{2}y^{-2}(dy^{2}+dr^{2}+r^{2}d\varphi^{2}+dr'^{2}+r'^{2}dS_{3}), $ 
$F_{7}= \star F_4 = d C_6=  6L^{6}\vol_{\ads_{7}} = 6L^{6}y^{-7}rr'^{3}\ dy\wedge dr\wedge d\varphi\wedge dr' \wedge \vol_{S^{3}}$,  
 and  have chosen 
$C_{6} = -L^{6}y^{-6}rr'^{3}\ dr\wedge d\varphi\wedge dr' \wedge \vol_{S^{3}}$. 
This gauge choice does not respect  the $\ads_{3}$  symmetry  but is  justified by the condition that the resulting M5 solution with
$\ads_{3}$ having flat $\mathbb R^{2}$ boundary has zero action, as expected for a flat defect. Then  transformed to the coordinates used  here 
one gets (\cf\ \rf{215})  $ 
C_{6} = L^{6}\kappa^{4}(1+\kappa^{2})\ \vol_{\ads_{3}}\wedge \vol_{S^{3}}+\delta C_{6}$
where $\delta C_{6}$ vanishes on the classical M5 brane solution.
This leads to the  value of the on-shell action in \rf{226}.} 
\ba\la{226} 
S_{\rm E} &= \T_{5}\kappa^{2}(1+\kappa^{2}) \vol(\ads_{3})\vol(S^{3})
= -2N^{2}\kappa^{2}(1+\kappa^{2})\,\beta = - N k\big(1+\tfrac{1}{2} k N^{-1}\big)\,\beta\,.
\ea
Comparing to \rf{1.3}  this  reproduces 
 the leading semiclassical ($N,k\gg 1$,  fixed $k/N$)   contribution 
to  the $\text{d}_{2}$  anomaly coefficient  that matches the large $N$ term
 in  ${\rm d}_{2\,(k)} $ in  \rf{1.8},\rf{1.12}.\foot{The value of the M5   probe action
   in the case of AdS$_3$  with $S^1_\b\times S^1$  boundary 
  was matched to the  leading-order  matrix model result
  (the ``Casimir energy'' or linear in $\b$  part of the  WL 
expectation value)   in \ci{Mori:2014tca}  but it  was not  explicitly acknowledged earlier 
    that this
   coefficient is   the  same as the $\rm d_2$    coefficient (as expected for  the  $S^1_\b\times S^1$  defect  \ci{Chalabi:2020iie})
  for  both symmetric  and antisymmetric   representations    \ci{Estes:2018tnu,Jensen:2018rxu}.}

\subsubsection{Probe \Ib \la{s212}} 

The  M5   brane  wrapped on  $\ads_3 \subset \ads_7$ and $S^3= S'^3  \subset S^4$ was 
 discussed in \cite{Lunin:2007ab,Chen:2008zzr,Mori:2014tca}. 
From $F_{4}$ in (\ref{2.6}) and \rf{2.10} we get 
\be\la{227}
F_{4} = dC_{3} = \tfrac{3}{8}L^{3}\sin^{3}\theta\ d\theta\wedge \vol_{S'^{3}}
\,, \qquad \qquad 
C_{3} = \tfrac{1}{8}L^{3}(2-3\cos\theta+\cos^{3}\theta)\ \vol_{S'^{3}}.
\ee
The M5  brane solution is represented by  (\cf\ \rf{2.13}; we do not put prime on world-volume $S^3$) 
\be
u=u_{0}=0\,, \ \ \ \ \  \theta=\theta_{0}\,, \ \ \ \ \ \cos\theta_{0}=1-4\kappa^{2}\,, \qquad \ \ \
H_{3}=\kappa^{2}L^{3}\vol_{S^{3}}\,, \qquad \ \  \kappa^{2}=\tfrac{k}{2N}  \,. \la{228}
\ee
Here $k \leq N$ is an integer fixed again from the  quantization condition
of the M2 brane charge  carried by the  M5 brane (\cf\ \rf{2.13},\rf{213}). 
The induced metric is 
\be\la{229}
ds^{2} = L^{2}\big(ds^{2}_{\ads_{3}}+\tfrac{1}{4}\sin^{2}\theta_{0}ds^{2}_{S^{3}}\big)
=  L^2 \big[  ds^{2}_{\ads_{3}}+  2 \ka^2 ( 1 -2 \ka^2) ds^{2}_{S^{3}}\big]\,.
\ee
Here 
\be
\la{2.37}
{\sfH}_{3} = H_{3}-C_{3} = 
L^{3}f(\kappa) \vol_{S^{3}}\,, \qquad \ \ \ \ \  f(\kappa) = \kappa^{2}(1-2\ka^2) (1-4 \ka^2) \,. 
\ee
On this solution $C_{6}=0$, $C_{3}\wedge H_{3}=0$ and the classical action \rf{2.1}   is  given only 
 by  the  first ``volume'' term  (the  ${\star \sfH} {\sfH}$ term in \rf{2.1}  again  does not contribute). 
As in  \rf{2.21},\rf{2.24} we then find    
\ba
&\sqrt{-|G_{ij}+\hat\sfH_{ij}|} = L^{6}\kappa^{2}(1-2\kappa^{2})\sqrt{-g_{A}}\sqrt{g_{S}}\,, \\
&
S = -T_{5}L^{6}\kappa^{2}(1-2\kappa^{2})\int \vol_{\ads_{3}}\we\vol_{S^{3}} \,. \la{232} 
\ea
As a result (\cf\ \rf{2.241}) 
\be\la{2.38} 
F^{(0)}_{\bf Ib} = S_{\rm E} =  \T_{5} \kappa^{2}(1-2\kappa^{2}) \vol(\ads_{3})\vol(S^{3})
 = \tfrac{4}{\pi}N^{2} \kappa^{2}(1-2\kappa^{2}) \vol(\ads_{3})\,.
\ee
 In the  case when   $\partial \ads_{3}=S^{2}$  comparing to \rf{1.2} we then get 
\be\la{2.39} 
\bb^{(0)}_{\bf Ib}= 24 \kappa^{2}(1-2\kappa^{2}) =  12Nk\big(1-{k}{N}^{-1}\big) \,, 
\ee
which is thus  in agreement with the leading  large $N$  term in \rf{6},\rf{150}. 

In the case when   $\partial \ads_{3}=S^{1}_{\beta}\times S^{1}$   the action  has the same form \rf{232}
(here there is no subtlety  with  a different  gauge choice of $C_6$). 
Using again that $\vol(\ads_{3}) = -\frac{1}{2}\pi \beta$  and comparing   with \rf{1.3},\rf{120}  we 
reproduce the first leading large $N$ term in  ${\rm d}_{2\,[k]}  $ in \rf{120}
which   happens to be   the same as in  ${\rm b}_{[k]}  $ in \rf{120},  \ie\ is equal to $\bb^{(0)}_{\bf Ib}$ in \rf{2.39}.

\subsection{M5 brane  wrapping  $\ads_{3}\times S^{3}\subset \ads_{4}\times S^{7}$: probe \II  
\la{s32}}

The $\ads_{4}\times S^{7}$   background is described by (\cf\ \rf{2.5},\rf{2.6})
\ba
\la{2.45}
ds_{11}^{2} &= \LL^{2}(du^{2}+\cosh^{2}u\, ds^{2}_{\ads_{3}})+4\LL^{2}(d\theta^{2}+\cos^{2}\theta\, ds^{2}_{S^{3}}+
\sin^{2}\theta\, ds^{2}_{S'^{3}})\,, \\
\la{2.46}
F_{4} &= dC_{3} = 3\LL^{3}\,\cosh^{3}u\, du\wedge\vol_{\ads_{3}}\,, \qquad \qquad \ \ \  \LL^{6}=\tfrac{1}{2}\pi^{2}N\,\, \ell_{P}^{6} \,,  \\
\la{2.47}
C_{3}&= \LL^{3} \ff(u) \vol_{\ads_{3}}\,, \qquad\qquad  \ff(u) = 3\sinh u+\sinh^{3}u \,, 
\ea
where $N$ is the number of M2 branes that  form the $\ads_{4}\times S^{7}$ background.
The  effective 2-brane and 5-brane tensions   here are  (\cf\ \rf{27}) 
\be
\la{2.48}
 \T_{2} =\LL^3 T_2=  \tfrac{1}{4\sqrt 2 \pi} \sqrt N \,, \qquad \qquad 
 \T_{5} =\LL^6 T_5=  \tfrac{1}{2\pi} (\T_2)^2= \tfrac{1}{64\pi^{3}} N \,.  
\ee
In this case  we shall  label the coordinates in \rf{2.45}  as (\cf\ \rf{lab1})
\be
\def\arraystretch{1.3}
\begin{array}{ccccc}
\toprule
\ads_{3} (t,w,\phi ) & u &  \theta & S^{3} &   S'^{3}  \\
 0,1,2 & 3 & 4 &  5,6,7 &  8, 9, \natural\\
\bottomrule
\end{array}\la{lab2}
\ee
The  corresponding   solution for  the  $\half$-BPS  M5 brane  probe wrapping $\ads_{3}\times S^{3} $ 
 was  found  in \cite{Lunin:2007ab}; an equivalent 
solution (in different coordinates) was  constructed 
in \cite{Arean:2007nh}. 
Here one has (\cf\ \rf{2.13},\rf{228})\footnote{For simplicity, we set to zero the  free parameter multiplying   a possible
 ``electric'' ($\sim \vol_{\ads_{3}}$)  term in $H_{3}$ \cite{Lunin:2007ab}.}
\ba
\la{2.49}
u&=u_{0}\,,\qquad \sinh u_{0} = \varkappa\,, \qquad  
\theta=0\,,\qquad H_{3} = -8\LL^{3}\vk\vol_{S^{3}}\,.
\ea
Like in \rf{213}  the free parameter $\vk$   can be   expressed in terms of an  integer 
M2 brane  charge $\kk$ carried by  the M5 brane (\cf\  \rf{2.48})\footnote{The  notation  for the parameter  determining $u_0$ 
 used in  
\cite{Lunin:2007ab} was  $b = 4\varkappa$ and we corrected a typo there.} 
\be\la{240} 
\vk = {\kk \ov \sqrt{2N}} \,, \qquad \qquad 
8  \vk \T_{2} \vol(S^{3}) = 2 \pi \kk \,. 
\ee
Then the induced metric  and ${\sfH}_{3}$ are given by 
\ba\la{241}
& ds^{2}=\LL^{2}\big[ (1+ \vk^2) \ ds^{2}_{\ads_{3}}+4ds^{2}_{S^{3}}\big] \,, \\
&
\la{2.51}
{\sfH}_{3} = H_{3}-C_{3} =- \vk\,  \LL^{3}\,\big[(3 +  \vk^2) \vol_{\ads_{3}} + 8\vol_{S^{3}}\big] \,. 
\ea
Here the projection of $F_4$  and thus of $dC_6$ to the brane are  trivial   so that 
 we  may  take $C_{6}=0$. 
 Then the  M5 brane action \rf{2.1}  is given by (here $\ff(u_{0})=\vk(3 + \vk^2)$)
\ba
\la{2.53}
S = -\T_{5}\int\Big(  \big[8 (1+\vk^{2})^2  - 4\vk\, \ff(u_{0})\big] - 
{4\vk\, \ff(u_{0})}
\Big) \ \vol_{\ads_{3}}\wedge \vol_{S^{3}} \,, 
\ea
where the first term is the  ``volume'' part contribution,  the second is that of the ${\star \sfH} {\sfH}$  term 
 and the 
 last $4\vk\, f(u_{0})$  one   comes from the WZ  part $ C_{3}\wedge H_{3}$. 
The  resulting Euclidean action or the classical contribution to the  M5 brane free energy is  (\cf\ \rf{2.24},\rf{2.38}) 
\ba
\la{2.54}
F^{(0)}_{\bf II}= S_{\rm E} &=\tfrac{1}{4\pi}  N (1  -  \vk^2) \vol(\ads_3)  =  \tfrac{1}{4\pi} ( N - \ha \kk^2) \vol(\ads_3) 
\,.
\ea 
This expression was already given in \rf{1.9}. 


\section{Quadratic fluctuations and one-loop  partition function \la{s3} }  

In this  section we  will    study the 
quadratic fluctuations  near the  three  M5 probe  solutions  {\bf Ia, Ib}  and {\bf II}
and derive the general expressions for the corresponding one-loop  fluctuation determinants. 
We will then expand the  6d fluctuation  fields  in modes on $S^3$  and present the corresponding mass spectra
for the KK modes in $\ads_3$. 

Here we  will   give details about  the  bosonic fluctuations while the  fermionic fluctuation operator 
  will be discussed in  Appendix \ref{fer}.



\subsection{Probe \Ia \la{s31}}

We shall assume the static gauge in which the M5  coordinates along the 
$\ads_3\times S^3$ are identified with the  world-volume ones $\xi^i$, \ie\  they will not be  fluctuating. 
The   bosonic fluctuations   will be those  of the coordinate $u$  in \rf{2.5},  the  coordinates of  $S^{4}$, and of  the 
 2-form  field  defining  $H_{3}$ (\cf\ \rf{2.13}). 

We shall denote the fluctuations of the five   transverse coordinates  as $U$ and $\z_p$ where 
 \be \la{3.2} 
 u= u_0 + U \,, \qquad \qquad 
 ds^{2}_{S^{4}} = \frac{d\z_{p}d\z_{p}}{(1+\frac{1}{4}\z^{2})^{2}}=  d \z_p d\z_p + ...\,, \qquad\ \ \  p=1,2,3,4 \,. 
\ee
Then the expression for the  induced metric $G_{ij}$  in \rf{21}
including the  second-order terms  in  fluctuations 
may be written as 
\ba
 ds^{2} ={} & G_{ij} d \xi^i d\xi^j = L^2 \Big\{[(1+\kappa^{2})ds^{2}_{\ads_{3}}+\kappa^{2}ds^{2}_{S^{3}}]
 \notag \\* &
+dU^{2} + [2\kappa\sqrt{1+\kappa^{2}}\,U+(1+2\kappa^{2})\,U^{2}]\, (ds^{2}_{\ads_{3}}+ds^{2}_{S^{3}})
+\tfrac{1}{4} d\z_p d \z_p 
\Big\}+ ...\,, \la{32} 
\ea
where $dU^{2}$ stands for $\del_i U \del_j U d\xi^i d\xi^j$ and similarly for $d\z_p d \z_p$. 

The  $C_{3}$  field has support in $S^4$ (\cf\ \rf{2.6})   while the   classical solution is in  $\ads_3\times S^3$ part of 
 $\ads_7$  and    thus  the pull-back   of $C_3$  will not contribute at  the quadratic fluctuation level. 
 As a result, we  have from  \rf{2.3} 
 \be\la{34}
 {\sfH}_{ijk} = H_{ijk} = H_{ijk}^{(0)}+\f_{ijk} = \kappa^{2}L^{3} [\vol_{S^{3}}]_{ijk}+\f_{ijk}\,,
\ee
where $H_{ijk}^{(0)}$  is the classical value in \rf{2.13} and 
 $\f_3=d\tilde B_2 $ is the contribution  of the  2-form fluctuation.
In the gauge \rf{211}  we have 
\be
\la{3.4}\te 
\hat\sfH^{ij} = \frac{1}{6\sqrt{-|G|}}\frac{1}{\sqrt{-G^{11}}}\,\eps^{ij1\ell mn} {\sfH}_{ijk}\,, 
\ee
which  will thus depend  on both $\f_3$ and the coordinate fluctuations  via $G_{ij}$ in \rf{32}.
Expanding $\sqrt{-|G_{ij}+\hat\sfH_{ij}|}$ to quadratic  fluctuation order we  find (\cf\ \rf{2.21}) 
\ba
\la{3.6}
 \sqrt{-|G_{ij}+\hat\sfH_{ij}|} = &L^{6}\kappa^{2}(1+\kappa^{2})^{2} \ \sqrt{-\bg} \  \Big\{
1+\tfrac{6\kappa}{\sqrt{1+\kappa^{2}}}\,U+\TermF\lp
+\tfrac{1}{8(1+\kappa^{2})}\bg^{ij}\partial_{i}\z_{p}\partial_{j}\z_{p}
+\tfrac{1}{2(1+\kappa^{2})}\big[\bg^{ij}\partial_{i}U\partial_{j}U+18(1+2\kappa^{2})U^{2}
\big]+ \TermFF\Big\} + ...\,. 
\ea
Here $\TermF$ and $\TermFF$  stand for terms    linear and quadratic in $\f_{ijk}$ (see below). 
We observe that the fields $U$ and $\z_p$ have kinetic terms  corresponding to the equal-radius \adsst metric
 $\bg= (g_A, g_S)$  (with each factors normalized to  have unit radius)\foot{The 
 $\ka$-dependent prefactors in the quadratic  terms in \rf{3.6}  can be   rescaled away and  will not
  contribute to the  one-loop free energy.}
\be
\la{3.7}
\bg_{ij}(\xi)\,  d\xi^{i}d\xi^{j} = ds^{2}_{\ads_{3}}+ds^{2}_{S^{3}}\,, \ \ \ \ \ \ \ \ \ \ \ \ 
\sqrt{-\bg}= \sqrt{-g_{A}}\sqrt{g_{S}} \,, \qquad \ \   L_A=L_S=1\,. 
\ee
While the standard  induced metric in \rf{2.15}  had unequal radii, 
 the effective metric that
governs the  quadratic fluctuation propagation  receives contribution from  the
non-trivial  background of the $H_{ijk}$ field   and as a result 
 turns out to be an equal-radii one. The equality of the two radii is a  special feature of the solution \Ia\
where \adsst  belongs to $\ads_7$. 
In the other two  cases $\Ib$ and $\II$   the effective metric will  have the ratio of the $\ads_3$ and $S^3$  radii  again 
being parameter-independent and equal to 2  and $\ha$ respectively.\foot{Let 
us note that here 
the effective metric   is not  simply  related to  $G_{ij}+\hat\sfH_{ij}$ in the M5 brane action. This is 
different from the case of the D3-brane probe with $\ads_2 \times S^2$   geometry discussed in \ci{Faraggi:2011bb,Buchbinder:2014nia}  where the  equal-radii effective $\ads_2 \times S^2$  metric  was related to the inverse of  the 
symmetric part of $(G_{ij}+F_{ij})^{-1} $.  
 Note also that a similar  equal-radii effective \adsst  M5 brane world-volume metric 
appeared in a different context in \ci{Berman:2001fs}.}


The  linear term $\TermF$    originates from a  product 
of  $H_{3}^{(0)}$  and $\f_{3}$  in terms with  higher-order powers  of  $ {\sfH}_{ijk}$   and is given by 
\be
\la{3.8}
\TermF = \frac{1}{6L^{3}\kappa^{2}(1+\kappa^{2})\sqrt{g_{S}}}\, \eps^{012\ell m n }\f_{\ell mn} = 
\frac{1}{L^{3}\kappa^{2}(1+\kappa^{2})\sqrt{g_{S}}}\ \f_{345}\,.
\ee
It vanishes after the  integration  over $\xi$  in  (\ref{3.6})  being a total derivative ($\f_3=dB_2$)  as there are no 
 factors   depending on $S^3$ coordinates remaining  in (\ref{3.6}).
 
 The expansion of the action \rf{2.1}  contains also 
another  contribution linear in $\f_3$ that comes from  the  second  $\star \sfH \sfH $ term  that may  be written as 
\ba
\la{3.9}\te
  \frac{\sqrt{-|G|}}{4(\del a)^2}\partial_{i}a \,  {\star \sfH}^{ijk}\, {\sfH}_{jk\ell}\partial^{\ell}a &= 
\tfrac{1}{4}\sqrt{-|G|} (G^{11})^{-1}\, {\star\sfH}^{1jk}\, {\sfH}_{jk1}\,  G^{11} 
=\tfrac{1}{4!}\eps^{1ijk\ell m}\sfH_{1ij}\sfH_{k\ell m}\,.
\ea
The linear in $\f_3$ term in its expansion is 
\ba
\la{3.10}
 \tfrac{1}{4!}\eps^{1ijk\ell m}\sfH_{1ij}\sfH_{k\ell m}\ \ \to \ \ 
\tfrac{1}{4!}\eps^{1ijk\ell m}\, \f_{1ij}H^{(0)}_{k\ell m} = \tfrac{1}{2}L^{3}\kappa^{2}\ \sqrt{g_{S}}\ \f_{012}\,.
\ea
It again  gives a total derivative in the action  as its prefactor does not depend on $\ads_3$ coordinates. 


In addition,  we need to   include terms coming from the expansion of   the  $C_6$ term in \rf{2.1}. 
Using  (\ref{2.12}) we get 
\be\la{310}
C_{6} = L^{6}\Big[\kappa^{4}(\kappa^{2}+\tfrac{3}{2})+6\kappa^{3}(1+\kappa^{2})^{3/2}\, U+9\kappa^{2}(1+\kappa^{2})(1+2\kappa^{2})\, U^{2}+\cdots
\Big]\, \vol_{\ads_{3}}\we\vol_{S^{3}} \,. 
\ee
We observe that the linear in $U$  contributions in \rf{3.6} and in \rf{310}  cancel each other in the total action \rf{1.2}, 
which is a manifestation of the fact  that the background \rf{2.13} is  indeed an extremum of  the M5 brane action.

Moreover,  the quadratic  $U^2$ terms  in \rf{3.6}  and in \rf{310} also cancel each other
so that $U$ is also a massless  fluctuation  like $\z_p$ in \rf{3.6}.
  This is consistent with the fact that $u_0$ (or $\ka$)    is a free parameter of   the solution \rf{2.13}
  so that $U\to  U + $const  should be  a symmetry of the fluctuation action. 

The  terms  quadratic in $\f_3$ come directly from \rf{3.10}, \ie\ 
$\frac{1}{4!}\eps^{1ijk\ell m}\sfH_{1ij}\sfH_{k\ell m}\to \frac{1}{4!}\eps^{1ijk\ell m}\f_{1ij}\f_{k\ell m} $
and also from \rf{3.6}. The latter 
may be written as  (the indices are  contracted with the same effective  metric  $g_{ij}$  as  in (\ref{3.7}))
\ba
\la{3.14}
\TermFF &= \frac{1}{12L^{6}\kappa^{2}(1+\kappa^{2})^{2}}\sum_{i,j,k\neq 1}\f_{ijk}\f^{ijk}\,. 
\ea
In total, the expansion of the integrand in the PST action \rf{1.2}  then  contains the following $\f_3 \f_3$ terms 
\be\la{312} 
L^{(+)}_2  = \tfrac{1}{24} \eps^{1ijk\ell m} \f_{1ij} \f_{k\ell m}  -  \tfrac{1}{12}\sqrt{-\bg}  \sum_{i,j,k\neq 1}\f_{ijk}\f^{ijk}\,.
\ee
Eq. \rf{312}   has the form of the  non-covariant  Lagrangian  describing propagation of a free (anti) self-dual 3-form  field 
on a curved 6d background $g_{ij}$ \cite{Henneaux:1988gg,Schwarz:1993vs,Schwarz:1997mc,Aganagic:1997zq}.\foot{With  index 1  interpreted as the  time-like one it has a ``phase-space''  form 
with  the spatial $B_{rs}$ components   as coordinates and  $\partial_{1}B_{rs}$ as momenta (\cf\ \cite{Henneaux:1988gg}).}
The equations of motion following from \rf{312} imply (doing one integration under proper boundary conditions) 
the 6d (anti)self-duality  condition $\f^{1ij} =-  {1\ov 3!\sqrt{-g}} \eps^{1ijk\ell m}\f_{k\ell m}$  
or $\star \f_3=-\f_3$  
(\cf\ 
 \cite{Ko:2013dka}). 
 
 Using the  (anti) self-duality condition to replace  the   factor $\f_{k\ell m}  $   with $\f_{1ij} $ 
   with  in the first term in \rf{312}   one finds that \rf{312} it  takes 
 the standard covariant form  for the 2-form  Lagrangian, \ie\  
 \be \la{300}
 L_2= -  \tfrac{1}{12}\sqrt{-\bg}\,  \f_{ijk}\f^{ijk}   \,.  \ee 
  The corresponding partition  function evaluated under the self-duality constraint on $\f_3$   should then be 
  given by the {\it square root}
   of the standard gauge 2-form partition function on a curved 6d background 
  (discussed   in \ci{Obukhov:1982dt,Fradkin:1982kf,Bastianelli:2000hi}  and refs. there).
  This   can be shown, \eg, at a diagrammatic level  \ci{Alvarez-Gaume:1983ihn}.\foot{This approach was applied in \cite{Bastianelli:2000hi} to   find the   conformal anomaly coefficients of the $(2,0)$ tensor
multiplet. It extends to the full partition function  $Z$  as a functional of curved
metric.  $Z$  can be  computed as  a sum of diagrams with external graviton lines and  implementing the 
projection to the (anti) self-dual 2-form component  in each internal propagator  \cite{Alvarez-Gaume:1983ihn}. 
  Note that here we will be  interested only 
   in the  
  real part of the partition function, \ie\ will ignore the  phase part related to the  gravitational anomaly \ci{Alvarez-Gaume:1983ihn}.
  In general,  both the  self-dual 3-form and  chiral 6d fermions of the   M5 brane  action    will contribute to its gravitational anomaly
  (the  cancellation of the M5 brane anomalies  in the M-theory context 
  was discussed in \ci{Freed:1998tg}  and refs. there). 
  }
  
  The same  conclusion  should  follow  directly  from the PST action  or 
   \rf{312}  
  provided  the  corresponding path integral over $B_{ij}$  is defined with an appropriate measure factor (containing 
  a determinant of a particular 1st-order 
  differential operator not involving $\del_1$ derivative).\foot{Eq.    \rf{312} (with  index ``1'' in  \rf{312}   relabelled as ``0'')   may be viewed as 
     a 6d   generalization  of 
 the Floreanini-Jackiw   Lagrangian  \ci{Floreanini:1987as}  for a chiral scalar  $\varphi^{(+)}$ in 2 dimensions
    given (in flat 2d space)  by 
$L= \partial_{0} \varphi^{(+)} \partial_{1} \varphi^{(+)}  - \partial_{1} \varphi^{(+)} \partial_{1} \varphi^{(+)} = \partial_{1}\varphi^{(+)}\partial_{-}\varphi^{(+)}$. 
The corresponding equation of motion $\partial_{1}\partial_{-}\varphi^{(+)} = 0$  implies (assuming $\partial_{-}\varphi^{(+)} = 0$  holds at spatial infinity) 
that $\partial_{-}\varphi^{(+)}=0$, \ie\ gives  the 2d self-duality  condition. 
Integrating over $\varphi^{(+)}$ with a measure  containing the $[{\det \del_1}]^{1/2}$ factor  
 gives  the chiral scalar partition function as   $[\det \partial_{-}]^{-1/2}$.  
 Generalized to curved space this is the same  as  the   square root 
 of the real scalar   partition function 
 ${[\det(\partial_{-}\partial_{+})]}^{-1/2}$ (up to pure-phase gravitational anomaly factor in the Euclidean case).
 The reason for   the  $[{\det \del_1}]^{1/2}$  measure factor  can   be understood as   follows  \ci{Tseytlin:1990nb}.
 Starting with the  real scalar  Lagrangian $L_0 = \partial_{0} \varphi \partial_{0} \varphi  - \partial_{1} \varphi \partial_{1} \varphi$
 and writing  the corresponding path integral   in phase-space form  with 
 $L=2 p\del_0 \vp -  p^2 -  \del_1 \vp \del_1 \vp $ one can then   set $p=\del_1  \td \vp$
getting duality-symmetric Lagrangian $L =\del_0 \td \vp \del_1 \vp  +   \del_1 \td \vp \del_0 \vp -  \del_1 \td \vp  \del_1 \td \vp 
 -  \del_1 \vp \del_1 \vp $ (ignoring total derivative). The path integral over  $p$  and $\vp$ had canonical measure, so 
 then one over $\vp$ and $\td \vp$  should  contain the Jacobian  factor  ${\det \del_1}$.
 Introducing $\vp^{(\pm)} =  \vp \pm \td \vp$  one  gets 
 $L= \partial_{1}\varphi^{(+)}\partial_{-}\varphi^{(+)} + \partial_{1}\varphi^{(-)}\partial_{+}\varphi^{(-)}$
 and  thus the  original path integral factorises   into $Z^{(+)}  Z^{(-)}  $  with $Z^{(\pm )}= {[\det\partial_{\pm})]}^{-1/2}$   each originating from  path integral  containing the measure factor
 $[{\det \del_1}]^{1/2}$.
}
In more detail,  starting with the  covariant Lagrangian \rf{300}  for $B_{ij}$  and writing it  in the   phase-space form (considering flat space case,  fixing  $B_{0r}=0$ gauge,  ignoring some  trivial factors  and using  here $r,s,q=1,...,5$)   we get  
$L= \half  p^{rs} \del_0 B_{rs}  -  {1\ov 4} p^{rs} p_{rs}  - {1\ov 12}  \f_{rsq} \f^{rsq}$. We   may then introduce  a new (``dual'') field 
$\td B_{rs}$  by  setting $p^{rs} = {1\ov 2} \eps^{rsquv} \del_q \td B_{uv} \equiv ({\rm D} \td B)^{rs}$.
Here $ {\rm D}$ is a 1st-order operator containing only spatial derivatives. 
One can easily see    that its  square is $(\del)^2_\perp$   defined on $B^\perp_{rs}$, i.e. is 
 the same  operator that  appears in the  spatial part $\f_{rsq} \f^{rsq}$  of the Lagrangian.
The resulting path integral over $B_{rs}$ and $\td B_{rs}$ 
 will  have the Jacobian factor  $\det {\rm D} $.  The   action  for $B_{rs} $  and $\td B_{rs}$ 
 can be re-written as a sum of the decoupled actions for $B^{(\pm)}_{rs}  = B_{rs} \pm \td B_{rs} $ 
 each  similar to \rf{312} (with the  role of the indices 1 and 0 interchanged and gauge fixed as $B_{0r}=0$), i.e.  having kinetic operators $\OO_\pm {\rm D}$ with 
   $\OO_\pm= \del_0 \pm {\rm D}$  (which are direct analogs of $\del_0 \pm \del_1$ in the 2d chiral scalar cases). 
 The original path integral  for the $B_{ij}$ field   then formally factorizes as  $Z^{(+)} _2  Z^{(-)}_2  $ 
 where  each  factor is defined by  the integral   over  the  ``chiral''  $B^{(\pm)} $ field   with the measure containing 
 the $[\det {\rm D} ]^{1/2}$  factor. This  factorization is then equivalent to $\det (\del_0^2 -  \del^2_\perp) = \OO_+ \OO_-$.

This ``square root'' prescription  for the one-loop partition function $Z^{(+)} _2$ 
 of  a self-dual 2-form in the M5 brane action was used
 in the $S^5 \times S^1$   geometry  case in  \ci{Beccaria:2023cuo}   and in  the $\ads_5 \times S^1$   case  in \ci{Jiang:2024wzs}. 
 In the present \adsst case   the expression for $Z_2$   in terms of the determinants  of the relevant   2nd-order   operators
  is presented in Appendix \ref{apz}.

In total,  the scalar  $U$  and  $\z_p$  terms in \rf{3.6},  \rf{310}   and  \rf{300}  lead to 
 the following action for the  quadratic  fluctuations of the bosonic fields 
 \be
\la{3063}
S_2 = -\int d^{6}\xi\ \sqrt{-\bg}\, \Big(  \tfrac{1}{2} \bg^{ij}\partial_{i} U \partial_{j} U +\tfrac{1}{2}\bg^{ij}\partial_{i}\z_p\partial_{j}\z_p
 +\tfrac{1}{12}\f_{ijk}\f^{ijk}  \Big) \,.
\ee
To arrive at \rf{3063}  we rescaled the fluctuation fields  by constant $\ka$-dependent factors.
In contrast to the D3-brane case discussed in  \ci{Buchbinder:2014nia}
here the Seeley coefficient  $b_6$ of each of  the relevant differential operators, \ie\ 
 the coefficient of the logarithmic UV 6d divergence  vanishes (see  Appendix \ref{apdi})  and thus 
 the rescaling  of the fields (or local  measure factors) do not contribute to the finite part of the resulting free energy.
 The same  observation  applies to the fermionic field contribution discussed in  Appendix \ref{fer}
 and also  is true in the two other cases  \Ib\  and \II.

 Combining   the $Z^{(+)} _2$   contribution  for self-dual part of 
 $\f_{ijk} $ with  the contribution of the  5 massless scalars $U$ and $\z_p$ and the fermions 
  we  end up with the partition function $Z$  of  the (2,0) tensor multiplet on  the  equal-radii \adsst  space.  
 As this space 
 is conformally flat and its 6d Euler density vanishes 
 and as each field    has Weyl-invariant Lagrangian, 
 we  conclude  as in  \ci{Bastianelli:2000hi}  that there is  no  conformal anomaly or no  logarithmic UV divergence. 


Explicitly, using the expression for $Z^{(+)}_{2}$  in \rf{z9}  in  Appendix \ref{apz}  the bosonic part of 
 $Z_{\rm B}$ may be written as
\ba
Z_{\rm B}= Z^{(+)}_{2} Z_0^5  &=
{\big[\det \bm\Delta_{1\perp,1\perp}(0)\big]^{-1/4}  \ \big[\det\bm\Delta_{0,0}(0)\big]^{-1/2}} \ 
{\big[ \det\bm\Delta_{0,0}(0)
\big]^{-5/2}}\notag\\
& =
\big[\det \bm\Delta_{1\perp,1\perp}(0)\big]^{-1/4}  \ \big[\det\bm\Delta_{0,0}(0)\big]^{-3}\,. \la{327}
\ea
We   used  \rf{z7} and that  $L_A=L_S$. 
Here $\bm \Delta_{1\perp,1\perp}=   -\nabla^{2}_{A}-\nabla^{2}_{S}      $   is defined on the $B_{ar}$  field  
on \adsst   which is separately transverse in the $\ads_3$ index 
$a$ and  the $S^3$ index $r$.  $\bm \Delta_{0,0}(0)= \nabla^2  $ is a massless scalar Laplacian.   
The number of the bosonic degrees of freedom is thus the expected one: 
 $\frac{1}{2} (2\times 2) + 6 = 8$.

  In addition, we have 8 fermionic degrees of freedom in the M5-action.  
 A detailed analysis of the fermionic fluctuation
 contribution $Z_\text F$  to the partition function 
is   delegated to the  Appendix  \ref{fer}.  $Z_\text F$   is given by the determinant 
of a  massive   Dirac operator on \adsst  defined  using  the  same  equal-radii  effective metric  as in  \rf{3.7} 
\begin{equation}\label{3.36}
Z_\text F=
\big[\det \mc D   \, \big]^{1/2}\, , \qquad \qquad \mc D = i \slashed \cd + \mathcal M \,. 
\end{equation}
Here  we assume that  $\mc D   $ acts on 32 component spinor
and thus \rf{3.36} describes 
8 real degrees of freedom. 
It turns out that in   all the  cases {\bf Ia, Ib, II} 
 the Dirac operator  $\mc D $ in \rf{3.36} 
  has the form (see Appendix \ref{fer})
\ba
\la{3.39}
&\mc D =   i \slashed \cd +\mc M  =
  i\slashed{\nabla}_{A}+i\slashed{\nabla}_{S}+\, m_{_{\rm F}}\,\Gh\,,\qquad \qquad \mc M =  m_{_{\rm F}}\,\Gh\,, \\
&\la{3399}
\Gh^2=1\,,  \ \ \qquad   [\slashed\nabla_{S}, \Gh] = \{\slashed\nabla_{A}, \Gh\}=0\,. 
\ea
In the \Ia\ case  one finds that $m_{_{\rm F}}=0$ (see  (\ref{D.13})). 

To summarize, in the \Ia\ case the one-loop  partition function  is  the same as for the  (2,0) 
multiplet (self-dual tensor, 5 massless  scalars   and 4 massless Weyl  6d fermions)
defined on the equal-radius  \adsst space.

The spectra of the operators  in \rf{327},\rf{3.36}  can be found  by first expanding in modes on $S^3$ and thus  
getting a tower of massive KK fields on $\ads_3$ labelled by level $\ell$. 
As a result, their determinants  can be expressed in terms of  products of determinants of  operators on $\ads_3$. 
For the scalar field  in \rf{3.2}  we get (here  for generality we assume that  the radii  of the two factors in \adsst metric are 
$L_A$ and $L_S$)
\be \la{340}
\bm\Delta_{0,0}(0) \ \   \to \ \  \bm\Delta_{0}= - \nabla^2_A + M^2_{0,\, \ell} \,, \qquad \ \ 
M^2_{0,\, \ell }=L_S^{-2}  \ell(\ell + 2) \,, \ \ \  \ \ \  d^{(0)}_\ell =( \ell+1)^2   \,,  \ee
where $d_\ell$ is the scalar degeneracy. 
Similarly, 
for the  transverse vector operator in \rf{327} (see \rf{B.8})
\be \la{341}
\bm\Delta_{1\perp,1\perp}(0) \ \   \to \ \  \bm\Delta_{1\perp}= - \nabla^2_A + M^2_{1,\, \ell} \,, \qquad \ \ 
M^2_{1,\, \ell}=L_S^{-2} (\ell^2 + 4 \ell+2)   \,, \ \ \  \ \ \  d^{(1)}_\ell =2 (\ell+1)( \ell + 3) \,.  \ee
Using a split basis for the 11d gamma matrices  and expanding the   fermions in the 
$S^3$  spinor spherical harmonics  we  get 
\be\la{319}
\slashed\nabla_{S} \  \to \ \  \pm  i M_{{1\ov2},\, \ell }\,, \qquad \ \ 
M_{ {1\ov 2},\, \ell } =  L_S^{-1} (\tfrac{2\ell+1}{2}+1)\,, \qquad \qquad 
d^{({1\ov 2})}_\ell =  (\ell+1)(\ell+2) \,, 
\ee 
where $ d_\ell $  is the degeneracy of the single fermion  mode on $S^3$ 
 (see, \eg, Eq.~(3.44) in \cite{Camporesi:1995fb}).
Squaring  the resulting  Dirac operator on $\ads_3$   gives 
\be\la{320}
\bm\Delta_{\frac{1}{2}} = -\nabla^{2}_{A}+\tfrac{1}{4}R_{A}+ {\hat M}^2_{{1\ov 2},\, \ell }  \,, \ \ \ \ \ \ \ \ \ 
\ \ \ { \hat M}_{ {1\ov 2},\, \ell } =M_{ {1\ov 2},\, \ell}\pm m_{_{\rm F}}  \,, \ \ \ \ \ \ 
R_A = - 6 L^{-2}_A \,. 
\ee
The corresponding  dual  conformal dimensions of the $\ads_3$  scalar ($s=0$)  and the   transverse 
vector ($s=1$) with mass $M_s$ 
 can be found using  the standard relations\footnote{To recall, the 
   dual-field dimensions for  a massive $p$-form field  in $\ads_{d+1}$ is  $
\Delta= \tfrac{d}{2}\pm\sqrt{(\tfrac{d}{2}-p)^{2}+m^{2}L^2_A} 
$ (see, \eg\ \ci{Metsaev:2003cu}). 
Here $m^2$ is the  term in addition to the standard Hodge-deRham  structure of the Laplacian
For $d=2$   and  $p=1$ the  dimension satisfies 
$\Delta(\Delta-2) = -1+m^{2}L^2_A$.
For a vector $(\bm \Delta_{1\perp})_{ab} = - (\nabla^2_A)_{ab}  + R_{ab} +m^2 g_{ab}  =( - \nabla^2_A   + M^2)g_{ab} $, 
where  $M^2 = -2L^{-2}_A + m^2 $  and thus  $\Delta(\Delta-2) =  1 + M^2 L^2_A $.  } 
\be
\la{3.37}
\Delta^{(0)} (\Delta^{(0)}-2) = L^{2}_{A}\,M^{2}_0\, , \qquad\qquad  \ 
\ \ \ \ \Delta^{(1)}(\Delta^{(1)}-2) =1 +  L^{2}_{A}\,M^{2}_1\,.
\ee
As a result,  from \rf{340},\rf{341}   for $L_A=L_S$ we get   for the  dimensions of the $\ads_3$  fields in the scalar   and vector  KK towers 
\be  
\Delta^{(0)}= \ell + 2 \, , \qquad \qquad  \ \ \  
 \Delta^{(1)}= \ell + 3 \,. \la{339}\ee
For spin $s=\ha$  fermions with the squared Dirac operator $\bm\Delta_{\frac{1}{2}} = -\nabla^{2}_{A}+\tfrac{1}{4}R_{A}+ M^2_{\frac{1}{2}} $
one has the corresponding $\ads_3$ dimension given by 
\be
\la{3.38}
 \  \ \ \Delta^{(\frac{1}{2})} =  1+\sqrt{1+s+ L^{2}_{A}\,M^{2}_{ \frac{1}{2} }} = 1+\sqrt{\tfrac{3}{2}+ 
 L_{A}^{2}\, M^{2}_{\frac{1}{2}}}\,. 
\ee
Then for the value of the mass in  \rf{320} we get
\be
\la{3.42}
 \ \ \   \ \ \Delta^{(\frac{1}{2})} = 1+L_{A}L_{S}^{-1}(\ell +\tfrac{3}{2})\pm  L_A m_{_{\rm F}}  \,. 
\ee
In present case {\bf Ia} with  
 $m_{_{\rm F}}=0$  and  $L_{A}=L_{S}$ 
the  dimensions of the $\ads_3$ fermionic fields are  found to be 
\be \la{333} 
\ \ \ \ \ \ \Delta^{(\frac{1}{2})} =\ell+\tfrac{5}{2}\,.
\ee

\subsection{Probe \Ib  \la{s321}}

Here (see \rf{228})  we may  use  again the static gauge, setting to zero fluctuations of the coordinates along $\ads_3 \subset \ads_7$
and $S'^3 \subset S^4$. The remaining transverse fluctuations will then be of $u$  and $S^3$  directions 
in the $\ads_7$ metric  and of $\theta$ in \rf{2.10}. Since the expansion goes around $u=0$ we  may parametrize these 
5 fluctuations  like in \rf{3.2} as $\z_p$   ($p=1,2,3,4$)   and $\Te $
\be \la{342}
ds^{2}=du^{2}+\sinh^{2}u\,  ds^{2}_{S^{3}} = {d \z_p d\z_p \ov (1- \four \z^2)^2} = 
d \z_p d\z_p  + ... \,, \ \ \ \ \ \  \qquad  \theta= \theta_{0}+ \Te \,. \ee
The  perturbed induced metric may  then  be written  like in \rf{32} as (\cf\ \rf{229}) 
\ba
ds^{2} = &L^{2}\Big\{ ds^{2}_{\ads_{3}}+ 2 \ka^2 (1- 2 \ka^2)  ds_{S^{3}}^{2}  + 
  d\z_p d\z_p  +\z_p \z_p ds^{2}_{\ads_{3}}+\tfrac{1}{4}  d\Te ^{2} \no  \\ &\ \ 
+\Big[\, \kappa(1-4\kappa^{2})\sqrt{2-4\kappa^{2}}\, \Te + (\four -4\kappa^{2}+8\kappa^{4})\,  \Te ^{2}\, \Big] \, ds^{2}_{S^{3}}\Big\}
+\cdots \,. \la{343}
\ea
Here we again  denote   the world volume  sphere  $S'^3 $ as $S^3$.  
Using  the expression for $C_3$ in \rf{227}  the  expansion of ${\sfH}_{ijk}$ in \rf{2.3}  may be written as 
(\cf\ \rf{34}) 
\ba
{\sfH}_{ijk} &= H^{(0)} _{ijk}-C_{ijk}+\f_{ijk}  = L^{3}\U(\Te )\  [\vol_{S^{3}}]_{ijk}+\f_{ijk}, \la{445}
\\
\U(\Te ) &= \kappa^{2} (1- 2 \ka^2) \Big[ (1-4\kappa^{2})-6\kappa \sqrt{2-4\kappa^{2}}\, \Te 
-\tfrac{9}{2}(1-4\kappa^{2})\,\Te ^{2}\Big] +\cdots\,. \la{446}
\ea
The  analog of the expansion of  $\sqrt{-|G_{ij}+\hat\sfH_{ij}|}$ in \rf{3.6} is found to be 
\ba
\la{3.53}
\sqrt{-|G_{ij}+\hat\sfH_{ij}|} = & L^{6}\, 8\kappa^{2}(1-2\kappa^{2}) \, \sqrt{-\bg}    \ \Big[1+\TermF \lp 
+\tfrac{1}{8}(\bg^{ij}\partial_{i}\Te \partial_{j}\Te +24\,\Te ^{2})
+\tfrac{1}{2}(\bg^{ij}\partial_{i}\z_p\partial_{j}\z_p+3\z_p \z_p)
+\TermFF+\TermMix\Big]+...  \,. 
\ea
Here the effective \adsst  metric $\bg_{ij} $ which the fluctuations are propagating in  turns out to be   independent  of $\ka$ 
and when   written in terms of the unit radius $\ads_{3}$ and $S^{3}$ metrics $g_{A}, g_{S}$  is simply 
\be
\la{3.54}
\bg_{ij}d\xi^{i}d\xi^{j} = ds^{2}_{\ads_{3}}+\tfrac{1}{4}\, ds^{2}_{S^{3}}\,,\qquad 
\ \ \ \ \sqrt{-g} = \tfrac{1}{8} \sqrt{-g_{A}}\sqrt{g_{S}}\,, \qquad \ \ \   L_A=1\,, \ \   L_S=\ha \,.
\ee
 The linear  $\TermF$  and quadratic   $\TermFF$   and $\TermMix$  terms are 
 \ba
\TermF & = -\frac{1-4\kappa^{2}}{6L^{3}\kappa^{2}(1-2\kappa^{2})\sqrt{g_{S}}}\ \eps^{021\ell m n}\ \f_{\ell m n },\qquad \qquad 
\TermFF = \frac{1}{96L^{6}\kappa^{2}(1-2\kappa^{2})}\sum_{i,j,k\neq 1}\f_{ijk}\f^{ijk}\,, \la{1333}\\
& \qquad \qquad \TermMix   = \frac{1}{L^{3}\kappa\sqrt{\frac{1}{2}-\kappa^{2}}\, \sqrt{g_{S}}}\, \eps^{021\ell m n}\ \f_{\ell m n }\, \Te \,.
\la{135}
\ea
The  indices  in  $\TermFF$
 are contracted using $\bg_{ij}$. 
The linear term  $\TermF$ is again a total  derivative (as in \rf{3.8}), \ie\ it  vanishes after the integration over 6d space. 
From \rf{445} we find also that   the second term in \rf{2.1}   here is (\cf\ \rf{3.10}) 
\ba\la{1356}
\tfrac{1}{4!}\eps^{1ijkmn}\sfH_{1ij}\sfH_{kmn} \ \ \to \ \  
  \ha {L^{3}} \sqrt{g_{S}}\, \f_{012}\, \U (\Te ) + \tfrac{1}{4!}\eps^{1ijkmn}\f_{1ij}\f_{kmn}   \,.
\ea
The $C_6$ term  in the WZ part of the action \rf{2.1}   does not  contribute at the quadratic fluctuation level 
while (\cf\ \rf{445}) 
\be\la{0335}
\tfrac{1}{2}\int \sfH_{3}\we C_{3}  =  \tfrac{1}{2}L^{3} \int \sfH_{3}\we \big[\kappa^{2}-\U(\Te )\big] \vol_{S^{3}}  
\ \to \  \ha L^3
\int d^{6}\xi\, \sqrt{g_{S}}\, \f_{012}\, \big[\kappa^{2}-\U(\Te ) \big]\,.
\ee
Combining the integral of \rf{1356}   with \rf{1356} we conclude that the $\U(\Te )$ term cancels, \ie\ there are no linear  and  quadratic  in $\Te $ terms remaining  in the  fluctuation action
(and the integral of the  remaining $\f_{012}$ term again vanishes). 

The resulting  bosonic fluctuation   Lagrangian contains  (i)  5 massive scalar fields $\Te $ and $\z_p$  (\cf\ \rf{3.53});  
(ii)  $\f_3\f_3$ term given  by the  combination of the $\TermFF$ term in \rf{1333} and the 
second term in \rf{1356}; (iii) the $\TermMix$   mixing term. 
The $\f_3\f_3$ term  is  the same as in  \rf{312}  in the \Ia\ case  describing the (anti) self-dual 2-form field. 
We may formally use  the (anti) self-duality condition to rewrite  it in the covariant form \rf{300} 
 assuming that the path integral is to be  carried out  over the (anti) self-dual fields only
or, equivalently, using   the ``square root'' prescription for the corresponding contribution to the partition function.

Rescaling the  fields  by 
constant factors  the resulting action for the quadratic fluctuation
fields on \adsst with the metric \rf{3.54}  may be written  in the following $\ka$-independent form (\cf\ \rf{3063})
\be
\la{3.63}
S_2 = -\int d^{6}\xi\ \sqrt{-\bg}\, \Big[
 \tfrac{1}{2}\big(\bg^{ij}\partial_{i}\Te \partial_{j}\Te +24 \Te ^{2}\big)+\tfrac{1}{2}(\bg^{ij}\partial_{i}\z_p\partial_{j}\z_p+3\z_p \z_p\big)+\tfrac{1}{12}\f_{ijk}\f^{ijk}
-\tfrac{48}{\sqrt{g_{S}}}\, \f_{345}\, \Te  \Big] \,.
\ee
The remaining non-trivial problem is to diagonalize the  last term representing 
the mixing \rf{135} between the scalar $\Te $ and the $S^3$ components of the  field  $\f_3=dB_2 $.

Using the  explicit form of the  gauge-fixed  Lagrangian for the  $B_{ij}$ field  on \adsst given in \rf{3.25} 
and specifying it to the present case of  the metric \rf{3.54}  with $L_{A}=1$, $L_{S}=\frac{1}{2}$
we get 
\be
\la{3.64}
\tfrac{1}{12}\f_{ijk}\f^{ijk} \  \to\  \four\big[B_{ab}(-\nabla^{2}-2)B^{ab}+B_{rs}(-\nabla^{2}+8)B^{rs}+2B_{ar}(-\nabla^{2}+6)B^{ar}\big]\,,
\ee
where the indices  $a,b$  correspond to  $\ads_{3}$, the indices  $r,s$   to  $S^{3}$  and $\nabla^2= \nabla^{2}_{A}+ \nabla^{2}_{S}$. 
The ``mixed''   part of the Lagrangian in \rf{3.63} that involves  $\Te$ and $B_{rs}$      may be written as 
\be\la{1364}
L_{\rm mix} = -\tfrac{1}{2}\sqrt{-g_{A}}\, \sqrt{g_{S}}\, \big[ g^{ij}\partial_{i} \Te \partial_{j}\Te +24\Te ^{2}+\tfrac{1}{2}
B_{rs}(-\nabla^{2}_{A}-\nabla^{2}_{S}+8)B^{rs}
\big]+24\sqrt{-g_{A}}\, \Te \,\eps^{rst}\,\partial_{r}B_{st}\,.
\ee
We   may replace $B_{rs}$  by a 3-vector  $V^r$ given by 
 \be\la{555}
B_{rs}= \tfrac{8}{\sqrt{\bg_{S}}} \eps_{rsu}  V^{u}\,.
\ee
Here  $\eps_{rst} $ is defined using  the $S^3$ metric with radius $\ha$.
Then \rf{1364} becomes 
\ba\la{1340}
L_{\rm mix} &= -\tfrac{1}{2} \sqrt{-g_{A}}\,\sqrt{g_{S}}\, \Big[  g^{ij}\partial_{i} \Te \partial_{j}\Te +24\Te ^{2}+V_{r}
(-\nabla^{2}_{A}-\nabla^{2}_{S}+8)V^{r}+\ 12\, \Te \,\nabla_{r}V^{r}\Big] \,, 
\ea
where $\nabla_r$ is defined with respect to the $S^3$ part of $\bg_{ij}$, \ie\ $\four g_S$
where $g_S$ is the unit radius metric. 
Next,  it is useful to split $V_r$ into the  transverse and scalar part as 
\ba
\la{3.69}
V_{r} = V_{r}^{\perp}+(\nabla_{S})_{r}\, P \,, \ \qquad 
\ea
so that \rf{1340} becomes ($\nabla^2 = \nabla^{2}_{A}+\nabla^{2}_{S}$)
\ba
\la{3.73} 
&L_{\rm mix} =  -\ha  \sqrt{-g_{A}}\, \sqrt{g_{S}} \  V_{r}^{\perp}(-\nabla^{2} +8)V^{\perp r}  + L_{\Te ,P} \,,\\
&\la{373} L_{\Te ,P}=
 - \ha \sqrt{-g_{A}}\, \sqrt{g_{S}}\, \big[  \Te (-\nabla^{2} )\Te +24\Te ^{2}
+ P(-\nabla^{2})(-\nabla^{2}_{S}) P+12\, \Te \,\nabla^{2}_{S}P\big] \,. 
\ea
Thus 
 we are left  only with the  scalar $(\Te ,P)$  mixing described by  $L_{\Te ,P}$. 
One may further redefine  $P$ to $P'= (-\nabla^{2}_{S})^{1/2}P$ to get a  2-derivative mixed scalar action. 
The associated  Jacobian  cancels the one  that is introduced by \rf{3.69}  and one finds that 
the partition function corresponding to \rf{373}  is 
\be\la{77}
Z_{\Te ,P} = \big[\det(\nabla^{4}-24\nabla^{2}+36\nabla_{S}^{2})\big]^{-1/2}\,.
\ee
It is convenient to separate  the contribution of the scalar $\Te $ and its mixing to $B_{rs}$ from the 
total  contribution of  the $B_{ij}$   field in \rf{1364}.  This amounts to  normalizing  \rf{77} to  the contribution 
of the $(-\nabla^2)$ factor in the $P$ kinetic term in \rf{373}, \ie\ 
to replacing  \rf{77} by 
\be
\la{375}
Z_{\Te ,\,\rm mix} = { [\det (-\nabla^2) ]^{1/2}}\ 
{ [\det (\nabla^{4}-24\nabla^{2}+36\nabla^{2}_{S})]^{-{1}/{2}}}\,.
\ee
Then in the limit of  no mixing, \ie\ when the $\nabla^2_S$ term in \rf{373} and thus \rf{375}  is  dropped, the partition  function \rf{375}
reduces to the one of the decoupled  massive scalar $\Te $ only. 

The total  bosonic partition function will then be given by \rf{375}  combined   with 
the contribution of the 
self-dual  tensor  and 
 4 massive scalars $\z_p$ 
  in \rf{3.63},  which is analogous to the one in \rf{327}.  Cancelling the common scalar det 
  factor against the  determinant of  $\bm\Delta_{0,0}(0) =-\nabla^2=- \nabla^2_A-\nabla^2_S$ gives 
\be\la{3466}
Z_{\rm B}= Z^{(+)}_{2} \, Z_0^4\,  Z_{\Te ,\,\rm mix} =  \big[\det \bm\Delta_{1\perp,1\perp}(6)\big]^{-1/4} 
\big[ \det\bm\Delta_{0,0}(3) \big]^{-2} \ 
\big[ \det (\nabla^{4}-24\nabla^{2}+36\nabla^{2}_{S}) \big]^{-{1}/{2}} \,.
\ee
Here we specialised  the general \adsst  self-dual  field contribution   $Z^{(+)}_{2} $ in \rf{z9} 
to the present case of $L_{A}=1$, $L_{S}=\frac{1}{2}$
(so that $\bm\Delta_{1\perp,1\perp}(6)$ originates from the $B_{ar}$ term in \rf{3.64}, etc.).

The  expression is similar  to  \rf{327} in the \Ia\  case  apart from the  last  
factor \rf{375} involving a more complicated 4-derivative operator factor.
Once we expand in harmonics  on $S^3$   the latter can be written in  terms of a  product of 2-derivative 
scalar operators on $\ads_3$. 
Indeed,  doing the  replacement  $-\nabla^{2}_{S}\to  {L_{S}^{-2}}\ell(\ell+2)$ 
 as in \rf{340} 
we get 
\ba  & 
\nabla^{4}-24\nabla^{2}+36\nabla^{2}_{S}= (\nabla^{2}_A + \nabla^2_S)^2+ 24\nabla^{2}_A + 60\nabla^{2}_{S}  \ \ \to \ \ 
(-\nabla^{2}_A+M_{0_+,\, \ell}^{2})
(-\nabla^{2}_A+M_{0_-,\, \ell}^{2})\,, \no \\
& \qquad M_{0_+,\, \ell}^{2}=4(\ell+2)(\ell+3) \,, \qquad \qquad  \ \  M_{0_-,\, \ell}^{2}=4\ell(\ell-1) \,.\la{777}
\ea
Each of these  factors enters with   the   scalar degeneracy $d_\ell^{(0)}=(\ell+1)^{2}$.
Note   that this  simple factorization is a consequence of  particular coefficients that appeared   in the fluctuation Lagrangian in \rf{1340}.

As a result, the last factor in \rf{3466} or \rf{77}   may be written as 
\be\la{88}
Z_{\Te ,\,\rm mix} = \prod_\ell \Big[\det\bDelta_{0}(M_{0_+,\, \ell}^{2} )\ \det\bDelta_{0}(M_{0_-,\, \ell}^{2})\Big ]^{- {1\ov 2} d_{\ell}^{(0)}}\,, 
\ee
where $\bDelta (M^2)=-\nabla^2_A + M^2 $,  \cf\ \rf{340}.
Similarly,  we get (\cf\ \rf{341}) 
\ba \la{888}
&\big[\det \bm\Delta_{1\perp,1\perp}(6)\big]^{-1/4}  \
\big[ \det\bm\Delta_{0,0}(3) \big]^{-2}= \prod_\ell 
\big[\det \bm\Delta_{1\perp}(M^2_{1,\, \ell})\big]^{-{1\ov 4} d_{\ell}^{(1)} }\
\big[ \det\bm\Delta_{0}(M^2_{0,\, \ell}) \big]^{-2d_{\ell}^{(0)} }\,,\\
&  \qquad  
M^2_{1,\ell} = 6 + 4(\ell^2 + 4 \ell +2) \,, \ \ \qquad  \ \  M^2_{0,\ell} = 3  + 4 \ell(\ell+2) \,.\la{0881} \ea
The fermion contribution to the partition function  has the same general  form as in the \Ia\  case, \ie\
is given by \rf{3.36}, \rf{3.39}  (see Appendix  \ref{fer}). In the  
  present case we  get   $  m_{_{\rm F}}=\frac{3}{2}$, see  (\ref{D.17}),\rf{D177}. 
  Thus after expanding in spinor harmonics on $S^3$ 
   the effective  mass of the fermions 
 in $\ads_3$ appearing in the squared Dirac operator  in \rf{320}  here is 
 \be\la{99}
  { \hat M}_{{1\ov 2},\, \ell } =M_{{1\ov 2},\, \ell}\pm m_{_{\rm F}} = (2\ell + 3) \pm \tfrac{3}{ 2} \,.  
  \ee
The conformal dimensions  corresponding to the above $\ads_3$ mass spectrum are found as in \rf{3.37}--\rf{333}
using \rf{3.54}. For the four   scalars $\z_p$  with mass $M^2_{\ell,0}$ in \rf{088}   and the  two 
mixed scalars  with masses in \rf{777}  we get 
\be\la{999} 
\Delta^{(0)}_{\z_p} = 2 \ell +3 \,, \qquad \ \ \    \Delta^{(0)}_{+} = 2\ell+6\,, \qquad \ \ \ \Delta^{(0)}_{-} = 2\ell\,. \ee
For the transverse vector   we get 
from  \rf{341},\rf{3.37}   and  \rf{0881} 
\be\la{1991} 
\Delta^{(1)} = 2  \ell +5 \,.\ee
For the fermions 
from \rf{3.42} and  \rf{99}  we get $\Delta^{({1\ov 2})}_{\pm} = 1 + (2 \ell + 3) \pm \tfrac{3}{2}$, \ie\ 
\be \la{000}
\Delta^{({1\ov 2})}_{+} = 2 \ell + \tfrac{11}{2} \,, \ \ \ \ \ \ \ \ \ \qquad
\Delta^{({1\ov 2})}_{-} = 2 \ell + \tfrac{5}{2} \,. 
\ee

\subsection{Probe \II  \la{s33}}

The derivation  of fluctuations near the solution \rf{2.49} 
 describing M5 brane  wrapped on $\ads_3 \subset \ads_4$ and $S^3\subset S^7$ 
is very similar to the one in the  \Ib\  case (the two cases are, in fact, closely related by an analytic continuation, see 
Appendix  \ref{s34}).\foot{Bosonic  M5 brane fluctuations  near   this solution  written in different coordinates \ci{Arean:2007nh} were previously studied  in \cite{Fiol:2010wf,Fiol:2010un}.}

In the static gauge where the  fluctuations of $\ads_3$   and $S^3$  coordinates in \rf{2.45} are  set to zero we are left with the  fluctuations of $u$  and of $\theta $  and $S'^3$ (and also of the 2-form field $B_{ij}$  and fermions).
  We parametrize them  like in \rf{3.2}  or  in \rf{342} ($p=1,2,3,4$) 
\be \la{3420}
ds^{2}=d^{2}\theta  +\sin^{2}\theta\,  ds^{2}_{S'^{3}} = {d \z_p d\z_p \ov (1+ \four \z^2)^2} = 
d \z_p d\z_p  + ... \,, \ \ \ \ \ \  \qquad   u = u_{0}+ U \,. \ee
The perturbed induced metric   corresponding to (\ref{2.45}) is then (\cf\ \rf{343}) 
\ba
&ds^{2} = \LL^{2}\Big\{ (1+ \vk^2) ds^{2}_{\ads_{3}}+ 4  ds_{S^{3}}^{2}  \lp
\qquad \ \ \ + dU^{2} +\big[\vk^{2}+2\vk\sqrt{1+\vk^{2}}\, U+(1+2\vk^{2})\, U^{2}\big]ds^{2}_{\ads_{3}} + 
  4 d\z_p d\z_p  - 4 \z_p \z_p ds^{2}_{S^{3}}\Big\}+... \,.
  \la{3431}
\ea
From \rf{2.47} and \rf{2.49} we get  (\cf\ \rf{445},\rf{446}) 
\ba\la{4451}
&\sfH_{ijk} = -\LL^{3}\ff(u)\, [\vol_{\ads_{3}}]_{ijk}-8\LL^{3}\vk\, [\vol_{S^{3}}]_{ijk}+\f_{ijk},\\
&\la{4461}
\ff(u) = 3\sinh u+\sinh^{3}u = \vk(3+\vk^{2})+3(1+\vk^{2})^{3/2}\, U+\tfrac{9}{2}\vk\,(1+\vk^{2})\, U^{2}+\cdots.
\ea
Computing $\sqrt{-|G_{ij}+\hat\sfH_{ij}|}$ we find  for 
 the quadratic fluctuation  terms (linear terms  will eventually drop out  as in the two  previous cases and  so 
 we omit them here)
\ba
&\sqrt{-|G_{ij}+\hat\sfH_{ij}|} = 4\LL^{6}(1+\vk^{2})^{2}\sqrt{-g_{A}}\sqrt{g_{S}}\ \Big\{
1+
\frac{1}{8(1+\vk^{2})}\big[\bg^{ij}\partial_{i}U\partial_{j}U+12(1+3\vk^{2})\,U^{2}\big]\no\\
&\qquad \qquad \ \ \ \ \ \ \ \qquad \ \  \ \ \ \ +\frac{1}{2(1+\vk^{2})}\big(\bg^{ij}\partial_{i}\z_p \partial_{j}\z_p -3\z_p \z_p\big)
+\TermFF+\TermMix\Big\}+... \,.  \la{3611}
\ea
Here the effective \adsst metric  $g_{ij}$  is  again independent of the parameter  $\vk$, and is given by  
\be
\la{3.92}
\bg_{ij}d\xi^{i}d\xi^{j} = \tfrac{1}{4}ds^{2}_{\ads_{3}}+ds^{2}_{S^{3}}, \qquad \ \ 
\sqrt{-g}=\tfrac{1}{8} \sqrt{-g_{A}}\sqrt{g_{S}}\,, \qquad \ \ \ 
L_{A}=\ha \,, \ \ \  L_{S}=1\,. 
\ee
Also, we  get   (\cf\ \rf{1333},\rf{135})
\ba
\TermFF = \frac{1}{768\LL^{6}(1+\vk^{2})^{2}}\sum_{i,j,k\neq 1}\f_{ijk}\f^{ijk}, \quad \quad
\la{3.94}
\TermMix = \frac{\vk^{2}}{256\LL^{3}(1+\vk^{2})^{3/2}\, \sqrt{g_{S}}}\, \eps^{021\ell m n}\f_{\ell mn}\, U\,.
\ea
The contribution of the  WZ term in the M5 brane action \rf{2.1} is (\cf\ \rf{0335})
\ba
& \tfrac{1}{2}\int \sfH_{3}\we C_{3} = 
\tfrac{1}{2}
\int \big(-8\vk\LL^{3}\vol_{S^{3}}+\f_{3}\big)\we \LL^{3} \ff(u)\vol_{\ads_{3}}
\no \\
&\to \ \  \LL^{3}\int d^{6}\xi\, \sqrt{-g_{A}}\, \sqrt{g_{S}}\, \Big[18\vk^{2}(1+\vk^{2})U^{2}
+\tfrac{3}{2\sqrt{g_{S}}}
(1+\vk^{2})^{3/2}\f_{345} U + ...\Big]\,. \la{3641}
\ea
The $U^2$ term  here cancels  against 
the $\vk$-dependent part of the $U^2$  term in \rf{3611}.
As a result,   putting $\f_3\f_3$ terms  into the  covariant form as in \rf{3.63}   and rescaling the fields we get 
for  the bosonic part of the quadratic fluctuation action\foot{Closely related expression was found in \ci{Fiol:2010un}.}
\be
\la{3.99}
S_{2} =- \int d^6 \xi \sqrt{-g} \Big[   \tfrac{1}{2} ( \bg^{ij}\partial_{i}U\partial_{j}U+12\,U^{2})
+ \tfrac{1}{2} (\bg^{ij}\partial_{i}\z_p\partial_{j}\z_p-3\z_p\z_p )
+ \tfrac{1}{12}\f_{ijk}\f^{ijk}+\tfrac{6}{\sqrt{g_{S}}}\, \f_{345}\, U\Big]  \,.
\ee
This  action  has the same structure as in the \Ib\ case in \rf{3.63}   
so the   derivation  of the corresponding  bosonic partition function follows the same steps as in \rf{3.64}--\rf{3466}
with the difference  being in the values of the coefficients.  Namely, taking into account that here 
 $L_{A}=\frac{1}{2}, \ L_S=1 $  we get 
 (setting  $B_{rs}= \tfrac{1}{\sqrt{\bg_{S}}} \eps_{rsu}  V^{u}, \  V_{r} = V_{r}^{\perp}+(\nabla_{S})_{r}\, P$ and
   $\nabla^2 = \nabla^{2}_{A}+\nabla^{2}_{S}$)
 \ba
\la{3.640}
&\tfrac{1}{12}\f_{ijk}\f^{ijk} \  \to\  \four\big[B_{ab}(-\nabla^{2}-8)B^{ab}+B_{rs}(-\nabla^{2}+2)B^{rs}+2B_{ar}(-\nabla^{2}-6)B^{ar}\big],
\\[0.5em]
\no
&L_{\rm mix} = -\tfrac{1}{2}\sqrt{-g_{A}}\, \sqrt{g_{S}}\, \big[\partial^{i} U\partial_{j}U+12 U^{2}+\tfrac{1}{2}
B_{rs}(-\nabla^{2}_{S}-\nabla^{2}_{A}+2)B^{rs}
\big]+3\sqrt{-g_{A}}\, U\,\eps^{rst}\,\partial_{r}B_{st} \\
&\ \ \ \ \ \ = -\tfrac{1}{2} \sqrt{-g_{A}}\,\sqrt{g_{S}}\, \Big[ \partial^{i} U\partial_{j}U+12U^{2}+V_{r}
(-\nabla^{2}_{A}-\nabla^{2}_{S}+2)V^{r}-12 \, U\,\nabla_{r}V^{r}\Big]  \no \\
&\ \ \ \ \ \  =  -\ha  \sqrt{-g_{A}}\, \sqrt{g_{S}} \  V_{r}^{\perp}(-\nabla^{2} +2)V^{\perp r}  + L_{U,P} \,,\la{3650}\\[0.5em]
&\la{3730} L_{U,P}=
 - \ha \sqrt{-g_{A}}\, \sqrt{g_{S}}\, \big[  U(-\nabla^{2} )U+12U^{2}
+ P(-\nabla^{2})(-\nabla^{2}_{S}) P-12\, U\,\nabla^{2}_{S}P\big] \,. 
\ea
The resulting contribution to the partition  function is also analogous to \rf{77}--\rf{3466}
(we  use again  the expression for  $Z^{(+)}_{2} $ in \rf{z9}  now with $L_{A}=\ha $, $L_{S}=1$)
\ba \no 
&Z_{U,P} = \big[\det(\nabla^{4}-12\nabla^{2}+36\nabla_{S}^{2})\big]^{-1/2}\,, 
\\[0.5em] \la{3750}
&Z_{U,\,\rm mix} = { [\det (-\nabla^2) ]^{1/2}}\ 
{ [\det (\nabla^{4}-12\nabla^{2}+36\nabla^{2}_{S})]^{-{1}/{2}}}, \\[0.5em]
\no 
&Z_{\rm B}= Z^{(+)}_{2} \, Z_0^4\,  Z_{U,\,\rm mix}\no \\
 & \ \ \ \  =  \big[\det \bm\Delta_{1\perp,1\perp}(-6)\big]^{-1/4} 
\big[ \det\bm\Delta_{0,0}(-3) \big]^{-2} \ 
\big[ \det (\nabla^{4}-12\nabla^{2}+36\nabla^{2}_{S}) \big]^{-{1}/{2}}\,. \la{3751}
\ea
Expanding in harmonics on $S^3$ we get the following counterparts of  the \Ib\ relations  \rf{777}--\rf{088}:
\ba  & 
\nabla^{4}-12\nabla^{2}+36\nabla^{2}_{S}\ \ \to \ \ 
(-\nabla^{2}_A+M_{0_+,\, \ell}^{2})
(-\nabla^{2}_A+M_{0_-,\, \ell}^{2})\,,\no   \\
&
 \qquad M_{0_+,\, \ell}^{2}=(\ell+2)(\ell+6) \,, \qquad \qquad  \ \  M_{0_-,\, \ell}^{2}=\ell(\ell-4) \,, \la{779}
\\
\la{880}
& Z_{U,\,\rm mix} = \prod_\ell \Big[\det\bDelta_{0}(M_{0_+,\, \ell}^{2} )\ \det\bDelta_{0}(M_{0_-,\, \ell}^{2})\Big ]^{- {1\ov 2} d_{\ell}^{(0)}}\,,\\
\no 
&\big[\det \bm\Delta_{1\perp,1\perp}(-6)\big]^{-1/4}  \
\big[ \det\bm\Delta_{0,0}(-3) \big]^{-2}= \prod_\ell 
\big[\det \bm\Delta_{1\perp}(M^2_{1,\, \ell})\big]^{-{1\ov 4} d_{\ell}^{(1)} }\
\big[ \det\bm\Delta_{0}(M^2_{0,\, \ell}) \big]^{-2d_{\ell}^{(0)} } ,\\
&  \qquad  
M^2_{1,\ell} = -6 + (\ell^2 + 4 \ell +2) \,, \ \ \qquad  \ \  M^2_{0,\ell} =- 3  +  \ell(\ell+2) \,.\la{088} \ea
The fermion contribution to the partition function is 
 given by \rf{3.36}, \rf{3.39}   with the same  value  $  m_{_{\rm F}}=\frac{3}{2}$ as in the \Ib\ case
 (see   (\ref{D.21}),\rf{D178}).
Hence    the effective  mass of the fermions 
  in the squared Dirac operator in $\ads_3$  in \rf{320}  here is (\cf\ \rf{319},\rf{99}) 
 \be\la{990}
  { \hat M}_{{1\ov 2},\, \ell } =M_{{1\ov 2},\, \ell}\pm m_{_{\rm F}} =  (\ell + \tfrac{3}{2}) \pm \tfrac{3}{ 2} \,.  
  \ee
The  dimensions  corresponding to the above $\ads_3$ mass spectrum are found as in \rf{3.37}--\rf{333}
(here $L_A=\ha$ according to \rf{3.92}).
For the scalars   we get   from  \rf{0881}, \rf{779} 
\be\la{9991} 
\Delta^{(0)}_{\z_p} = \ha  \ell +\tfrac{3}{2} \,, \qquad \ \ \    \Delta^{(0)}_{+} = \ha \ell+3\,, \qquad \ \ \ \Delta^{(0)}_{-} = \ha \ell\,,  \ee
while for the vector in  \rf{088}  we have (\cf\ \rf{3.37},\rf{3.92})
\be\la{19910} 
\Delta^{(1)} = \ha  \ell +2 \,.\ee
For the fermions  \rf{3.42} and  \rf{990}   give  $\Delta^{({1\ov 2})}_{\pm} = 1 + ( \ell + \tfrac{3}{2}) \pm \tfrac{3}{2}$, \ie\ 
\be \la{0001}
\Delta^{({1\ov 2})}_{+} = \ha  \ell + \tfrac{5}{2} \,, \ \ \ \ \ \ \ \ \ \qquad
\Delta^{({1\ov 2})}_{-} = \ha  \ell + 1  \,. 
\ee


\section{One-loop free energies of  \Ia, \Ib\ and \II\ probes}
\la{sec:free-energy}

To compute the free  energy $F=-\log Z$  corresponding to the one-loop partition functions  found 
above in the  cases  \Ia, \Ib\ and \II\  it remains to sum  up the contributions of each of the (scalar, vector, spinor) 
 field  on $\ads_3$ of  the corresponding mass, or, equivalently, scaling dimension $\Delta$     (see \rf{339},\rf{333}, \rf{999}--\rf{000}  and \rf{9991}--\rf{0001}). The  single-field contributions  are given by the standard relations
 (see, \eg,  \cite{Camporesi:1994ga,Giombi:2016pvg})\footnote{These are for a  scalar, 
 transverse $\ads_{3}$ vector and a  single fermion  counted as one real degree of freedom. We have total of 6  scalars 
 in \rf{327},\rf{3466}  and \rf{3751}.
 The vector  corresponds to  two degrees of freedom  and the fermions carry a total of 8 degrees of freedom.
 }
\bea
\la{4.2}
 F_{0}(\Delta) &= -\tfrac{1}{12\pi}(\Delta-1)^{3}\,\vol(\ads_{3})\,,  \\
 F_{1}(\Delta) &= -\tfrac{1}{12\pi}(\Delta-1)[(\Delta-1)^{2}-3]\,\vol(\ads_{3})\,, \\
 F_{\half}(\Delta) &= -\tfrac{1}{12\pi}(\Delta-1)[(\Delta-1)^{2}-\tfrac{3}{4}]\,\vol(\ads_{3})\,. 
 \eea
Each $\ads_3$ field  contribution  should  be  taken with its  $S^3$ degeneracy factor
(given in  \rf{340},\rf{341},\rf{319}).

For consistency with the underlying supersymmetry  (see  Table \ref{tab:preserved-susy})
the 
$\ads_3$   fields  should be organized into  short 
supermultiplets  (labelled  by  $\ell$) 
of the $OSp(4^{*}|2)\times OSp(4^{*}|2)$   superalgebra in the cases {\bf Ia} and {\bf Ib}, 
and of the $OSp(4|2, \mathbb R)\times OSp(4|2,\mathbb R)$ in the  case {\bf II}. 
A detailed way  how  to do this  is discussed in Appendix \ref{mult}.

In all the  three  cases the resulting supermultiplets 
 contain individual   fields  with  masses or dimensions  corresponding to  particular 
 shifted  values of  the $S^3$ level $\ell$ (or, equivalently, they include   fields from different levels $\ell$). 
  Remarkably,  these shifts turns out to be 
same as in the spin-1 short supermultiplet of the $SU(1,1|2)\times SU(1,1|2)$ supergroup
that describes the standard massless 
 $(2,0)$ tensor multiplet  on  the supersymmetric $\ads_{3}\times S^{3}$ vacuum 
 of $(2,0)$ 6d supergravity  \cite{Deger:1998nm,deBoer:1998kjm}. 
 The corresponding spectrum is formally the same  as  we found  above 
  in the \Ia\ case,  but  the  organization of  states in the supermultiplet (\ie\ the assignment of the $\Delta$-values) 
  is different  as  the corresponding   superalgebras are  different
  (see Appendix \ref{mult} for details). 
 

In Table \ref{muld}    we present the   summary  of the corresponding 
spectra in the three  cases   and also, for comparison,  for the massless (2,0)  tensor multiplet on \adsst  corresponding to 
the $SU(1,1|2)\times SU(1,1|2)$  symmetry. 

There $A^{(\ell)}_\perp$  stands for the transverse   vector originating  from  the $B_{ar}$ 
 component of the 2-form field  on \adsst  or the operator $\bm \Delta_{1\perp,1\perp}$ upon reduction  on $S^3$
 (\cf\ \rf{341}).  Also,  $\zeta^{(\ell)}_p$   represents the modes of the 4 scalar transverse fluctuations, 
$\vp_\pm $    denote the two scalar  modes corresponding to the  scalar  part of $B_{rs}$ component 
of the  2-form field  mixed  with the    transverse scalar $U$ or $\Te$  in \rf{88} or \rf{880},  
 and $\psi_\pm$ are the two  sets of  the fermions  having different masses in  \rf{99},\rf{000}  and  \rf{990},\rf{0001}. 

In  \Ia\ case there is no scalar mixing so  that $\vp^{(\ell)}_+ $   and $\vp^{(\ell)} _-$  are both massless  and thus have equal   dimension 
(given in  \rf{339}) 
for the equal  values of $\ell$; 
 the same applies  to the  corresponding massless fermions $\psi_+$  and $\psi_-$. 
 $\dd_{S^3}$ denotes the  total $S^3$  degeneracy   including the  factor of  the  relevant  number $n$  of the number of fields, 
 \ie\ $\dd_{S^3} \to  n \times d^{(s)}_\ell$. 

\begin{table}[h]
\be
\begin{array}[t]{c|c|cccc}
 &  & {\bf (2,0)}   & \text{\bf Ia} & \text{\bf Ib} & \text{\bf II} 
  \\
\ads_{3}\ \text{field} &  \dd_{S^{3}}   & \Delta & \Delta & \Delta & \Delta \\
\midrule
A_{\perp}^{(\ell)} & 2(\ell+1)(\ell+3) &
\ell+3  &  \ell+3  & 2\ell+5  & \frac{1}{2}\ell+2 \\
\midrule 
\vp^{(\ell+2)}_{-} & (\ell+3)^{2}  
& \ell+2 &\ell+4& 2\ell+4 & \frac{1}{2}\ell+1   \\
{\z}^{(\ell+1)}_p & 4\times (\ell+2)^{2} 
& \ell+3  &\ell+3&2\ell+5 & \frac{1}{2}\ell+2  \\
\vp^{(\ell)}_{+} & (\ell+1)^{2} 
&  \ell+4 &\ell+2 &2\ell+6 & \frac{1}{2}\ell+3  \\
\midrule
\psi_{+}^{(\ell+1)} & 4\times(\ell+2)(\ell+3)  
& \ell+\frac{5}{2} &\ell+\frac{7}{2}&2\ell+\frac{9}{2}& {\frac{1}{2}\ell+\frac{3}{2}}  \\
\psi_{-}^{(\ell)} & 4\times(\ell+1)(\ell+2) 
 & \ell+\frac{7}{2}&\ell+\frac{5}{2}&2\ell+\frac{11}{2}& {\frac{1}{2}\ell+\frac{5}{2}}   \\
\bottomrule
\end{array}\notag
\ee
\caption{$\ads_3$   supermultiplet    structure of the fluctuation spectra in  \Ia, \Ib, \II\  cases
and also  for the massless (2,0) tensor multiplet on the \adsst space.
}
\la{muld}
\end{table}

In all these cases we can check the  balance of the numbers of bosonic and fermionic states 
and  also  the   validity of the two additional   sum rules for  each supermultiplet  
\be
\la{4.1}
\sum_{\rm multiplet}  (-1)^{\rm F} \dd_{S^{3}}=0\,, \qquad\ \ \ 
\sum_{\rm multiplet} (-1)^{\rm F} \dd_{S^{3}}\, \Delta=0\,, \qquad\ \ \ 
\sum_{\rm multiplet}  (-1)^{\rm F} \dd_{S^{3}}\, \Delta^{2}=2 \,.
\ee
Here  the sum  is over the set of fields   in the   supermultiplet  corresponding to a  fixed  value of $\ell$.

 To compute  the total value of  the free energy $F$  we may  first   find the contribution $F_\ell $ of each 
 supermultiplet with fixed $\ell$  and then sum over  $\ell$, \ie\
 $F=\sum_\ell F_\ell $.
 For example,  in the case of the (2,0)  multiplet in Table \ref{muld}  we get, using the expressions in \rf{4.2}, 
 \ba
 F^{\bf (2,0)}_\ell= & \ d_{\ell}^{(1)}F_{1\perp}(\ell +3)+d_{\ell+2}^{(0)}F_{0}(\ell +2 )
+4\,d_{\ell+1}^{(0)}F_{0}(\ell + 3 )
+d_{\ell}^{(0)}F_{0}(\ell + 4 )  \no \\
& - 4\, d_{\ell+1}^{(\half)}\, F_{\half}(\ell+\tfrac{5}{2}) -  4\, d_{\ell}^{(\half)}\, F_{\half}(\ell +\tfrac{7}{2}) \,, \la{4.3} \ea
where  according to  \rf{340},\rf{341},\rf{319} the degeneracies are given by 
\ba  d^{(0)}_{\ell} =  (\ell+1)^{2}\,,\qquad \ \ \ d^{(1)}_{\ell} = 2(\ell+1)(\ell+3)\,, \qquad\ \ \  d^{(\half)}_{\ell} = (\ell+1)(\ell+2) \,. \la{4.4}
\ea
Using \rf{4.2}   we then find  from the data in Table \ref{muld} that 
\be 
 F^{(\bf Ia)}_\ell  = F^{(\bf Ib)}_\ell  = -\tfrac{3}{2\pi}(\ell+2)\, \vol(\ads_{3})\,, \qquad \ \  F^{\bf (2,0)}_\ell = \tfrac{1}{3} 
 F^{(\bf Ia)}_\ell \,, 
 \qquad \qquad F^{(\bf II)}_\ell=0
   \,. \la{4.5}
 \ee
 These surprisingly simple expressions  follow from cancellations  that are due to the underlying supersymmetry.
 
 Indeed, if  one  computes the free energy 
 without taking into account the shifts of $\ell$   required  by the supermultiplet structure, 
 \ie\ by  just directly combining together the contributions of  all modes with a given $\ell$, 
 one gets the following  quartic polynomial expressions in $\ell$ (we use the notation  $F'_\ell$  for these
 ``no-shifts'' values) 
 \ba
F'^{\bf (2,0)}_\ell &= \tfrac{1}{6\pi}(1+\ell)(33+34\ell+15\ell^{2}+5\ell^{3})\, \vol(\ads_{3})\,,  \la{577} \\
F'^{(\bf Ia)}_\ell  &= \tfrac{1}{6\pi}(1+\ell)(9+16\ell+15\ell^{2}+5\ell^{3})\, \vol(\ads_{3})\,, \la{46}\\
F'^{(\bf Ib)}_\ell  &= \tfrac{1}{3\pi}(1+\ell)(63+91\ell+60\ell^{2}+20\ell^{3})\, \vol(\ads_{3})\,,\la{47}\\
F'^{(\bf II)}_\ell  &= \tfrac{5}{48\pi}(1+\ell)(18+14\ell+3\ell^{2}+\ell^{3})\, \vol(\ads_{3})\,.\la{48}
 \ea
 Summing  these expressions  over $\ell $ with a  UV cutoff  at large $\ell$  one concludes that
 total free energy contains  quintic (and lower-order)  UV divergences. The logarithmic divergences cancel out 
 as can be   shown independently  by starting with the general expression for the  partition function on \adsst (see  Appendix \ref{apdi}). 
 At  the same time, the sums over $\ell$ in the first three  cases in \rf{4.5}  are only quadratically divergent, which is  obviously a  consequence of maintaining supersymmetry-implied  multiplet structure at each  value of level $\ell$. 
 
 Let us note that the multiplet structure presented in  Table \ref{muld}  applies  for 
  $\ell\ge 0$  when all states are present (have positive $\dd_{S^3}$). 
For $\ell =-1$  there are additional states that also form a short supermultiplet (we include the factors of their degeneracies
from Table 3  taken at $\ell=-1$) 
\be \la{4.10} 
4\times   \vp^{(1)}_+ , \ \ \ \qquad   4 \times \z^{(0)}_+, \ \ \ \qquad  4\times 2\times    \psi_+^{(0)} \,. 
\ee 
This is again a balanced multiplet since the total number of states is $4+4-8=0$. 
Note that the contribution of 
\rf{4.10} to the total free energy is the same as  of a  general multiplet evaluated at $\ell=-1$ 
(at this  value all extra states have zero degeneracy and thus do not contribute).

Taking this into account,  the total    free energy should be given by summing the expressions in \rf{4.5} 
from $\ell=-1$ to $\infty$, or, equivalently, after shifting $\ell$ so that the sum starts from $1$, we get 
\ba\la{4.11} 
F_{\bf II}=\sum_\ell F^{(\bf II)}_\ell= 0 \,, &\qquad\ \ \ \  \ \ F_{\bf Ia}=F_{\bf Ib} = 3  F_{\bf (2,0)}=\sum_{\ell=-1} ^\infty F^{(\bf Ia)}_\ell 
=- \tfrac{3}{2\pi }{\mc C} _1  \vol(\ads_{3})\,, \\  \la{4.12}
&{\mc C}_1= \sum_{\ell=-1}^{\infty}(\ell+2) = \sum_{\ell'=1}^{\infty}\ell' 
\,. 
\ea
We conclude   that  while the one-loop free energy is manifestly zero in the probe \II\ case, 
it  requires   a  particular regularization to define it  in the other  two cases. 

The familiar Riemann  $\zeta$-function (with $\zeta(-1) = -{1\ov 12}$) is not appropriate here as $\ell$ is the radial 
quantum number of $S^3$ (and not, e.g, an $S^1$ mode number). 
A more natural 
 alternative is to  use  a   prescription  like $\sum_{\ell'=1}^{\infty}\ell' z^{\ell'}\big|_{z\to 1}$
or a  sharp  cutoff $\ell' < \La$ as   in   
\ci{Mansfield:2003gs,ArabiArdehali:2013vyp,Beccaria:2014xda,Bae:2017spv,Liu:2017ruz}.
Dropping singular terms one then finds that ${\mc C}_1=0$.


\section{Open questions \la{op}}

The main  open question   that  remains to be understood is if the subleading in $N$   terms in the b-coefficients in \rf{15},\rf{150} 
can be reproduced  in the framework of the semiclassical  M5  brane  \Ia\ and \Ib\  probes.\foot{As already mentioned in the Introduction, 
it  is   not  clear a priori   if an effective description of a defect or multiple M2 brane  in large $SU(N)$ 
  representation in terms of    an effective  M5 brane  may  apply  beyond the classical approximation.}
One speculative  suggestion  is that  a classical   probe action   should actually 
 be given by a  combination of the M5 and M2  brane actions, 
with  an  on-shell value  of the latter (wrapping $\ads_3$)  reproducing the required order $N$  terms in \rf{15},\rf{150} (recall that according to \rf{27} the M2  brane tension   is proportional to $N$).
 A problem    with this idea  is that it is not  clear  how to see why  both the 
 M5 and M2  contributions   should have the same 
$\ka$-dependent  factors that are required to match the expressions in \rf{15},\rf{150}. 

Assuming there is some ``classical action''  explanation  for the exact expressions in \rf{15},\rf{150}
 one would  then   to prove   that there are  no  further   quantum   M5 brane  corrections to the
 free energy and thus to the  b-coefficient. 
 At the  one-loop level discussed above  that  would   require 
the  choice of a particular regularization that sets to zero  the infinite-sum coefficient in \rf{18},\rf{4.12}. 
Such  a regularization  was  required in similar  supersymmetric conformal anomaly contexts 
 \ci{Mansfield:2003gs,ArabiArdehali:2013vyp,Beccaria:2014xda}.
  The   choice of regularization 
  should be motivated by  some  underlying symmetries of the  quantum 
M5 brane   theory on the  \adsst  space  that  may not be   manifest  after the explicit expansion in modes on $S^3$. 
 We already accounted for  the $\ads_3$ supersymmetry  by combining the contributions 
  of all states in a supermultiplet  at level $\ell$  before introducing a cutoff. Which are additional symmetries 
  that select  a specific  choice of the subtraction procedure  that leads  to ${\mc C}_1=0$
  remains to be understood.

 As for the  probe \II\  case  when the M5 brane is wrapped on \adsst in $\ads_4 \times S^7$,  
 the one-loop correction  was found to be manifestly zero (\cf\  \rf{20},\rf{4.5}). However,    
 it remains to find  the interpretation of the corresponding classical action \rf{1.9} in the dual $k=1$ ABJM  theory, 
 \ie\ to see which is the corresponding  $\half$-BPS  spherical defect   and  which is its b-anomaly coefficient.
 To try to shed  light on this, it  would be useful to generalize  the discussion of the probe \II\ case 
 to the  M5 brane probe  wrapping \adsst  in the M-theory background 
 $\ads_4 \times S^7/\mathbb Z_k$   dual to the level $k$ ABJM theory. 
 
 Similar issues  appear in trying to go beyond the classical M5 brane probe  action values  \cite{Mori:2014tca}
   in order to match the exact  expressions for d$_2$ 
 in \rf{1.12},\rf{120}. As was mentioned in the Introduction, that will involve considering quantum M5 branes in
 a   ``twisted'' $\ads_{7,\beta} \times \tilde S^4$  background 
 with  the $\ads_{3,\b}$   boundary being $S^1_\b \times S^1$.  It would be of interest to  compute the 
 corresponding   one-loop M5 brane correction 
 that   should be related to  order $N^0$ term in free energy and thus in d$_2$
  and  check that it again vanishes. 


\section*{Acknowledgements}
AAT   would like to thank S. Giombi for a collaboration at an early stage of this project  and many  discussions. 
We are   grateful to  N. Drukker,  J. Estes, B. Fiol, H. Jiang, O. Lunin, 
R. Rodgers,  D. Sorokin,  J. van Muiden and S. Yamaguchi  for important clarifications and discussions.
Part of this work was done during the  meeting ``Integrability in low-supersymmetry theories'' held in  Trani in July 2024 (funded by  COST Action CA22113, INFN and  Salento University). 
MB  is supported by the  INFN grant GAST.
AAT  is supported by the STFC Consolidated Grant ST/X000575/1.

\small 

\appendix

\section{Fermionic fluctuation operator
\la{fer}}

In this Appendix we will determine the structure of the   analog of the Dirac operator 
appearing in the quadratic fermionic part of the M5 brane action. This fermionic part  
  complements \ci{Bandos:1997ui} the bosonic part of the PST  action in \rf{2.1}.
 The  determinant of this Dirac   operator 
  contributes (after a  $\kappa$-symmetry gauge fixing)   to the one-loop  M5 brane partition function (\cf\ \rf{3.36}).
  
  This fermionic operator may be  found from the covariant M5 brane equations of motion 
  \ci{Howe:1997fb,Kallosh:2005yu}.\foot{The fermionic equations are given in
  an  implicit form in \cite{Howe:1997fb}. The explicit Dirac-like equation 
  for the spinor variable \(\vartheta\) was obtained  (with manifest  \(\kappa\)-symmetry  and for generic fluxes)  
  to linear order in $\vartheta$   in \cite{Kallosh:2005yu}. Let us 
  note that the PST formalism produces a different \(\kappa\)-symmetry generator \cite{Bandos:1997ui}, but as was shown in 
   \cite{Bandos:1997gm}, the two  corresponding \(\kappa\)-symmetry 
   transformations are equivalent up to a redefinition of the gauge   parameter.}
  The presence of a  non-trivial  3-form $H_3$   background for the  three  solutions  \Ia\ \rf{2.13}, \Ib\ \rf{228}   and \II\ \rf{2.49}
   makes the determination of the explicit form of this  Dirac  operator more complicated 
    than in the previously discussed $S^1 \times S^5$ and $S^1 \times \ads_5$   M5-brane  cases \cite{Beccaria:2023cuo,Jiang:2024wzs}  where the  background $H_3$  field   was  zero. 

 In this Appendix we will  denote the 11d target space indices by capital letters
 \(M,N,P=0,1,2,...,9,\XX\) and the  6d world-volume indices by lower-case Latin letters \(i,j,m,n,.. = \hat 0, \hat 1,.., \hat 5\).
 We will   use   ``hat''   to indicate  quantities induced on the world-volume, 
 introducing, in particular,  the notation $\hat g_{ij}$  for the induced metric which is equivalent to $ G_{ij}$  used in \rf{21}. 
 We will  underline the  tangent space indices  so that 
 \be
 \label{D.1}
 ds^2_{11} 
 = G_{M N } dx^M dx^N\,, \qquad G_{M N } 
  = \eta_{\ul M \ul N}e^{\ul M} _M e^{\ul N}_N
  \,,
  \qquad\qquad
  ds^2_{\text M5} 
  = \hat g_{ij } d\xi^i d\xi^j \,, \ \ \ \   \hat g_{ij } 
   = \eta_{\ul i \ul j} \hat e^{\ul i} _i \hat e^{\ul j}_j
   \,.
 \ee
 We will use the following definitions 
 \ba 
& \{\Gamma_{\ul M } , \Gamma_{\ul N}\} 
 =
 2 \eta_{\ul M\ul N}
 \,, \qquad \Gamma_{\ul 0 \ul 1 \ldots \ul 9 \, \ul \XX} = 1
 \,,
 \qquad
 \cd_M=\partial_M
 + \tfrac14  \omega_{M}^{\ul  R \ul S}\Gamma_{\ul R \ul S}  
 \,,\label{D12}\\
 &D_M =  
  	\cd_M  	+   \TF_M	\,,\qquad \qquad 
  	  \TF_M = -   \tfrac1{288} \left(
  	 	 			\Gamma_{PNKLM} 
  	 	 		- 8G_{MP} \Gamma_{NKL} 
  	 	 		\right) F^{PNKL} \,.  \la{D.2}
 \ea
 For the two backgrounds of interest  (I)  $\ads_7 \times S^4$ in \rf{2.5},\rf{2.6}  and  (II)  $\ads_4 \times S^7$ in \rf{2.45},\rf{2.46}   we have 
 \begin{equation}\label{D.3}
 \TF^{\text (I)}_M = e_M^{\ul M }\ \Gamma_ {\ul M} \  \Gamma^{\ul 7 \ul 8 \ul 9 \ul \XX} 
 \cdot 
 \begin{cases}
 - \frac{ 1 }{ 2 }  {L }^{-1}
 &
 \ads_7
 \\
\phantom{-} {   L }^{-1}
 &
 S^4
 \end{cases}
 \,,\qquad\qquad
 \TF^{\text {(II)}  }_M = e_M^{\ul M }\ \Gamma_ {\ul M} \  \Gamma^{\ul{0123}} 
 \cdot 
 \begin{cases}
 \phantom{-}  \frac{ 1 }{ 2 }{\LL}^{-1}
 &
 \ads_4
 \\
 - \frac{ 1 }{   4}{ \LL }^{-1}
 &
 S^7
 \end{cases}
\,,   \end{equation}
 where   we indicated the overall scaling factors on the two subspaces 
 (\ie\ for index $M$ in  $\ads_d$ or $S^d$).\foot{In our notation (\cf\ \rf{2.5},\rf{2.45}) 
 the  coordinates and thus  derivatives like $\nabla_M$  are dimensionless
  and these   factors of $L^{-1}$ or $\LL^{-1}$  cancel against  similar factors 
  in the vielbein $e_M^{\ul M }$ in \rf{D.1}.}


Let us first  recall  the connection between the  bosonic part of the 
PST  action \rf{2.1}  and the covariant  formulation of the M5 brane  equations   of motion 
  in terms of  the  {(anti) }self-dual 3-form \(h_{ijk }\)
  \ci{Howe:1997fb}.\foot{It should not be confused with the fluctuation field $\f_{ijk}$ used in the text (as, \eg, in  \rf{34}).} 
  One  defines 
  (\cf\ \eqref{2.3})
\begin{equation}\label{D.4}
h^{ij} 
 =   {h}^{ij k }\, V_{k}
 = {-} \frac{1}{6\sqrt{-\hat g }} \eps^{ijks  mn} \,  {h}_{s  m n} \, V_{k}
 \,,\qquad\qquad 
 V_{ k} \equiv  
 \frac{1}{\sqrt{-(\del a)^2 }}\del_{k} a  \,,
\end{equation}
where  we used the  anti-self-duality  condition  on ${h}^{ijk }$. As in   \rf{211}, we will use the gauge \  \(a= \xi^{ 1}\).
The field   $h^{ij} $   is  related  to \(\hat\sfH^{ij}\) in \eqref{2.3}  by  \cite{Bandos:1997ui,Bandos:1997gm} 
\begin{align}\label{D.5}
h_{ij} = \tfrac 14 \hat {\sfH}_{ij}
+ \tfrac 14 \frac{	
	\hat { \sfH}_{i m} \, 
	\hat { \sfH}^{m n} \,
	\hat { \sfH}_{n j}
+ \  \frac14 \hat { \sfH}_{m n }   
  \hat { \sfH} ^{m n  }  \hat { \sfH}_{ij} }{1  +
\sqrt{-|\hat g _{rs} + \hat{ \sfH}_{rs}  |}  + \frac 14  \hat { \sfH}_{rs }   
  \hat { \sfH} ^{ rs }   } 
  \,.
\end{align}
The general expression for the  linearized  Dirac-like    fermionic equation
 is  given by  \cite{Kallosh:2005yu}\footnote{We follow here 
the  notation for the M5-brane fermionic  operator in  \cite{Beccaria:2023cuo,Jiang:2024wzs}.
Note that the sign in front of the second term in $\Gamma_*$ is convention-dependent  and the final 
results will not depend on its  choice.}
\ba\label{D.6}
&
{\bre}^{i} _{ \ul i} \ 
\Gamma^{\ul i} \ \hat D_{i} \big[ (1 +\Gamma_*)   \vartheta\big] =0
\,,\qquad  \qquad \ \ \ 
\Gamma_*  =  \Gamma_{ \ul 0 \ul 1\ul  2\ul  3\ul  4\ul  5 } + 
\tfrac13 
h^{\ul i \ul j \ul k }\Gamma_{ \ul i \ul j \ul k}
\,,
\\
&\Gamma_{i} = \partial_{i} x^M \Gamma_M
\,,\qquad 
\hat D_{i} =  \partial_{i} x^M D_M  \,, \qquad \quad 
(\Gamma_{ \ul 0 \ul 1\ul  2\ul  3\ul  4\ul  5 })^2=1 = (\Gamma_*)^2 \,, \la{D7}
\ea
where $D_M$ was defined in \rf{D.2}.
Note that   the  second term in $\Gamma_* $  in \rf{D.6} 
 squares to zero ($h_{ijk}$ is  assumed to be anti-self-dual). 
  $ {\bre}^{i} _{ \ul i}$ in \rf{D.6}
 is  the  \emph{effective} sechsbein  corresponding  to the effective metric \( g_{ij}\)  appearing in the  Lagrangian  for the 
  bosonic fluctuations  in   \rf{3.7},\rf{3.54},\rf{3.92}.
There is  also the following relation  between $ {\bre}^{i} _{ \ul i}$,  
 the induced  metric  \(\hat g_{ij} \) and \(h_{a c } \) 
 \cite{Bandos:1997ui}
\begin{align}\label{D.7}
\eta^{\ul i \ul j}  \, {\bre}^{ i }_{ \ul i }  \, \hat e^{ j }_{ \ul j  }
 & = \hat g^{ij} +2
\big[
\hat g ^{ij}   h^{ m n }  h_{ m n } 
+ 2 V^{ i} V^{j}h^{ m n }  h_{ m n } 
+ 4   h^{ ik }  h\indices{ _{k} ^ {j} } 
-   \tfrac 1 {\sqrt{-\hat g}}   V^{ (i  }  \varepsilon^{j) mnrs k   } V_{m} 
 h_{n r}  h_{s k}  
\big] \,.
\end{align}  
Below we will  derive the explicit expression for the  operator in \rf{D.6} 
in the  three cases \Ia, \Ib\  and \II.
We will show that this  effective Dirac  operator appearing  in the quadratic fermionic 
action after  a $\kappa$-symmetry gauge fixing  has indeed 
the form given in \rf{3.39},\rf{3399}   with $m^{(\bf Ia)}_{\rm F}=0$ and $m^{(\bf Ib)}=m^{ (\bf II)}_{\rm F}={3\ov 2}$.

Let us  note that in  the analysis of the bosonic fluctuations in section 3, when discussing the effective metric
in \rf{3.7},\rf{3.54},\rf{3.92} that enters the  quadratic  fluctuation Lagrangians  in  \rf{3063},\rf{3.63},\rf{3.99} 
 we  scaled out the overall factor of $L$ or $\LL$ (as well as numerical factors that can be absorbed into the fluctuation fields).
Similarly, in \eqref{D.6} the overall scale of the effective  $ {\bre}^{ i }_{ \ul i } $    will be  irrelevant (it can be absorbed
into $\vartheta$ in the  quadratic fermionic action) so we will also  drop it in the expressions below.\footnote{Note also that there is also no difference between the  covariant derivatives
for the effective and the induced \adsst   metrics   as they are different  only by the values of the radii
 which the spin connection is independent of.}
 We will use  the labelling of the  coordinates  as given in \rf{lab1} and \rf{lab2}.\foot{Let us note for completeness
that the  global supersymmetry preserved by the M5 brane probes  is determined 
  by the solutions of the Killing spinor equations  of the target space backgrounds
  (see, \eg,  \ci{Lunin:2007ab,Gupta:2021hko}) 
$\cd_M \varepsilon = - \mathrm F_M \varepsilon $
subject to the $\kappa$-symmetry gauge condition on the brane
$
\Gamma_* \varepsilon = \varepsilon.
$
For example, in the case \II\   we get $\varepsilon = 
e^{\frac 12 \theta \Gamma^{\ul{01234} }} \, \Sigma_{S^3}\, \Sigma_{S'^3}\,  \Sigma_{\ads_4} \varepsilon_0
$
where  \(\Sigma \)'s  are the matrices that  define  the Killing spinors on \(S^3\),  \(S'^3\) and \(\ads_4\) respectively.  
This   solution for $ \varepsilon$  preserves \(\mathcal N=8\) supersymmetry. 
Solving the \(\kappa\)-symmetry gauge condition imposes an additional projector   and therefore halves the amount of supersymmetry preserved by the brane configuration.
}

\subsubsection*{Probe \Ia}

From  the  form of the   background  in 
 (\ref{2.5}) and the evaluation of \eqref{D.7}, we obtain the induced and the  effective sechsbeins
  as 
\begin{equation}\label{D.8}
\hat e_{i }^{\ul i } = L \,  (\sqrt {1 + \kappa^2}  \ e_{\ads_3} , \    \kappa  \  e_{S^3 }   )
\,,
\qquad\qquad \qquad 
{\bre}^{\ul i}_i=  
  (   e_{\ads_3} , \ e_{S^3})
\,. \end{equation}
The corresponding metrics   are the induced  \adsst  one as  in \rf{32}   and 
 the effective one  of the equal-radii   \(\ads_3  \times S^3\)  space 
 as  in (\ref{3.7}).
The pull-back of the covariant derivative on the world volume is 
\begin{equation}\label{D.9}
\begin{aligned}
\hat \nabla_i= \partial_{i } x^M \nabla_M  
&
\equiv \cd_{i} 
+ \tfrac 12 \omega_{i} ^{ \ul  i \ul 6} \Gamma_{\ul  i \ul 6 } \,,
 \qquad \qquad \cd_i \equiv \del_i
+ \tfrac14 \omega_{ i }^{\ul   j  \ul  k}\Gamma_{\ul   j  \ul  k}\,, 
\end{aligned}
\end{equation}
where there is no sum over  the index \(\ul  i\) and $\nabla_i$  is the covariant spinor derivative  corresponding to the effective
${\bre}^{\ul i}_i$   in \rf{D.8}.

Using  \rf{D.3}    and \rf{D.9}  the two  terms  $\nabla_M$ and  $\TF_M$  in $D_M$ in \rf{D.2} lead  to the 
following two  contributions to the 
operator ${\bre}^{i} _{ \ul  i} \ 
\Gamma^{\ul  i} \ \hat D_{i}$ in \rf{D.6}:
\ba\label{D.10}
{\slashed \cd }
 			+ \tfrac 32   \big(  \sqrt{1+ \kappa^2} + \kappa\big)    \Gamma_{\ul 6}
\,, \qquad\qquad \ \ \ 
\bre _{\ul  i }^{ i } 
\  \Gamma^{\ul  i }
\ \TF_ { i }
= -
\tfrac 3 2   \big(  \sqrt{1+ \kappa^2} + \kappa\big)      \Gamma^{\ul{789\XX}}
\,,\qquad \ \ \  {\slashed \cd } \equiv  {\bre}^{i} _{ \ul i} \ 
\Gamma^{\ul  i} \nabla_i \,. 
\ea
As a result,  the fermionic equation of motion  in \rf{D.6} 
may be written as 
\begin{equation}\label{D.11}
\mathcal D\big[ (1 + \Gamma_* )\vartheta\big] =0
\,,
\qquad\qquad \ \ 
\mathcal D = 
i  
{ \slashed \nabla} 
-  \tfrac 3 2  i  \big(  \sqrt{1+ \kappa^2} + \kappa\big)       \,
	\big(    \Gamma_{\ul{012345}} - 1 \big)\Gamma_{\ul 6} \,, 
\end{equation}
where \(i {\slashed \cd }\) is the Dirac operator  corresponding to 
 the effective \adsst  sechsbein   in  \eqref{D.8}.  
 Here we  used that \(  \Gamma_{\ul{6789\XX}} =   \Gamma_{\ul{012345}}\), which follows from
  \(\Gamma_{\ul 0\ul 1...\ul 9\ul \XX}=1\) in \rf{D12}. 

From \eqref{D.5}  and the explicit form of $H_3$ in \rf{2.13}    we find  that the only nonzero components
 of \(h_{ijk}\)  correspond  to  $h_{\uh 0 \uh 2  } $ (we scale out $L^2$ factor and $\uh 0, \uh 1, ...$  are the values of the  world-volume indices)
\be 
h_{\uh 0 \uh 2  } = \ha \,  \big(  \sqrt{1+ \kappa^2} - \kappa\big) \,,  
\qquad \ \ \ \ \ \ \ \ \ 
 h_{\uh 0 \uh 2  }= h^{\uh 0 \uh 1 \uh 2}
= 
- h^{\uh 3 \uh 4 \uh 5} \,. \la{D111}\ee
{Then  $\Gamma_*$ in \rf{D.6} takes the form
\begin{equation}\label{D.12}
\Gamma_*
= \Gamma_{  \uh 0  \uh 1  \uh 2  \uh 3  \uh 4  \uh 5 }
+  \, h_{\uh 0   \uh 2} ( 
\Gamma_{   \uh 0  \uh 1  \uh 2  }  
- \Gamma_{   \uh 3  \uh 4  \uh 5  }  )
=
\Gamma_{  \ul  0  \ul  1  \ul  2  \ul  3  \ul  4  \ul  5 }
+  \, h_{\uh 0   \uh 2} \Gamma_{   \ul  0  \ul  1  \ul  2  }  ( 
1
- \Gamma_{\ul  0 \ul  1\ul  2   \ul  3  \ul  4  \ul  5  }  )
\,,
\end{equation} 
where we used that  \((\Gamma_{\ul  0 \ul  1 \ul  2})^2 =1\).}
Here all  terms in \(\Gamma_*\) anticommute with each other and the term proportional to \(h_{\uh 0 \uh 2  }\) squares to zero, consistently with \(( \Gamma_*)^2=1\) in \rf{D.7}. 

{We fix the \(\kappa\)-symmetry gauge  by the condition    \(\Gamma_* \vartheta = \vartheta\). 
Using  the explicit form  of $\Gamma_*$ in \eqref{D.12} this   condition  can be rewritten as 
\begin{equation}\label{dfw}
\big(1
- \, h_{\uh 0   \uh 2} \Gamma_{   \ul 0  \ul 1  \ul 2  }   \big)\, 
(\Gamma_{  \ul 0  \ul 1  \ul 2  \ul 3  \ul 4  \ul 5 }-1)\vartheta = 0\,, 
\end{equation}
and  is then  solved by
\((\Gamma_{  \ul 0  \ul 1  \ul 2  \ul 3  \ul 4  \ul 5 }-1)\vartheta = 0\). As a consequence,  the mass 
term in \eqref{D.11} vanishes 
and we are left  simply with      massless Dirac operator 
which is independent of the parameter $\ka$ of the M5 brane solution 
}
\begin{equation}\label{D.13}
\mathcal D = 
i 
{ \slashed \nabla} 
\,.
\end{equation}
This is the operator that  appeared in  \rf{3.36}, \rf{3.39} corresponding to $m_{\rm F}=0$ in \rf{3.42}.


\subsubsection*{Probe  \Ib}

In this case  from \rf{343} and \rf{3.54}  we find instead of \rf{D.8} 
\begin{equation}\label{D.14}
\hat e_{i }^{\ul i } = L \,  (  \, e_{\ads_3} , \   \ka \sqrt{2 - 4 \ka^2}   \  e_{S^3 }   )
\,,
\qquad\qquad 
{\bre}^{\ul i}_i= 
 (  L_A\,  e_{\ads_3} ,\   L_S \, e_{S^3}) \,, \ \ \ \ \ \  L_A=1, \ \  L_S=\ha 
\,. \end{equation}
Since the  resulting fermionic operator should  depend only on the effective \adsst\ geometry and thus  should not depend
on  the value of $\ka$ (after a  rescaling of the fermion field)\foot{Note that a similar 
conclusion  that like the bosonic fluctuation operators the 
 fermionic operator does not have a non-trivial   dependence on the $\ka$-parameter of the classical solution was 
 reached in  analogous  examples discussed in  \cite{Faraggi:2011bb,Jiang:2024wzs}.}
  to 
 simplify the discussion  we  may just 
  set   \(\kappa = \frac12 \)  from the start (corresponding to the choice of  \(\theta_0 = \frac {\pi} 2\) in \rf{228}). 
  Then 
   there will be  no extra term from the spin connection  compared 
    to \eqref{D.9}, \ie\  \(\partial_{i } x^M \nabla_M  
=  \cd_{i} \), the latter being the effective spinor  covariant derivative  on the world-volume.
Furthermore, in this  case  the expression in \eqref{2.37} implies that 
 \(\sfH_{ijk} = 0 \)    and then  from  \rf{D.5}  it follows  also that    \(h_{ij}=0\).

Again, we may  scale  out factors of $L$ but it is useful to keep  the values of $L_A$ and $L_S$  in \rf{D.14} generic
till the very end. 
Then we find for the  \(\TF_M\) term in \rf{D.2}  and the  matrix $\Gamma_*$ in \rf{D.6}
\begin{equation}\label{D.15}
e _{\ul i }^{  i } 
\  \Gamma^{\ul i }
\ \TF_ { i }
= 
\tfrac 3 2 (L_S^{-1} - L_A^{-1})\,    \Gamma^{\ul{789\XX}}
\,,
\qquad\qquad \qquad 
\Gamma_*
= \Gamma_{\uh 0 \uh 1 \uh 2 \uh 3 \uh 4 \uh 5 }
= \Gamma_{ \ul 0 \ul 1 \ul 2 \ul 8 \ul 9 \ul \XX }\,. 
\end{equation}
As a consequence, the fermionic equation  in \rf{D.6} 
takes the form 
\begin{equation}\label{D.16}
\mathcal D\big[ (1 + \Gamma_* )\vartheta\big] =0
\,,
\qquad\qquad \qquad 
\mathcal D = 
i  {\slashed \nabla} 
+  \tfrac 3 2  i  (L_S^{-1} - L_A^{-1})\,    \Gamma^{\ul{789\XX}}
\,,
\qquad
\end{equation}
where \(i {\slashed \nabla} \)   corresponds to the  effective  $e _{\ul i }^{  i } $ in  \eqref{D.14}.

Choosing  the $\kappa$-symmetry gauge \(\Gamma_* \vartheta = \vartheta\), 
we  get   \(\Gamma_{\ul{89\XX}} \vartheta  =  \Gamma_{ \ul {012}} \vartheta \) and  thus \rf{D.16} reduces to\foot{Note that 
here  \(\ul 0 ,\ul 1 ,\ul 2 \) are also indices on the world volume, while \(\ul 7 \) is  the  index from the transverse space.}
\begin{equation}\label{D.17}
\mathcal D \vartheta =0\,,
\qquad\qquad 
\mathcal D = 
 i {\slashed \nabla}  +  \mathcal M \,, \qquad \qquad 
 \mathcal M= 
-   \tfrac 3 2 \,   i  (L_S^{-1} - L_A^{-1})\,    \Gamma^{\ul{0127}}
\,,
\qquad
(\Gamma^{\ul{0127}})^{2}=-1 \,. 
\end{equation}
In the present case of  \rf{D.14}  with 
 \(L_A=1, \  L_S=\ha\)  we thus find that the mass matrix is given by  
 \be 
  \mathcal M = -   \tfrac 3 2 \,   i      \Gamma^{\ul{0127}} \,, \qquad \qquad    \mathcal M^2 = \tfrac 9 4 \,. \la{D177}
  \ee
 Thus  $\mathcal M$   has the eigenvalues $\pm {3\ov 2}$ 
 corresponding to   $m_{\rm F}={3\ov 2}$ used in \rf{99}.


\subsubsection*{Probe \II} 

In this case  from \rf{3431} and \rf{3.92}  we  get (\cf\ \rf{D.14})   
\begin{equation}\label{D.18}
\hat e_{i }^{\ul i } = \LL \,  ( \sqrt{1 + \vk^2}  \, e_{\ads_3} , \ 2   \  e_{S^3 }   )
\,,
\qquad\qquad 
{\bre}^{\ul i}_i= 
(  L_A\,  e_{\ads_3} ,\   L_S \, e_{S^3}) \,, \ \ \ \ \ \  L_A=\ha , \ \  L_S=1 
\,. \end{equation}
Again,  the  fermionic  fluctuation spectrum should  not depend on the  parameter $\vk$ of the solution   so  we may 
simplify the discussion  by setting $\vk=0$ (corresponding to  the choice of 
$u_0=0$ in \rf{2.49})  (and  also scale out factors of $\LL$). 
Then  according to \rf{2.51}  we get   \(\sfH_{ijk} = 0 \) and thus also  \(h_{ij}=0\) from  \eqref{D.5}.

Instead of \rf{D.15}  and \rf{D.17}  here we  find 
\ba\label{D.19}
&e _{\ul i }^{ i } 
\  \Gamma^{\ul i }
\ \TF_ { i }
= 
\tfrac 3 2 \big( L_A^{-1} - L_S^{-1}  \big) \,   \Gamma^{\ul{0123}}
\,,
\qquad\qquad \qquad 
\Gamma_*
= \Gamma_{\uh 0 \uh 1 \uh 2 \uh 3 \uh 4 \uh 5 }
= \Gamma_{ \ul 0 \ul 1 \ul 2 \ul 5 \ul 6 \ul 7 } \,, \\
&
\label{D.20}
\mathcal D\big[ (1 + \Gamma_* )\vartheta\big] =0
\,,
\qquad\qquad \qquad\qquad  \ \ 
\mathcal D = 
i {\slashed \nabla} 
+  \tfrac 3 2 i  \big(L_A^{-1} - L_S^{-1}  \big)    \Gamma^{\ul{0123}}\,, 
\ea
where \(i \overline{\slashed \nabla} \)  corresponds to the  effective $e^{\ul i}_i$ in \rf{D.18}. 
Using the $\kappa$-symmetry gauge \(\Gamma_* \vartheta =  \vartheta\)  we get 
\begin{equation}\label{D.21}
\mathcal D \vartheta =0
\,,
\qquad\qquad 
\mathcal D = 
 i { \slashed \nabla}    +   \mathcal M \,, \qquad \ \    \mathcal M =
   \tfrac 3 2 i     \big(L_A^{-1} - L_S^{-1}  \big)      \Gamma^{\ul{0123}}
\,,
\qquad
(  \Gamma^{\ul{0123}})^{2}=-1 \,. 
\end{equation}
In the   present  case of  \rf{D.18}, \ie\ for 
 \(L_A=1, \  L_S=\ha\),   we thus  conclude  that (\cf\ \rf{D.13},\rf{D177}) 
 \be 
  \mathcal M =    \tfrac 3 2 \,   i      \Gamma^{\ul{0123}} \,, \qquad \qquad    \mathcal M^2 = \tfrac 9 4 \,. \la{D178}
  \ee
 This  matrix $\mathcal M$    has the eigenvalues $\pm {3\ov 2}$ 
 corresponding to the value of  $m_{\rm F}={3\ov 2} $ used in \rf{990}.

To conclude, in  all the three cases  discussed above  $  \mathcal M$  happens at the end  to be proportional to 
$ (L_A^{-1} - L_S^{-1}) $  (\cf\ \rf{D.17},\rf{D.21}) thus 
  leading to the values $m_{\rm F}=0$  for  the  \Ia\   probe   and 
 $m_{\rm F}={3\ov 2} $ for the  \Ib\  and \II\  probes.


\section{Partition function of 2-form field on  $\ads_{3}\times S^{3}$
\la{apz} } 

Here we consider the partition function 
of  a   2-form field in 6d with the standard gauge-invariant 
  Lagrangian \rf{300} where $\f_{ijk}= 3 \del_{[i } B_{jk]}$
  defined  on the \adsst  space with  generic radii (\cf\ \rf{3.7}) 
  \be
\la{377}
ds^2 = g_{ij} d \xi^ d \xi^j= L^2_A ds^{2}_{\ads_{3}}+ L^2_S ds^{2}_{S^{3}}\,. \ 
\ee
The corresponding 6d curvature factorizes into the  $\ads_3$ and $S^3$ parts  in the obvious way 
\be\la{z2} 
 R_{ijk\ell} = \mp  {L^{-2}_{A,S}}(g_{ik}g_{j\ell}-g_{i\ell}g_{jk})\,,
\qquad 
\te R_{ij} = \mp{2}{L^{-2}_{A,S}}g_{ij} \,, 
\qquad R_{A,S}  = \mp {6}{L^{-2}_{A,S} }\,.  
\ee
Starting with $L_2=\tfrac{1}{3!}\f^{ijk} \f_{ijk}$ where $\f_3=dB_2$   and following  the standard procedure (see, \eg, \cite{Bastianelli:2000hi} and references there), 
we add to it  the  covariant gauge  fixing term $(\nabla^{i} B_{ij})^2$  to get\footnote{We  include  $1\ov n!$  factor in the  definition of  antisymmetrization, 
\ie\
$T_{[ab]} = \frac{1}{2!}(T_{ab}-T_{ba})$, etc. }
\be\la{z3}
L_2=\tfrac{1}{3!}\f^{ijk} \f_{ijk}\ \ \ \to \ \ \  L_2'= \tfrac{1}{2} B_{ij} (\Delta_{2})^{ij}_{kl } 
B^{k\ell}\,,
\ee   
where $\Delta_2$  is the standard Hodge-deRham operator  defined on 2-forms. 
Including the ghost factors,  the corresponding partition function is 
\ba
\la{z8}
& \qquad \qquad Z_{2} = { \det \Delta_{1}\,\over  (\det \Delta_{2})^{1/2}\,  (\det \Delta_{0})^{3/2}} \,, \\
&(\Delta_{2})^{ij}_{kl } 
=  - \nabla^{2}\delta^{ij}_{k\ell}+2R^{[i}_{[k}\delta^{j]}_{\ell]}-R\indices{_{k\ell}^{ij}}\,, 
 \qquad
(\Delta_{1})^{i}_{j} = -\nabla^{2}\delta^{i}_{j}+R^{i}_{j}\,, \qquad 
\Delta_{0} = -\nabla^{2}.\la{326}
\ea
Let  $a,b$   be  the $\ads_{3}$  and $r,s$  the $S^3$  indices. Then $L_2'$ in   \rf{z3} may be written as 
\be
\la{3.25}
L'_2=  \tfrac{1}{2}\Big[B_{ab}(-\nabla^{2}-{2}{L_{A}^{-2}})B^{ab}+2B_{ar}(-\nabla^{2}+
{2}{L_{S}^{-2}}-{2}{L_{A}^{-2}})B^{ar}
+B_{rs}(-\nabla^{2}+{2}{L_{S}^{-2}})B^{rs}\Big]\,.
\ee
One can do a similar split  for the ghost $\Delta_1$ operator.
Then $Z_2$ in \rf{z8}  may be written as 
\be
\la{z4}
Z_{2} = 
\frac{\det \bm\Delta_{1,0}(-{2}{L_{A}^{-2}})\ \det \bm\Delta_{0,1}({2}{L_{S}^{-2}})}
{\big[ \det \bm\Delta_{2,0}(-{2}{L_{A}^{-2}})\   \det \bm\Delta_{1,1}({2}{L_{S}^{-2}}-{2}{L_{A}^{-2}})\ \det \bm\Delta_{0,2}({2}{L_{S}^{-2}})\big]^{1/2} \ \big[\det\bm\Delta_{0,0}(0)\big]^{3/2}}
\,, 
\ee
where  
\be\la{z7}
\bm{\Delta}_{p,q}(M^{2})\equiv  -\nabla^2 + M^2 = -\nabla^{2}_{A}-\nabla^{2}_{S}+M^{2} \,,
\ee 
 denotes an operator defined on a field in \adsst  which has  $p$-form
indices   in $\ads_{3}$ and $q$-form indices   in $S^{3}$. 
Since in 3d a rank 2 antisymmetric tensor is algebraically equivalent to a vector we have 
\be
\la{3.31}
\det\bm{\Delta}_{2,0} = \det\bm{\Delta}_{1,0}\,,\qquad\qquad 
\det\bm{\Delta}_{0,2} = \det\bm{\Delta}_{0,1} \,, 
\ee
 and thus   (\ref{z4})   may be rewritten as 
\be\la{z6}
Z_{2} = \Big[
\frac{\det \bm\Delta_{1,0}(-{2}{L_{A}^{-2}})\ \det \bm\Delta_{0,1}({2}{L_{S}^{-2}})}
{\det \bm\Delta_{1,1}({2}{L_{S}^{-2}}-{2}{L_{A}^{-2}})\ \big(\det\bm\Delta_{0,0}(0)\big)^{3}}
\Big]^{1/2}\,. 
\ee
We may further split the fields with a vector index in $\ads_3$  or $S^3$ into the transverse and the longitudinal (scalar) parts
which gives  \ba
&\det \bm\Delta_{1,0}(-{2}{L_{A}^{-2}}) 
= \det\bm\Delta_{1\perp,0}(-{2}{L_{A}^{-2}})\ \det\bm\Delta_{0,0}(0)\,, \qquad 
\det \bm\Delta_{0,1}({2}{L_{S}^{-2}}) = \det\bm\Delta_{0,1\perp}({2}{L_{S}^{-2}})\ \det\bm\Delta_{0,0}(0)\,, \notag
 \\
& \det \bm\Delta_{1,1}({2}{L_{S}^{-2}}-{2}{L_{A}^{-2}})
=  \det\bm\Delta_{1\perp,1\perp}({2}{L_{S}^{-2}}-{2}{L_{A}^{-2}}) \det\bm\Delta_{0,1\perp}({2}{L_{S}^{-2}})\det\bm\Delta_{1\perp,0}(-{2}{L_{A}^{-2}}) \det\bm\Delta_{0,0}(0) \,.    \la{z5}
\ea
Then  \rf{z6} becomes  simply 
\be
\la{z99}
Z_{2} =
{\big[\det \bm\Delta_{1\perp,1\perp}({2}{L_{S}^{-2}}-{2}{L_{A}^{-2}})\big]^{-1/2}  \ \big[\det\bm\Delta_{0,0}(0)\big]^{-1}}
\,. 
\ee
The  (modulus of) partition function of a  self-dual 2-form field is  given by the square root of \rf{z8} or of \rf{z99}, \ie\ 
\be
\la{z9}
Z^{(+)}_{2} =
{\big[\det \bm\Delta_{1\perp,1\perp}({2}{L_{S}^{-2}}-{2}{L_{A}^{-2}})\big]^{-1/4}  \ \big[\det\bm\Delta_{0,0}(0)\big]^{-1/2}}
\,. 
\ee

\section{Short supermultiplets of fluctuation fields in $\ads_3$  \la{mult}}

Here we will discuss the   supermultiplet    structure  of the  fluctuation modes as fields on $\ads_3$ 
presented  in  Table \ref{muld}.

In Table \ref{muld}   some fields    have  shifted values of level $\ell$. 
These  shifts are the same as in Table 4 of \cite{Deger:1998nm}
for spin-1 short supermultiplet of $SU(1,1|2) \times SU(1,1|2)$ corresponding to 
the tensor product $[\ell+3]_{L}\otimes [\ell+3]_{R}$.  Here 
$[k+1]$  (with $k=\ell +2$)  denotes  a  spin-1 short multiplet of $SU(1,1|2)$ (see  Eq.~(4.14) in \cite{deBoer:1998kjm}). 
This supermultiplet describes states of a (2,0)  tensor  multiplet in the 
supersymmetric \adsst\ vacuum of 6d supergravity. 

The bosonic subgroup of $SU(1,1|2)$ is  $SL(2,\mathbb R)\times SU(2)$.
Supercharges are in a doublet of R-symmetry automorphism $SU(2)_\Ss$\foot{Here we use sub-index 
 $\Ss$  for R-symmetry      to  distinguish it  from  the $R$-index  (for ``right'')
  below.}
and mix different representations of $SL(2,\mathbb R)\times SU(2)$.
Using labels  
$|\Delta ;  j, j'\rangle$ of $SO(2,2)\times SU(2)\times SU(2)_{\Ss}$ the spin-1 short multiplet of $SU(1,1|2)$ contains four  $SL(2,\mathbb R)\times SU(2)$
representations
\be
[k+1]^{SU(1,1|2)} \qquad = \qquad 
\def\arraystretch{1.3}
\begin{array}{cccc}
\toprule
\textsc{states}& j & j' & \Delta \\
\midrule
|0\rangle & \frac{k}{2} & 0 & \frac{k}{2} \\
Q^{\pm}|0\rangle & \frac{k-1}{2} & \frac{1}{2} & \frac{k+1}{2} \\
Q^{\pm}Q^{\pm}|0\rangle & \frac{k-2}{2} & 0 & \frac{k+2}{2} \\
\bottomrule
\end{array}
\ee
One can check that  taking the tensor product $[\ell+3]_{L}\otimes [\ell+3]_{R}$  (see, \eg,  Table I in \cite{Nicolai:2003ux})
one 
reproduces the  data  for the (2,0) tensor multiplet entry in Table \ref{muld}  (namely, 
the  
spin $|\Delta_{L}-\Delta_{R}|$,  the 
$S^{3}$ degeneracy $(2j_{L}+1)(2j_{R}+1)$, 
field multiplicity $(2j_{L}'+1)(2j_{R}'+1)$, and $\Delta= \Delta_{L}+\Delta_{R}$). 
For example,  the four  $\z_p$ scalar states   correspond to 
\be
|\tfrac{k+1}{2}; \tfrac{k-1}{2}, \tfrac{1}{2}\rangle \otimes
|\tfrac{k+1}{2}; \tfrac{k-1}{2}, \tfrac{1}{2}\rangle \  \stackrel{k=\ell+2}{=} \
 |\tfrac{\ell+3}{2}; \tfrac{\ell+1}{2}, \tfrac{1}{2}\rangle \otimes
|\tfrac{\ell+3}{2}; \tfrac{\ell+1}{2}, \tfrac{1}{2}\rangle\,,
\ee
that have spin $\Delta_{L}-\Delta_{R}=0$, the $S^{3}$ degeneracy $(\ell+2)^{2}$, multiplicity $(2\times\frac{1}{2}+1)^{2} = 4$, and $\Delta = \ell+3$.
Another example is the ``mixed'' scalar  $\vp_{-}$  that corresponds to 
\be
|\tfrac{k}{2}; \tfrac{k}{2}, 0\rangle \otimes
|\tfrac{k}{2}; \tfrac{k}{2}, 0\rangle \     \stackrel{k=\ell+2}{=} \ 
|\tfrac{\ell+2}{2}; \tfrac{\ell+2}{2}, 0\rangle \otimes
|\tfrac{\ell+2}{2}; \tfrac{\ell+2}{2}, 0\rangle\,,
\ee
which  has, indeed,   the  spin  $\Delta_{L}-\Delta_{R}=0$, the $S^{3}$ degeneracy   $(\ell+3)^{2}$,  the 
multiplicity $(2\times 0+1)^{2} = 1$, and $\Delta  = \ell+2$.

However,   in the  cases \Ia, \Ib\  and \II\  the dimensions  of states with  values of $\ell$  given 
 in  Table \ref{muld} (as found from the explicit mass spectrum of fluctuations) 
do not match those of the (2,0) tensor multiplet. 
This is due to the fact that  in these cases the   corresponding 
supergroups  are not $SU(1,1|2)\times SU(1,1|2)$  but the ones  given in Table \ref{tab:preserved-susy}. 
Let us discuss the  three cases \Ia, \II\  and \Ib\  in turn. 

\paragraph{Case Ia:}
Here  the relevant short-multiplet  factor is   not  of $SU(1,1|2)$   but of   $OSp(4^{*}|2)$  
discussed in \cite{Gunaydin:1990ag}. 
The bosonic subgroup  is  $SL(2, \mathbb R)\times SU(2)\times SU(2)_{\Ss}$ and fermions 
are doublets of $SU(2)_{\Ss}$  (we  use that  $SU(1,1)=SL(2,\mathbb R)$).
The corresponding 
   multiplet is\foot{See Eq.(5.12) in \cite{Gunaydin:1990ag} where  we  corrected a typo.}
\be
[\, n\, ]^{OSp(4^{*}|2)}
\qquad = \qquad 
\def\arraystretch{1.3}
\begin{array}{ccc}
\toprule
j & j' & \Delta \\
\midrule
\frac{n}{2} & \frac{1}{2} & \frac{n}{2}+1 \\
\frac{n-1}{2} & 0 & \frac{n+1}{2} \\
\frac{n+1}{2} & 0 & \frac{n+3}{2} \\
\bottomrule
\end{array}
\ee
Taking the product of   such  two multiplets
 with $n=\ell+1$, \ie\ 
$[\ell + 1 ] \times [\ell +1] $, 
 one finds   the same states  as in  the above (2,0) case 
 but with some interchanged values of $\Delta$, in agreement with the  \Ia\ entries  in Table \ref{muld}. 
For example,  to get  $\vp_-$  we are to consider 
\be\la{E5}
|\tfrac{n+3}{2}; \tfrac{n+1}{2}, 0\rangle \otimes
|\tfrac{n+3}{2}; \tfrac{n+1}{2}, 0\rangle  \   \stackrel{n=\ell+1}{=} \ 
|\tfrac{\ell+4}{2}; \tfrac{\ell+2}{2}, 0\rangle \otimes
|\tfrac{\ell+4}{2}; \tfrac{\ell+2}{2}, 0\rangle \,. 
\ee
This  state   has the 
 spin $\Delta_{L}-\Delta_{R}=0$, the $S^{3}$ degeneracy $(\ell+3)^{2}$, the multiplicity $(2\times 0+1)^{2} = 1$, and $\Delta \equiv\Delta_{L}+\Delta_{R} = \ell+4$ as  for the $\vp^{(\ell +2)}_-$  entry  in the case of \Ia\ in Table \ref{muld}. 

\paragraph{Case II: }
Here the building block is the 
short multiplet of the real form $OSp(4|2,\mathbb R)$
 discussed in \cite{Gunaydin:1988kz,Schmitt:1989qz}.
 The  bosonic  subgroup is $USp(2)\times SO(4^{*})$. We have $USp(2) = SU(2)_{\Ss}$
 and $SO(4^{*}) = SU(1,1)\times SU(2) = SL(2,\mathbb R)\times SU(2)$, so we find again the symmetry 
$SL(2, \mathbb R)\times SU(2)\times SU(2)_{\Ss}$
with the supercharges being doublets of $SU(2)_{\Ss}$.
The general unitary multiplet is presented in Eq.~(3.2) in \cite{Schmitt:1989qz}.\foot{The notation there is related to the present one as  $(K, R, S) \equiv (\Delta, j, j')$.} 
The shortening condition is $2\Delta-j-j'=0$. If we   solve it by setting 
$\Delta = \ell+2$ and $j=2\Delta$, $j'=0$  we  get  the states  of the  following short  multiplet 
\be\la{E6}
[\, \ell\, ]^{OSp(4|2,\mathbb R)}
\qquad = \qquad 
\def\arraystretch{1.3}
\begin{array}{ccc}
\toprule
j & j' & \Delta \\
\midrule
\frac{\ell+1}{2}  & \frac{1}{2} &  \frac{\ell}{4}+1 \\
\frac{1}{2}\ell+1 &  0 & \frac{\ell}{4}+\frac{1}{2}  \\
\frac{1}{2}\ell &  0 & \frac{\ell}{4}+\frac{3}{2} \\
\bottomrule
\end{array}
\ee
One can check that  the states in the corresponding  product $[\ell] \times [\ell]$ 
 all  have  the same 
 quantum numbers as before, with the exception of the  total $\Delta$  values 
 which are  in agreement with  the case \II\  entries in Table \ref{muld}. 
For  example,  $\vp_{-}^{(\ell+2)}$ originates from 
\be
|\tfrac{\ell}{4}+\tfrac{1}{2}; \tfrac{\ell}{2}+1, 0\rangle \otimes
|\tfrac{\ell}{4}+\tfrac{1}{2}; \tfrac{\ell}{2}+1, 0\rangle\,,
\ee
that has the spin $\Delta_{L}-\Delta_{R}=0$, the $S^{3}$ degeneracy $[2( \tfrac{\ell}{2}+1)+1]^{2} = (\ell+3)^{2}$, 
the multiplicity $(2\times 0+1)^{2} = 1$, and the total $\Delta =\frac{1}{2}\ell+1$.

\paragraph{Case Ib:} 
This case  should be  related by an analytic continuation to case {\bf II}
 and  the corresponding states should fit the tensor product of  the two short multiplets of $OSp(4^{*}|2)$.
To show this we  may  exploit the low-rank isomorphism $OSp(4^{*}|2)\simeq D(2,1; c)$ with $c= -2$ or $-\ha$ (the two cases  are equivalent). 
Shortened  representations of $D(2, 1; c)$ 
 play a role in determining  the BPS spectrum of states in 
string theory in $\ads_{3}\times S^{3}\times S^{3}\times S^1$ (where $c$ is a ratio of two $S^{3}$ radii)  and 
were discussed, \eg,  in
 \cite{deBoer:1999gea,Eberhardt:2017fsi}.\foot{The relation 
  of their notation to  ours is $(h_{0}, j^{+}, j^{-}) \equiv (\Delta, j, j')$.}
The multiplet in Eq.~(A.13) of \cite{Eberhardt:2017fsi} obeys the shortening condition 
\be\te 
h_{0} = \frac{1}{1+c}\, j^{-}+\frac{c}{1+c}\,j^{+}\qquad\stackrel{c=-2}{\longrightarrow} \qquad \Delta = 2j-j'\,,
\ee
and has the following content 
\be\la{E9}
[j,j']^{D(2,1; -2)}
\qquad = \qquad 
\def\arraystretch{1.3}
\begin{array}{ccc}
\toprule
j & j' & \Delta \\
\midrule
j^{+} & j^{-} & 2j^{+}-j^{-} \\
j^{+}-\frac{1}{2} & j^{-}-\frac{1}{2} & 2j^{+}-j^{-}+\frac{1}{2} \\
j^{+}-\frac{1}{2} & j^{-}+\frac{1}{2} & 2j^{+}-j^{-}+\frac{1}{2} \\
j^{+}+\frac{1}{2} & j^{-}-\frac{1}{2} & 2j^{+}-j^{-}+\frac{1}{2} \\
j^{+}-1 & j^{-} & 2j^{+}-j^{-}+1 \\
j^{+} & j^{-}-1 & 2j^{+}-j^{-}+1 \\
j^{+} & j^{-} & 2j^{+}-j^{-}+1 \\
j^{+}-\frac{1}{2} & j^{-}-\frac{1}{2} & 2j^{+}-j^{-}+\frac{3}{2} \\
\bottomrule
\end{array}.
\ee
The following tensor product of such multiplets  
$[\frac{\ell}{2}+1, 0] \times [\frac{\ell}{2}+1, 0]$
contains states   with all the quantum numbers  being the same as  in the above cases, 
with the exception of  the total  values of $\Delta$ which indeed  match the \Ib\ entries in Table \ref{muld}.
To illustrate this let  us consider again the example of $\vp_{-}$
that  originates from  the tensor product of two copies of the first state in  \rf{E9}
\be
|2j^{+}-j^{-}; j^{+}, j^{-}\rangle \otimes
|2j^{+}-j^{-}; j^{+}, j^{-}\rangle \  \stackrel{j^{+}=\frac{\ell}{2}+1,\, j^{-}=0}{=} \ 
|\ell+2; \tfrac{\ell}{2}+1, 0\rangle \otimes
|\ell+2; \tfrac{\ell}{2}+1, 0\rangle\,. \la{c10}
\ee
It has  the spin $\Delta_{L}-\Delta_{R}=0$, the  $S^{3}$ degeneracy $(\ell+3)^{2}$,  the 
multiplicity $(2\times 0+1)^{2} = 1$, and  the  total $\Delta = 2(\ell+2) = 2\ell+4$.
The same agreement can be checked for all other states.


\section{Analytic continuation between   \Ib\ and \II\ cases   \la{s34}}

The two 11d backgrounds $\ads_{7}\times S^{4}$  in \rf{2.5},\rf{2.6}  and  $\ads_{4}\times S^{7}$
in \rf{2.45},\rf{2.46}  are related by an   analytic continuation  
(see,  \eg,   \cite{Sezgin:2020avr}). This is a consequence of the relation between  the metrics of  $S^d$  and $\ads_d$, 
up to an overall sign change and an  interchange of the  values of the radii. 

  This  implies also a formal  relation between the 
M5 brane   solutions  wrapped on $\ads_3 \subset \ads_7$ and $S^3\subset S^4$  in $\ads_{7}\times S^{4}$  case  (probe \Ib) 
and on  $\ads_3 \subset \ads_4$ and $S^3\subset S^7$  in the  $\ads_{4}\times S^{7}$   (probe \II). 

Scaling out the overall  factors of  $L$   in \rf{2.5}  and $\LL$ in \rf{2.45} 
these two M5 brane  \adsst cases   have  the effective  metrics in \rf{3.54}  and \rf{3.92}  also related by  interchanging  the  two  factors  and  the radii  $L_A$  and  $L_S$.  
This translates into a relation between the M5 brane fluctuation 
spectra   discussed in sections \ref{s32} and \ref{s33}. 

The  scalar   operator    $-\nabla^{2}_{A}-\nabla^{2}_{S}+M^{2} $ on \adsst 
is mapped   by this analytic continuation into itself   with $\nabla^{2}_{A}\leftrightarrow  -\nabla^{2}_{S}$ 
provided  also $M^2 \to - M^2$. Equivalently,  since 
its   eigenvalues    are  given   by 
  \be\la{e1}
\mc O_{2}=  -\nabla^{2}_{A}-\nabla^{2}_{S}+M^{2} \ \ \to \ \  -{\Delta(\Delta-2)}{L_{A}^{-2}}+{\ell(\ell+2)}{L_{S}^{-2}}+M^{2} \,, 
\ee
 the     transformation  
  \be\la{e2}
L_{A}\leftrightarrow L_{S}\,, \qquad\ \ \qquad   \Delta\leftrightarrow -\ell \,, 
\ee
is a formal   symmetry   provided one also  reverses   the sign of the mass term.

In the case of the 4 scalar fluctuations   $\z_p$ in \rf{342},\rf{3.53} in \Ib\ case and in \rf{3420},\rf{3611} 
their masses  are indeed  related by $M^{2}\to -M^{2}$ (this   has to do with the reversed sign
 of the curvature of the ``transverse'' subspaces  $\ads_4\subset \ads_7$  in the \Ib\ case  and of 
$S^4\subset S^7$ in the  \II\ case). 

For  the mixed scalars, the quartic operators  in \rf{77} (for the scalar $\Te$ representing
the  transverse fluctuation in $S^4$  in the  \Ib\ case  mixed with the  scalar  part $P$ of $B_{rs}$) 
 and \rf{3750}  (for the scalar $U$ representing
the  transverse fluctuation in $\ads_4$  in the  \II\  case  mixed with the scalar  $P$) 
 are       
\be \mc O_{4}^{(\text{\bf Ib})} = \nabla^{4}-24\nabla^{2}+36\nabla^{2}_{S} \,, \qquad \qquad 
\mc O_{4}^{(\text{\bf II})} = \nabla^{4}-12\nabla^{2}+36\nabla^{2}_{S} \,. \la{e3}  
\ee
Here  $\nabla^2=\nabla^2_A + \nabla_S^2$  so that like in \rf{e1} 
 their \adsst spectra   may be represented as 
\ba\mc O_{4}^{(\text{\bf Ib})} &
\to\ \  \big[{\Delta(\Delta-2)}{L_{A}^{-2}}-{\ell(\ell+2)}{L_{S}^{-2}}\big]^{2}-24
{\Delta(\Delta-2)}{L_{A}^{-2}}-12{\ell(\ell+2)}{L_{S}^{-2}}\,, \la{e7}
\\
\mc O_{4}^{(\text{\bf II})} &\to  \ \ 
\big[{\Delta(\Delta-2)}{L_{A}^{-2}}-{\ell(\ell+2)}{L_{S}^{-2}}\big]^{2}-12
{\Delta(\Delta-2)}{L_{A}^{-2}}-24{\ell(\ell+2)}{L_{S}^{-2}}\,.  \la{e6} 
\ea
Thus they are also related by the  transformation \rf{e2}. 

Similar   formal   analytic continuation applies  also to the fermionic  Dirac operator on  \adsst
which turns out to have the  structure (see \rf{D.17},\rf{D.21})
\be\la{e8}
 i\slashed \nabla_{A}+ i\slashed \nabla_{S}\ + \tfrac{3}{2}\, \big(L_{A}^{-1}-L_{S}^{-1}\big) \widehat \Gamma\,,  \qquad \qquad  \widehat \Gamma^2=1 \,, 
\ee
 and thus is covariant  under  $\slashed \nabla_{A}\leftrightarrow \slashed \nabla_{S}$
 and $L_{A}\leftrightarrow L_{S}$.
 
 This  formal   correspondence  between the  spectra of fluctuations in the  \Ib\  and \II\ cases does not, of course, 
 imply that the  values  of the corresponding  one-loop  free energies  given by the sums over the spectrum 
 should be  the same (\cf\ section 4).  

\section{Comments on one-loop divergences}\la{apdi} 

Using heat kernel  regularization  the UV divergent term of one-loop  free energy  in 6d 
has the following form 
\be\la{50}
F^{(1)}_{\infty} = -\frac{1}{(4\pi)^{3}}\int d^{6}\xi\, \sqrt{g}\, \big(\tfrac{1}{6}b_{0}\Lambda^{6}+\tfrac{1}{4}b_{2}\Lambda^{4}
+\tfrac{1}{2}b_{4}\Lambda^{2}+b_{6}\log\Lambda\big)\,,\qquad \ \ \  \Lambda \to \infty\,, 
\ee
where the   Seeley's coefficients $b_{2n}$ are local expressions in terms of the curvature of  the 6d metric,  gauge connection
and mass matrix (see, \eg, \cite{Bastianelli:2000hi}  and references there). 
In  all the cases   discussed  in section 4  there were   no   logarithmic divergences.
In this Appendix we will   discuss the structure of divergences  and their cancellation   in more detail.

\subsubsection*{Logarithmic  divergences}

For a massive scalar  with an operator $-\nabla^2 + M^2$  defined on
$\ads_{3}\times S^{3}$ with radii $L_{A}$ and $L_{S}$ we  find that 
\be\la{51} 
b_{6}^{(0)}(M^{2}) = -\tfrac{1}{6}\big({L_{A}^{-2}}-{L_{S}^{-2}}+M^{2}\big)^{3}.
\ee
For the  self-dual 2-form, taking half of  the expression in  Eq.(2.26) in \cite{Bastianelli:2000hi} we get 
\be\la{52}
b_{6}^{(2+)} = -\tfrac{1}{6}\big({L_{A}^{-2}}-{L_{S}^{-2}} \big)^{3} \,.
\ee
In the cases {\bf Ib} and {\bf II}  where there are mixed scalars 
 we have also 
 a ``compensating'' massless scalar contribution  in the  numerator of (\ref{375}) and (\ref{3750}) that, according to \rf{51}, 
is given by 
\be\la{53}
 -b_{6}^{(0)}(0)= \tfrac{1}{6}\big({L_{A}^{-2}}-{L_{S}^{-2}} \big)^{3}  \,.  
\ee
It  thus cancels  against the one in \rf{52}.

From    the above expressions   one finds that  in the \Ia\  case  where the  bosonic part of the 
one-loop partition function  is given by \rf{327} with 
$L_A=L_S=1$  the  corresponding coefficient of the log divergence is 
 $b^{(\bf Ia)}_6=0$. 
This  can be easily understood on general grounds: 
here we have a massless (conformally-invariant)   (2,0) multiplet defined 
on the equal-radii  $\ads_{3}\times S^{3}$  space. 
This space is  conformally-flat (has zero Weyl tensor) and also  
its 6d Euler density vanishes. This implies that its conformal  anomaly  vanishes. 

In the cases \Ib\ and \II\ we need to   account also for 
the  contribution to $b_6$   of the 4th order operators in (\ref{375}) and (\ref{3750}). These 
can be  found  using  the expression  for 
 this  Seeley coefficient  for a  general  covariant 4-th order operator in 6d space 
given in   \cite{Casarin:2019aqw,Casarin:2023ifl}. 
Indeed, given the  operator on \adsst of the same  form as in (\ref{375}),(\ref{3750}), \ie\ 
\be\la{54}
\OO_4= \nabla^{4}+c_{1}\nabla_{A}^{2}+c_{2}\nabla_{S}^{2},
\ee
one may   represent it  in the 6d covariant form   using that 
\be c_{1}\nabla_{A}^{2}+c_{2}\nabla_{S}^{2} = (p_{1} R^{ij}+p_{2} R\, g^{ij})\, \nabla_{i}\nabla_{j}\,, \quad 
c_1=  -2 ( p_1 + 3 p_2)  {L_{A}^{-2}} + 6 p_2 {L_{S}^{-2}} \,, \quad 
c_2 =  2 ( p_1 + 3 p_2)  {L_{S}^{-2}} - 6 p_2 {L_{A}^{-2}}  \,.\no  \ee
One then finds that for the operator \rf{54} 
\ba
b_{6}^{(\OO_4)} =& -\tfrac{1}{192} \big({4}{L_{A}^{-2}}-{4}{L_{S}^{-2}}-c_1-c_2\big)\Big[16 \big({L_{A}^{-2}}-{L_{-S}^{2}}\big)^2
+7 (c_1^2+ c_2^2)+2c_{1}c_{2}\lp
+16 c_1  \big({L_{A}^{-2}}+{2}{L_{S}^{-2}}\big)-16c_{2}\big({2}{L_{A}^{-2}}+{L_{S}^{-2}}
\big) 
\Big]\,. \la{56}
\ea
In   the \Ib\  case  where $L_A=1, \, L_S=\ha $    the 4 decoupled scalars 
and  the 2-form plus massless scalar contributions to $b_6$     separately vanish.
From \rf{3466} here the operator $\OO_4$  has  $(c_{1},c_{2})=(-24,12)$  and as a result \rf{56}  vanishes too.
Thus the total  $b^{(\bf Ib)}_6=0$. The same conclusion is reached in the  ``analytically-continued''   \II\ case where 
 $L_A=\ha, \, L_S=1 $  and 
the operator $\OO_4$  has  $(c_{1},c_{2})=(-12,24)$,   so that \ $b^{(\bf II)}_6=0$.

One can  show  that the fermionic contribution to the coefficient of the logarithmic divergences 
  also vanishes  separately
in all the three cases.  Since it is not straightforward to square the fermionic  Dirac 
operator in \rf{3.39}  one may first expand in modes on $S^3$ and then  see  if the resulting sum
of  individual mode  contributions (\cf\ \rf{320}) 
to the free energy  over   will not have logarithmic  divergence. 


The vanishing  of the coefficient of the logarithmic  divergence  can be understood 
as   being  a consequence of the associated  conformal dimensions $\Delta$ being  simply  linear  in the $S^{3}$
mode number $\ell$.  
For example,   for a scalar operator  $-\nabla^{2}+M^{2}$
on  $\ads_{3}\times S^{3}$ with radii $L_{A}$, $L_{S}$  
we get for the dimension  of the  $\ads_3$  field representing a  mode with fixed $\ell$ (\cf\ \rf{3.37}) 
\be\la{c77}
\Delta_{\ell}(M) = 1+\sqrt{1+L_{A}^{2}\big[M^{2}+{\ell(\ell+2)}{L_{S}^{-2}}\big]}\,.
\ee
The corresponding contribution to the  free energy  is given  by $F_0(\Delta_\ell) $ in \rf{4.2}
(including also the degeneracy factor $d^{(0)}_\ell = (\ell +1)^2$ in \rf{340}),  
\ie\ it is proportional to $(\Delta_\ell-1)^3=  [ 1+L_{A}^{2}(M^{2}+{\ell(\ell+2)}{L_{S}^{-2}})]^{3/2}$. 
Its  large $\ell$ expansion   contains the  term $\sim \big({L_{A}^{-2}}-{L_{S}^{-2}} + M^2\big)^{3}\, \ell^{-1}$ 
that  leads  to the  logarithmic divergence in the sum $\sum_\ell F_0(\Delta_\ell)$. This log divergence is absent if  
\be \la{c778}
M^2= {L_{S}^{-2}}-{L_{A}^{-2}} \,, \qquad \ \ \ \ \ \ 
 \Delta_\ell = 1 + L_A L^{-1}_{S} (\ell +1)\,. \ee
  This  relation is indeed what we  have found  in all the three cases in \rf{339},\rf{999}, \rf{9991}
 where  the resulting  sum over $\ell$ contained  only a quadratic divergence. 
 
 In general, one  can check that   the coefficient of the  log divergence in the sum over $\ell$ 
  is given by (ignoring  here power divergencies) 
 \be \la{E.5}
 \sum_{\ell}^\Lambda F_0(\Delta_\ell)  = -B_6 \log \Lambda + ... \,, \ \ \ \ \ \ 
 B_6 = \tfrac{1}{(4\pi)^{3}} \vol(\ads_{3}\times S^{3}) \,  b_6 \,, \ \ \ \   
 b_6= -\tfrac{1}{6}\big({L_{A}^{-2}}-{L_{S}^{-2}}+M^{2}\big)^{3} \,. \ee
This is in agreement 
with the direct 6d computation of $b_6$  in \rf{50},\rf{51} which was not  using the expansion in modes on $S^3$.\foot{Note that such a direct agreement between the two procedures applies only to the  coefficient of the universal log divergence; coefficients 
of  power divergent terms  are sensitive to a particular choice of a UV cutoff.} 

Similar conclusion is reached  also in the case of  the  mixed scalar  operator in \rf{54},\rf{56} where 
the special values of the coefficients   $c_1,c_2$   for which the conformal dimensions  are linear in $\ell$ 
(\cf\ \rf{0881},\rf{999} and \rf{088},\rf{9991}) imply also the vanishing of $b_6$ in \rf{56}. 

This  discussion applies also  to a 6d  fermionic   field on \adsst  with the squared Dirac operator 
taken in the form $-\nabla^2 + M^2$  where $\nabla^2$ acts on a 6d spinor. 
In this case the direct computation of the $b_6$ Seeley  coefficient (as, \eg,  in \cite{Bastianelli:2000hi}) gives 
\be
\la{E.6}
b_{6} = -\tfrac{1}{24}\big({3}{L_{A}^{-2}}-{3}{L_{S}^{-2}}+2M^{2}\big)
\big(
{3}{L_{A}^{-2}L_{S}^{-2}}-{3M^{2}}{L_{A}^{-2}}+{3M^{2}}{L_{S}^{-2}}-2M^{4}
\big)\,.
\ee
Here  we ignored the factor of the number of fermionic components.
Considering first the expansion in spinor harmonics on $S^3$   we get 
$-\nabla^2 = -\nabla_A^2 - \nabla_S  \to - \nabla^2_A  +  {L_{S}^{-2}}[(\ell+\tfrac{3}{2})^2-\tfrac{3}{2}]$
(with degeneracy $d^{(\half)}_\ell= (\ell+1)(\ell+2)$). 
Then the corresponding conformal dimension of the $\ads_3$  $\ell$-mode is 
(\cf\ \rf{3.38}) 
\be
\la{E.7}
\Delta_{\ell}(M) =  1+\sqrt{\tfrac{3}{2}+ L_{A}^{2} \big(M^2 + {L_{S}^{-2}}[(\ell+\tfrac{3}{2})^2-\tfrac{3}{2}]  \big)} \,. 
\ee
Computing  $ \sum_{\ell}^\Lambda F_{\half} (\Delta_\ell)$ using \rf{4.2}  (and accounting for the fermionic minus sign factor
that was not included in \rf{4.2}) 
we get 
again the same  first two relations in \rf{E.5} where  $b_6$ now  matches the expression in \rf{E.6}. 
We conclude that the  log divergence is absent if  
\be\la{c122}
M^{2} =M^2_{\half}\equiv \tfrac{3}{2}\big({L_{S}^{-2}}-{L_{A}^{-2}}\big)\,.
\ee
This is also the condition when the scaling dimension \rf{E.7} becomes linear in $\ell$ \foot{To be precise, the 
 observation is that 
whenever $\Delta_{\ell}$ is polynomial in $\ell$ we have no log divergence in the sum over $\ell$, \ie\ 
 vanishing $b_{6}$. The converse  need not be true.  Indeed, from 
(\ref{E.6})  one can  see that there are other values of $M^{2}$ 
such that $b_{6}=0$. 
For these values  $\Delta$ is not polynomial in $\ell$, but the large $\ell$ expansion of  $F_{\ell}$
still  happens to have  a vanishing coefficient of  the $\ell^{-1}$ term.}
\ba\la{c123}
\Delta_{\ell}(M_{\half}) = 1+ {L_{A}}{L_{S}^{-1}}\big(\ell+\tfrac{3}{2}\big) \,.
\ea
This is indeed the expression  we had (up to the $m_{\rm F}$  part) in \rf{3.42},\rf{99},\rf{990}. 



\subsubsection*{Power divergences}
Let us now  comment on the coefficients of  power divergences in \rf{50}. 
In general, in a supersymmetric theory  one finds $b_0=0$   due to balance of the numbers of the 
 bosonic   and fermionic   degrees of freedom. The coefficient $b_2$  for an operator like $-\nabla^2 + M^2$
  is given by $\tr (\frac{1}{6} R - M^2) $.  In the  \Ia\ case  or for a massless  (2,0) tensor multiplet on \adsst  it 
   also vanishes due to an  effective supersymmetric mass  sum rule.
Then the only potentially  non-zero divergences are the quadratic ones
controlled by the $b_{4}$   coefficient and they do not vanish in general.  In
the  case {\bf Ia}  
we may   directly compute $b_4$ as  for the (2,0) tensor  multiplet on \adsst   from  its known expression 
for  the 2nd order 6d Laplacian (see  Eq. (3.64) in  \cite{Jiang:2024wzs})
\be\la{c779}
b^{(2,0)}_{4} = \tfrac{1}{4}R_{mnk\ell}R^{mnk\ell}-\tfrac{1}{2}R_{mn}R^{mn}+\tfrac{1}{10}R^{2} \,.
\ee
Specifying to the  equal-radii  \adsst  that gives
$b^{(2,0)}_{4} =-6$, \ie\ there  should be   a non-zero quadratic divergence.\foot{The same  conclusion was reached 
 in the  case of the (2,0) multiplet   defined on  $\ads_{5}\times S^{1}$  \cite{Jiang:2024wzs}.}
This is consistent with the presence of a
 quadratic divergence in the  sum over $\ell$  in \rf{4.11},\rf{4.12}.

To analyse the quartic and quadratic  divergences in the  cases \Ib\ and \II\ 
we need to find  the  mixed-scalar  quartic operator \rf{54}  contributions  to them. 
   An important point to emphasize is that the standard heat kernel cutoff does not regularize power divergences for 
  the  operators of different orders (\eg\ 2nd and 4th) 
   in a homogeneous manner. As a result, the corresponding  Seeley coefficients cannot be simply combined together to
   find the total  values of the   coefficients of the   power divergences.

A way to  circumvent this problem is  by  evaluating the contributions of  the  2nd order operators
by the same algorithm as for the quartic ones  by  first squaring them.
Then all  the operators  will  have   the same order  and their contributions to divergences 
 can be combined together directly.  Thus, for the  standard scalar  operator $-\cd^2 +M^2$  we may  define 
 the divergence coefficients   in terms of those  of  its square as 
\begin{equation}\label{fst}
b^{(0)}_n (-\cd^2 +M^2) \equiv  \tfrac 12 b_n( \cd^4 - 2 M^2 \cd^2 + M^4) \,, 
\end{equation}
and similarly for the  vector and 2-form Laplacians.
 Here $b_n$   for  the 4th-order operator is to be computed  according to the  prescription in  \cite{Gusynin:1988zt}.
In general,  this will  give the same value of $b_6$ coefficient 
  due to  the universal nature of the logarithmic divergence.
From \rf{fst}   we  find  using the  expressions  for $\hb_2$ and $\hb_4 $
 for the quartic operator in  \cite{Gusynin:1988zt}\footnote{In this case we observe
  a  proportionality  between  the standard $b_2$ and $b_4$ for a 2nd order operator and their values defined via \rf{fst}. 
This  is  due to the simplicity of the scalar operator, 
  and  is not true in  general, \cf\ \cite{Gusynin:1988zt}.
 Note that unfamiliar $\sqrt \pi$ coefficient in $b_4$  is due to its definition for  a 4th-order operator that was used  in  \cite{Gusynin:1988zt}. }
\be\label{dka}
b^{(0)}_2   =
- \tfrac{1}{4} (M^2+ {L_A^{-2}}- {L_S^{-2}})\,,
\qquad\qquad 
b^{(0)}_4   =
\tfrac{ \sqrt{\pi } }{8}   (M^2+ L_A^{-2}- L_S^{-2})^2
\,. 
\ee
Similarly,  for  the   self-dual  2-form contribution  we get the same $b_6$ as in    \rf{52}  and also  
\be\label{dka2}
b_2^{(2+)}   =
\tfrac{3}{4} \left(L_A^{-2}   -   L_S^{-2}   \right), \qquad \qquad 
b_4^{(2+)}   =
\tfrac{ \sqrt{\pi } }{8}\left( -18  L_A^{-2} L_S^{-2}+   L_A^{-4} +    L_S^{-4}\right)\,. 
\ee
For the quartic operator in \rf{54}  we find in addition  to  $b_6$ in  \rf{56} 
\ba\label{dfj}
&b_2^{(\OO_4)}
  = \tfrac{1}{8} \left(c_1+c_2-4  L_A^{-2} +4 L_S^{-2} \right)\,,\\
& \la{c121}
b_4^{(\OO_4)}
  = 
\tfrac{ \sqrt{\pi } }{512}\left[15 c_1^2+18 c_1 c_2 +15 c_2^2
+32  ({3 c_1 + c_2 }) L_S^{-2}
-32 ( { c _1 + 3 c_2 }) L_A^{-2}
+128 \left( L_A^{-2} - L_S^{-2} \right)^2
\right]
\,. \ea
Using these values we  conclude that  the bosonic contributions to the  coefficients of the  quartic  and quadratic  
divergences in cases \Ia, \Ib\ and \II\ are given by 
\ba \la{c222}
&b^{({\bf Ia})}_2 = 0 \,, \qquad \qquad\qquad  \  b^{({\bf Ib})}_2 = -3 \,, \qquad \qquad\ \qquad \ \   \ b^{({\bf II})}_2 = 3 \,, 
\\
&b^{({\bf Ia})}_4 = -2 \sqrt \pi  \,, \qquad \ \quad  \ \ \  b^{({\bf Ib})}_4 =  - \tfrac{337}{32}\sqrt{\pi}  \,,
 \qquad \ \quad\ \ \ \  b^{({\bf II})}_4 =  - \tfrac{337}{32}\sqrt{\pi}  \,. 
\ea
The non-zero bosonic contributions to the quartic divergence   coefficient 
$b_2$   in  the \Ib\ and \II\ cases should be cancelled  against the fermionic contributions, while
the total  values of  $b_4$  should be non-zero 
for consistency  with  the quadratic divergence in the  sum in \rf{4.12}.


\section{Spectra  and decompositions of Laplacians on $p$-forms \la{pfo}}



Let us recall the  spectrum of the Hodge-de Rham  (HdR) operator $\Delta_{p}$ on 
 co-exact $p$-forms  on a   unit  sphere $S^d$, \ie\ of the corresponding 
 Laplacian on transverse antisymmetric tensors.
 For 
 $p>0$  the    eigenvalues and degeneracies are given by (see, \eg,   \cite{Copeland:1984qk})
\bea
\la{B.1}
\l_{\ell}^{(p)} = (\ell+p+1)(\ell+d-p)\,, \qquad \ \ 
d_{\ell}^{(p)} = \frac{(\ell+d)!(2\ell+d+1)}{p!\ell!(d-p-1)!(\ell+d-p)(\ell+p+1)} \, , \ \ \  \ell=0,1,2, ... \,. 
\eea
In  particular, in the scalar case  (including the contribution of the zero mode) 
\bea\la{d22}
\l_{\ell}^{(0)} = \ell(\ell+d-1)\,, \qquad \ \ \ \ 
d_{\ell}^{(0)} = \frac{(\ell+d-2)!\, (2\ell+d-1)}{(d-1)!\, \ell!} \,. 
\eea
From here we may also   find the  spectrum of the part  $(-\nabla^{2})_{p\perp}$  of the HdR  operator that does not include extra curvature terms. 
For a vector we have 
\be
V_{i}(\bDelta_{1})^{i}_{j}V^{i} = V_{i}\big[ -\nabla^{2}+(d-1)\big]V^{i}\,,
\ee
so that the corresponding eigenvalues of $(-\nabla^{2})_{1}$ are $(\ell+2)(\ell+d-1)-(d-1).$ 
For a 2-form 
\ba
V_{ij}\big(\bDelta_{2})^{ij}_{rs}V^{rs} &= V^{ij}\big(-\nabla^{2}\delta^{ij}_{rs}
+2R^{[r}_{[i}\delta^{s]}_{j]}-R\indices{_{ij}^{rs}}\big)V_{rs}
= V_{ij}\big[-\nabla^{2}+2(d-2)\big]V^{ij}\,,
\ea
and thus the eigenvalues of $(-\nabla^{2})_{2}$   are $(\ell+3)(\ell+d-2)-2(d-2).$
In case of  $S^{3}$ we then  have 
\be
\la{B.8}
\def\arraystretch{1.3}
\begin{array}{ccc}
\toprule
\text{Spectrum of $(-\nabla^{2})_{p\perp}$ on $S^3$}\\
\toprule
p & \l_{\ell}^{(p)} & d_{\ell}^{(p)}  \\
\midrule
0 & \ell(\ell+2) & (\ell+1)^{2} \\
1 & (\ell+2)^{2}-2 & 2(\ell+1)(\ell+3) \\
2 & (\ell+1)(\ell+3)-2 & (\ell+2)^{2} \\
\bottomrule
\end{array}
\ee


Let us also  recall 
some standard relations for the determinants $\det \Delta_{p}$ and $\det \Delta_{p\perp}$ 
defined on transverse tensors
(see, \eg, \cite{Casarin:2024qdn} and references there). 
Let us consider  $\ads_{d}$ ($\eps=-1$) or $S^{d}$ ($\eps=1$) with radius $L$, \ie\ with 
\be\la{d11} \te
R_{ijrs } =\frac{\eps}{L^{2}}(g_{i r}g_{j s }-g_{is}g_{jr})\,,\qquad
R_{ij} = \frac{\eps}{L^{2}}(d-1)g_{ij}\,, \qquad 
R = \frac{\eps}{L^{2}}d(d-1)\,, 
\ee
and the operator
\be\la{d10}
\bDelta_{p} = -\nabla^{2}_{p}+M^{2},\ \ \ \ \ \   \ \ \ \ \ p=0,1,2\,, 
\ee
acting on scalars, vectors or  antisymmetric 2-tensors. 
For a vector we  set $V_{i} = V_{i\perp}+\nabla_{i} V$ and then 
\ba
&\te V_{i}  (-\nabla^{2}+ M^{2})V^{i}
= V_{i\perp}(-\nabla^{2}+ M^{2})V^{i}_{\perp}+V(-\nabla^{2})\big[-\nabla^{2}+ M^{2}-\frac{\eps(d-1)}{L^{2}}\big]V\,, \\
&\te \det\bDelta_{1}(M^{2}) = \det\bDelta_{1\perp}(M^{2})\ \det\bDelta_{0}\big(M^{2}-\eps\frac{d-1}{L^{2}}\big) \,, 
\ea
where $\det(-\nabla^{2})^{1/2}$ is cancelled against the Jacobian of the  transformation $V_i \to (V_{i\perp}, V)$.\foot{To recall, $ 1 = \int DV_{i}e^{-\int V_{i}V^{i}} = \int DV_{i\perp}DV\ J_{1}\ e^{-\int[ V_{i\perp}V^{i\perp}+ V(-\nabla^{2})V}]$ gives $
J_{1} = \det(-\nabla^{2})^{1/2}.$}

Similarly,  in the 2-form   case setting 
$
V_{ij} = V_{ij\perp}+\nabla_{i}V_{j\perp}-\nabla_{j}V_{i\perp}$   leads to 
\ba\te 
V_{ij}V^{ij}  
=  V_{ij\perp}V^{ij\perp} + 2 V_{i\perp}[- \nabla^{2}  + \frac{\eps(d-1)}{L^{2}} ]  V^{i\perp}
\ea
 and thus the associated Jacobian is  
$
J_{2} = \big[\det \bDelta_{1\perp}\big(\eps\frac{d-1}{L^{2}}\big)\big]^{1/2}.
$
Also, we have 
\ba
V_{ij} (-\nabla^{2}+ M^{2})V^{ij} = &V_{ij\perp}(-\nabla^{2}+ M^{2})V^{ij}_{\perp}
+2\nabla_{i}V_{j\perp}(-\nabla^{2}+ M^{2})\nabla^{i}V^{j}_{\perp} \lp
-2\nabla_{j}V_{i\perp}(-\nabla^{2}+ M^{2})\nabla^{i}V^{j}_{\perp}.
\ea
Integrating by parts and commuting the covariant derivatives the last two terms may be written as 
\be\textstyle 
 -2V_{j\perp}\nabla_{i}(-\nabla^{2}+ M^{2})\nabla^{i}V^{j}_{\perp} 
+2V_{i\perp}\nabla_{j}(-\nabla^{2}+ M^{2})\nabla^{i}V^{j}_{\perp}
= 2V_{i\perp}\Big[\big(-\nabla^{2}+\frac{\eps(d-1)}{L^{2}}\big)\big(-\nabla^{2}+ M^{2}-\frac{\eps(d-3)}{L^{2}}\big)\Big]V^{i}_{\perp} \no
\ee
so that we get 
\be\te 
\det\bDelta_{2}(M^{2}) = \det\bDelta_{2\perp}(M^{2})\ \det\bDelta_{1\perp}\big(M^{2}-\eps\frac{d-3}{L^{2}}\big)\,,
\ee
where again one factor was  canceled against  $J_{2}$.
In the   special cases of $\ads_{3}$  and $S^3$ this gives 
\ba
\la{C.14}
&\det\bDelta_{1}(M^{2}) = \det\bDelta_{1\perp}(M^{2})\ \det\bDelta_{0}(M^{2}+{2}{L_{A}^{-2}}),\qquad \ \ \ \
\det\bDelta_{2}(M^{2}) = \det\bDelta_{2\perp}(M^{2}) , \\
\la{C.15}
&\det\bDelta_{1}(M^{2}) = \det\bDelta_{1\perp}(M^{2})\ \det\bDelta_{0}(M^{2}-{2}{L_{S}^{-2}}),\qquad \ \ \ \ 
\det\bDelta_{2}(M^{2}) = \det\bDelta_{2\perp}(M^{2}).
\ea

\section{Casimir energy  and ``thermal'' partition function of $\ads_3$ multiplets
 \la{cas}}  

In section \ref{sec:free-energy} we focused on the free energy for  the multiplets  of fluctuations  corresponding to 
the cases {\bf Ia, Ib, II}  defined on $\ads_3$ with $S^2$ boundary. 
Given a   collection of $\ads_{3}$ fields   corresponding to  $(\Delta, s)$ representation of $SO(2,2)$
we can  similarly compute their  Casimir energy (see, \eg, 
  Eq.~(F.2) in \cite{Beccaria:2014qea})\footnote{Here $s=1$   
corresponds to  a self-dual tensor. 
This  expression contains a factor  of  $-\half $ compared to 
 Eq.~(F.2) in \cite{Beccaria:2014qea} 
  to get the Casimir energy of the $\ads_{3}$ fields with Dirichlet boundary conditions.
 This genuinely $\ads_{3}$ quantity was
 denoted as  $E_{c}^{+}$ in \cite{Beccaria:2014qea} (the same applies  to the related c-anomaly mentioned below).}
\be
E_{c}(\Delta, s) = \tfrac{1}{24}(-1)^{2s}(\Delta-1)^{2}\big[2(\Delta-1)^{2}-1\big]\,.
\ee
Using  the data in Table \ref{muld}  we then get
 \be
 \la{G.2}
E_{c}^{(\bf Ia)} = E_{c}^{(\bf Ib)} = \tfrac{3}{4}(\ell+2)\,,\qquad\qquad   \qquad E_{c}^{(\bf II)}=0.
\ee
One may also consider 
 the 2d central charge associated to 
  a single   ``chiral'' component of  a spin $s$ field in 3d   which is   given by (see, \eg,  Eq.~(F.3) in \cite{Beccaria:2014qea})
\be
\la{G.3}
{\rm c}(\Delta, s) = -\tfrac{1}{2}(\Delta-1)\big[(\Delta-1)^{2}-3s^{2}\big] \,. 
\ee
This is to  be taken with minus sign in  the fermionic case.  $\rm c$  is  directly related to  the free energy 
in (\ref{4.2}), \ie\  
\be
\la{G.4}  
 F_{s}(\Delta)= \tfrac{1}{6\pi} {\rm c}(\Delta, s)\, {\vol(\ads_{3})}   \,.
\ee
Note that  in  this  case of $\ads_3$   the central   charge  c  in \rf{G.4} 
is formally  the same  as the  b-anomaly coefficient in (\ref{1.2}). 
One also  finds  that after summing over the contributions of states in  a $\ads_3$  supermultiplet $E_c$ and c are related by 
\be \la{G5}  E_{c} = -\tfrac{1}{12} {\rm c}  \,. \ee 
One may also  compute the value of the ``thermal'' single particle partition function 
or, equivalently,  of the  character  of a $\Delta>|s|$ massive $SO(2,2)=SL(2,\mathbb R)\times  SL(2,\mathbb R)$
  representation associated to  a field in $\ads_{3}$  having  dual-field 
   conformal dimension $\Delta$ and spin $s$. It is given by  
 $\Tr q^{L_{0}+\overline L_{0}} = \frac{q^{\Delta}}{(1-q)^{2}}$  (here $q= e^{-\beta} < 1$). 
 Summing over the over fields in a  fixed-$\ell$ supermultiplets in Table \ref{muld} 
including the $S^{3}$ degeneracy  we may   consider 
\be\la{G55} 
\mc Z_{\ell}(q) = \sum_{\rm multiplet} (-1)^{\rm F}\dd_{S^{3}}\frac{q^{\Delta}}{(1-q)^{2}}\,.
\ee
Then computing \rf{G55} for the 4 multiplets 
 in Table \ref{muld}  
  we find
\ba
\mc Z_{\ell}^{\rm \bf (2,0)}(q) &= \frac{q^{2+\ell}}{(1-q)^{2}}\big[3+\ell(1-\sqrt q)^{2}-4\sqrt q+q\big]^{2}, \\
\mc Z_{\ell}^{\rm \bf (Ia)}(q) &= \frac{q^{2+\ell}}{(1-q)^{2}}\big[1+\ell(1-\sqrt q)^{2}-4\sqrt q+3q\big]^{2} \,, \la{G6}\\
\mc Z_{\ell}^{\rm \bf (Ib)}(q)  &= \frac{q^{4+2\ell}}{(1-q)^{2}}\big[3+\ell(1-\sqrt q)^{2}-4\sqrt q+q\big]^{2} , \\
\mc Z_{\ell}^{\rm \bf (II)}(q) &= \frac{q^{1+\half \ell }}{(1-q)^{2}}\big[3+\ell(1-\sqrt q)^{2}-4\sqrt q+q\big]^{2}. \la{G66} 
\ea
One observes  the following curious 
 relations (they follow  also  directly from the relations between $\Delta$'s in the supermultiplets)
\ba
\mc Z_{\ell}^{\rm \bf (2,0)}(q) = q^{1+\half \ell }\, \mc Z_{\ell}^{\rm (\bf II)}(q) &= q^{-2-\ell}\, \mc Z_{\ell}^{\rm (\bf Ib)}(q)\,,  \qquad \ \ 
\mc Z_{\ell}^{\rm \bf (2,0)}(q) = q^{4+2\ell}\, \mc Z_{\ell}^{\rm (\bf Ia)}(q^{-1})\,, \la{G7}\\
&
[\mc Z_{\ell}^{\rm (\bf II)}(q)]^{2} = \mc Z_{\ell}^{\rm (\bf Ia)}(q^{-1})\ \mc Z_{\ell}^{\rm (\bf Ib)}(q)\,.  \la{G8}
\ea
We note that the sums over the whole $S^3$ tower of modes  
\be 
\mc Z(q) = \sum_{\ell=0}^{\infty}\mc Z_{\ell}(q)\,, \qquad \qquad q= e^{-\beta} \,, \la{G9}
\ee
are well-defined, \ie\ are  finite for $q <1$. 
 The total Casimir energy can be extracted from the small $\beta$ expansion   of  \rf{G9} 
  (see, \eg,  Appendix B in \cite{Beccaria:2024lbt})
\be\la{G10}
\mc Z(e^{-\beta}) = {C_{1}}{\beta}^{-1} +C_{2}-2E_{c}\,\beta+\mc O(\beta^{2}). 
\ee
We find
\ba\la{G11}
& \te \mc Z^{\rm \bf (2,0)}(e^{-\beta}) = \frac{13}{8}{\beta}^{-1}-\frac{3}{2}+\frac{101}{192}\, \beta+\cdots, \qquad \qquad 
\mc Z^{\rm (\bf Ia)}(e^{-\beta}) = \frac{5}{8}{\beta}^{-1}-\frac{3}{2}+\frac{313}{192}\, \beta+\cdots,
\\  \la{G12}
&\te 
\mc Z^{\rm (\bf Ib)}(e^{-\beta}) = \frac{41}{64}{\beta}^{-1}-\frac{3}{2}+\frac{2491}{1536}\, \beta+\cdots,\qquad \qquad 
\mc Z^{\rm (\bf II)}(e^{-\beta}) = {5}{\beta}^{-1}-\frac{3}{2}-\frac{1}{12}\, \beta+\cdots, \
\ea
that gives
\be
\la{G.13}
E_{c}^{\rm \bf (2,0)} = -\tfrac{101}{384}\,, \quad\qquad  E_{c}^{\rm (\bf Ia)} = -\tfrac{313}{384}\,, \quad\qquad 
E_{c}^{\rm (\bf Ib)} = -\tfrac{2491}{3072}\,, \quad \qquad E_{c}^{\rm (\bf II)} = \tfrac{1}{24}\,. 
\ee
The value for $E_{c}^{\rm \bf (2,0)} $    is the  same as  in Eq.~(5.16) in \cite{Beccaria:2014qea}.
However, the  values in  the other 3 cases do not appear to be  compatible with (\ref{G.2}) summed over $\ell$. 
That may be related  to the fact that the procedure based on  using \rf{G9}  may  not  be manifestly consistent with
underlying supersymmetry.\foot{Also,   first  introducing a finite $\beta $,  then performing  sum over $\ell$
(which is well defined for finite $\beta$) 
and then taking the limit $\beta \to 0$ may not  be equivalent to  first expanding in $\beta$ and then summing
 over $\ell$.}

Let us note also that in 
all four cases  the values of $E_c$ are reproduced by the procedure 
described in in section 5.2 of \cite{Beccaria:2014qea}.
Namely, writing $\frac{q^{\Delta}}{(1-q)^{2}}=\sum_{n=0}^{\infty}(n+1)q^{\Delta+n}$ and summing over zero-point energies
$\frac{1}{2}(n+1)(\Delta+n)$ with a factor of $\dd_{S^{3}}$ 
implies that one may 
compute $E_{c}$ from the finite part of the small $\eps$ expansion of the following  sum 
\be
E_{c} = \tfrac{1}{2} 
\sum_{n,\, \ell=0}^{\infty}\sum_{\rm multiplet}(-1)^{\rm F}\, \dd_{S^{3}} \ (n+1)\big[\Delta(\ell)+n\big] \, e^{-\eps[\Delta(\ell)+n]}
 \Big|_{\text{finite part}} \,.
\ee
The  simplicity of (\ref{G.2}) has a counterpart here: {if we fix $n,\, \ell$ and set $\eps=0$} we find 
\be
\sum_{\rm multiplet}(-1)^{\rm F}\dd_{S^{3}}\ (n+1)\big[\Delta(\ell)+n\big] =0\,,
\ee
in all four cases. This relation  follows also from the sum rules in  (\ref{4.1}).


\section{Wilson loop in 5d SYM  and d$_2$ defect anomaly}
\la{WL}

As  discussed in the Introduction, the free energy $F$  for the    M5 brane probes  {\bf Ia} and  \Ib\ 
with $\ads_{3}$ boundary being  $S^{1}_\b\times S^{1}$  
which  may  be related as in \rf{1.3}  to the  d$_2$  defect  anomaly coefficient  
of the dual (2,0)  theory 
may  be interpreted  as $F(\b) = -\log \WW  $ where $\WW=\langle W\rangle$
is the  expectation value of   supersymmetric  circular  Wilson loop in the 5d SYM theory related to 
(2,0) theory on $S^1_\b \times \mathbb R^5$.
  $\WW$ itself may be  computed exactly 
  by localization methods \ci{Pestun:2016zxk}  using  Chern-Simons matrix model \cite{Kim:2012ava}.

Here  we will  review and elaborate on  the large $N$ expansion of  $\WW$ 
   in the symmetric or antisymmetric representation of $SU(N)$.
In particular,  we  will  show that in the large $N, \beta\gg 1$ limit 
 the expression for the Wilson loop  agrees, up to exponentially small corrections,  
  with its  saddle-point evaluation in \cite{Mori:2014tca}, \ie\ the  latter   does not
receive $1/N^{p}$ corrections. 
This explains an
 apparent puzzle of why that  saddle-point calculation reproduced
  the exact  values  \ci{Estes:2018tnu,Jensen:2018rxu,Chalabi:2020iie} 
  of the defect anomaly ${\rm d}_{2}$ coefficients in (\ref{1.8}).


\subsubsection*{Wilson loop in general representation}

Following   \cite{Kapustin:2009kz}, 
we start with the exact expression for the  partition function of  level $k$ $U(N)$ CS theory on $S^{3}$  (here $a_n$ are real eigenvalues of a matrix $\bf a $) 
\ba
\la{I.1}
Z_{N} = & \frac{1}{N!}\int \prod_{n=1}^{N}da_{n}\, e^{-ik\pi a_{n}^{2}}\  \prod_{n\neq m}^{N} 2\sinh \big[\pi(a_{n}-a_{m})\big]\ 
\\  
\la{I.2}
= &(-1)^{\half N(N-1)}\ e^{-{1\ov 4} i\pi N^{2}}\ e^{-\frac{i\pi}{6k}N(N^{2}-1)}\ k^{-\half N}\ \prod_{n=1}^{N-1}\big(2\sin\tfrac{\pi n}{k}\big)^{N-n}.
\ea
The expectation value of the circular  Wilson loop in fundamental representation is obtained by inserting 
$\Tr(e^{2\pi \bf a}) = \sum_{n=1}^{N}e^{2\pi a_{n}}$ 
under the integral in \rf{I.1} which leads to 
\cite{Kapustin:2009kz} 
\be
\la{I.4}
\WW_{N} = e^{-{\frac{i\pi N}{k}}}\frac{\sin\frac{\pi N}{k}}{\sin\frac{\pi}{k}} \,.
\ee
Upon the analytic continuation 
\be
\la{I.5}
\frac{2\pi}{k} \ \ \to \ \  i\beta
\ee
this  gives, for large $N$, 
\be
\la{I.6}
\WW_{N}(\beta) = e^{\frac{N\beta}{2}}\frac{\sinh\frac{N\beta}{2}}{\sinh\frac{\beta}{2}} = \frac{1}{2\sinh\frac{\beta}{2}}\, e^{N\beta}+\mc O(1).
\ee
More generally, for the Wilson loop defined by a matrix $\bf a$  in a representation $\mathsf R$ of $U(N)$ the result can be  found in \cite{Marino:2004uf}
(ignoring the finite renormalization $k\to k+N$  which is  absent in supersymmetric case)
\be
\la{I.7}
\WW^{\mathsf R}_{N} = \exp(\tfrac{\beta}{2}C_{\mathsf R})\, s_{R}(x_{1}, \dots, x_{N})\,, \qquad x_{n} = \exp\big[-\tfrac{\beta}{2}(N-2n+1)\big]\,. 
\ee
Here $s_{\mathsf R}$ is the Schur polynomial associated with the Young tableau of $\mathsf R$ represented by 
 the partition $(\ell_{1}, \ell_{2}, \dots)$ with $\ell_{i}\ge \ell_{i+1}$, $i\ge 1$.\footnote{
The Schur polynomial $s_{\mathsf R}(x_{1}, \dots, x_{N})$ is simply the character of $\mathsf R$, \ie\  
$\text{ch}_{\mathsf R}[{\bf a}] = s_{\mathsf R}(x_{1}, \dots, x_{N})\big|_{x_{n}=e^{a_{n}}}$.}
The factor $\exp(\frac{\beta}{2}C_{\rm R})$ is  (the analytic continuation of) the framing phase \cite{Marino:2004uf} where
\be
\la{I.8}
C_{\mathsf R} = (N+1)|\mathsf R|+\sum_{r}(\ell_{r}^{2}-2r\ell_{r})\,,\qquad\qquad  |\mathsf R|=\sum_{r}\ell_{r}\,.
\ee
In particular,  for the fundamental representation
$
s_{(1)}(x_{1}, \dots, x_{N}) = \sum_{n=1}^{N}x_{n}, \  C_{(1)} = N, 
$
and thus 
\be
\WW^{(1)}_{N} = e^{\frac{\beta N}{2}}\sum_{n=1}^{N}\exp\big[-\tfrac{\beta}{2}(N-2n+1)\big] = e^{\frac{N\beta}{2}}\frac{\sinh\frac{N\beta}{2}}{\sinh\frac{\beta}{2}}\,,
\ee
in agreement with \rf{I.4}--(\ref{I.6}).

\subsubsection*{Case of symmetric and antisymmetric representations of $U(N)$}

The generating functions  and dimensions 
of the rank-$p$ symmetric/antisymmetric representations $[p]_{\pm}$
  are
\ba
\la{HH9}
&\qquad \qquad \sum_{p=0}^{\infty}s_{[p]_{\pm}}(x_{1}, \dots, x_{N})s^{p} = \prod_{n=1}^{N}(1\mp sx_{n})^{\mp 1} \,, \\
&
\te \dim [p]_{\pm} = s_{[p]_{\pm}}(1, \dots, 1) = (1\mp s)^{\mp N}\big|_{s^{p}} = (\mp 1)^{p}\binom{\mp N}{p}\,, \qquad 
\dim [p]_{-}  = \binom{N}{p}\,,\qquad \dim [p]_{+} =  \binom{N+p-1}{p}\,. \no 
\ea
As in the Introduction we will  use also the  notation
\be
[p]_{+}=(p)\,, \qquad \qquad [p]_{-}=[p]\,.
\ee
As is well known, 
\be
\la{I.15}
s_{(p)}(x_{1}, \dots, x_{N}) = \sum_{i_{1}\le i_{2}\le \cdots \le i_{p}}x_{i_{1}}\cdots x_{i_{p}}\,, \qquad
s_{[p]}(x_{1}, \dots, x_{N}) = \sum_{i_{1}< i_{2}< \cdots < i_{p}}x_{i_{1}}\cdots x_{i_{p}}\,.
\ee
Following \cite{Chen-Lin:2016kkk}  one can also  represent $s_{(p)}$ as 
\be
s_{(p)}(x_{1}, \dots, x_{N}) = \sum_{n=1}^{N}x_{n}^{p}\prod_{m\neq n}\ \big({1-{x_{m}\ov x_{n}}})^{-1} \,, 
\ee
and then \rf{I.7}  gives 
\ba
\WW_{N}^{(p)}(\beta) = e^{\frac{\beta}{2}C_{(p)}}\sum_{n=1}^{N}\exp\big[-\tfrac{\beta p}{2}(N-2n+1)\big] \ \prod_{m\neq n}\frac{1}{1-e^{\beta(m-n)}}\,.
\ea
From  (\ref{I.8}) we get  
\ba
\la{I.18}
C_{(p)} = (N+1)p+p^{2}-2p = Np+p(p-1)\,, \qquad 
C_{[p]} = (N+1)p+\sum_{r=1}^{p}(1-2r) = (N+1)p-p^{2}\,.
\ea
We can obtain an exact representation of $\WW_{N}^{(p)}$ and $\WW_{N}^{[p]}$ starting from the identities\footnote{
They can be proved using the generating function in (\ref{HH9}). For instance, in symmetric representation case, we need to prove
$\prod_{n=1}^{N}\frac{1}{1-s q^{n}}\big|_{s^{p}} = \prod_{n=1}^{p}\frac{q-q^{N+n}}{1-q^{n}}$.
In terms of the $q$-Pochhammer symbol $(a,q)_{n}=\prod_{k=0}^{n-1}(1-aq^{k})$ and using the $q$-binomial theorem we have
$\prod_{n=1}^{N}\frac{1}{1-s q^{n}} = \frac{1}{(sq,q)_{N}}=\frac{(s q^{N+1},q)_{\infty}}{(sq,q)_{\infty}} = \sum_{p=0}^{\infty}\frac{(q^{N},q)_{p}}{(q,q)_{p}}(sq)^{p}$.
Hence $\prod_{n=1}^{N}\frac{1}{1-s q^{n}}\big|_{s^{p}} = q^{p}\frac{(q^{N},q)_{p}}{(q,q)_{p}} = q^{p}\prod_{n=0}^{p-1}\frac{1-q^{N}q^{n}}{1-q^{n+1}} = 
 q^{p}\prod_{n=1}^{p}\frac{1-q^{N}q^{n-1}}{1-q^{n}} = \prod_{n=1}^{p}\frac{q-q^{N+n}}{1-q^{n}}$ as in the first relation in  
  (\ref{HH15}). A similar procedure proves the second 
 relation  for  the antisymmetric representation case. 
 }
\ba
\la{HH15}
\sum_{1\le i_{1}\le i_{2}\le \cdots \le i_{p}\le N}q^{i_{1}+\cdots +i_{p}} &= \prod_{n=1}^{p}\frac{q-q^{N+n}}{1-q^{n}}\,, \qquad
\sum_{1\le i_{1}< i_{2}< \cdots < i_{p}\le N}q^{i_{1}+\cdots +i_{p}} = \prod_{n=1}^{p}\frac{q^{n}-q^{N+1}}{1-q^{n}}\,.
\ea
We then   get 
\be
\WW_{N}^{(p)} = q^{-Np-\frac{1}{2}p(p-2)}\, \prod_{n=1}^{p}\frac{1-q^{N+n-1}}{1-q^{n}}\,, \qquad\quad 
\WW_{N}^{[p]} =  q^{-Np-\frac{1}{2}p(1-2p)}\prod_{n=1}^{p}\frac{1-q^{N-n+1}}{1-q^{n}}\,, \qquad   q \equiv  e^{-\beta}\,.
\ee
Expanded at large $N$ these give 
\ba
 \WW_{N}^{(p)} &=q^{-Np-\frac{1}{2}p(p-2)}\, \prod_{n=1}^{p}\frac{1}{1-q^{n}}\ \big[1-\frac{1-q^{p}}{1-q}q^{N}+\mc O(q^{2N})\big]\,,\\
 \WW_{N}^{[p]} &= q^{-Np-\frac{1}{2}p(1-2p)}\prod_{n=1}^{p}\frac{1}{1-q^{n}}\  \big[1-q^{1-p}\frac{1-q^{p}}{1-q}\,q^{N}+\mc O(q^{2N})\big]\,, 
\ea
where $q^N= e^{-\b N}$, etc., terms in the square brackets represent exponentially suppressed  corrections. 
Taking  also  the large 
 $\beta$  ($q\to 0$) 
  limit   we  find  the following 
  simple expressions
\be\la{I19}
\WW_{N}^{(p)} \stackrel{N,\beta\gg 1}{=}  q^{-N p - \half p (p-2)} 
+\cdots,\qquad \qquad \ \ \ 
\WW_{N}^{[p]} \stackrel{N,\beta\gg 1}{=}  q^{- N p - \half p (1-2p) }   
+\cdots,
\ee
where dots stand for  corrections that are exponentially suppressed in both $N$ and $\beta$.

\subsubsection*{
$SU(N)$  case and relation to the  d$_2$  anomaly coefficient}

It is easy to derive a simple relation between the Wilson loop in the 
$U(N)$ gauge theory and the one in the $SU(N)$ theory.  
We can split the matrix integration into the  trace and traceless parts as follows
($a_n$ are  eigenvalues of $\bf a$)  
\be
\underbrace{\int \prod_{n}da_{n}}_{U(N)} {\mc F}({\bf a})= \int dt \prod_{n}da_{n}\, \delta(\sum_{n}a_{n}-t)\  {\mc F}({\bf a}) = \underbrace{\int dt}_{U(1)}
\underbrace{\int \prod_{n}da_{n}\, \delta(\sum_{n}a_{n})}_{SU(N)} {\mc F}({\bf a} +t N^{-1}) \,, 
\ee
where we redefined  $a_{n}\to a_{n}+t N^{-1}$.
 This  translation does not change the functions of $a_{n }-a_{m}$ in the integrand like in  \rf{I.1}. Thus  we have only to  
 account for a change in  the  Gaussian  factors in 
  (\ref{I.1}) and  in $\Tr(e^{2\pi \bf a}) = \sum_{n=1}^{N}e^{2\pi a_{n}}$ in $\WW$\footnote{From the bi-alternant formula 
  of Jacobi for Schur polynomials one has
$\chi_{R}(s x_{1}, \dots, s x_{n}) = s^{|R|}\chi_{R}(x_{1}, \dots, x_{n})$.}
\be
e^{-\frac{2\pi^{2}}{\beta} \sum_{n}(a_{n}+\frac{t}{N})^{2}}\chi_{R}(e^{2\pi (a_{n}+\frac{t}{N})})= e^{-\frac{2\pi^{2}}{\beta}\frac{t^{2}}{N}}e^{2\pi\frac{t}{N}|R|}\ 
e^{-\frac{2\pi^{2}}{\beta} \sum_{n}a_{n}^{2}}\chi_{R}(e^{2\pi a_{n}})\,, 
\ee
where we have used the traceless condition  $\sum_n a_n=0$  in the $SU(N)$ case. 
This implies 
\be
\frac{\WW^{\mathsf R}_{U(N)}}{\WW^{\mathsf R}_{SU(N)}} =  \frac{\int dt \exp\big(-\frac{2\pi^{2}}{\beta}\frac{t^{2}}{N}+2\pi\frac{t}{N}|R|\big)}
{\int dt \exp\big(-\frac{2\pi^{2}}{\beta}\frac{t^{2}}{N}\big)} = e^{\frac{\beta}{2N}|R|^{2}} = q^{-\frac{1}{2N}|R|^{2}}\,.
\ee
For  the symmetric representation $\mathsf R=(p)$ one has $|\mathsf R|=p$ and thus, up to the exponential corrections, 
\ba\la{I23} 
\WW^{(p)}_{SU(N)} &\stackrel{N,\beta\gg 1}=q^{-Np-\frac{1}{2}p(p-2)+\frac{p^{2}}{2N}} 
= q^{-\frac{1}{12}{\rm d}_{2 (p)}}\,, \qquad \ \ \  {\rm d}_{2\, (p)}= Np(1+\tfrac{p}{2N})(1-\tfrac{1}{N})\,, 
\ea
where ${\rm d}_{2(p)}$  is thus the same as in \rf{1.8}.
Similarly,  for $\mathsf R=[p]=(1, \dots, 1)$ 
we have again $|\mathsf R|=p$ and 
\ba
\la{I.29}
\WW^{[p]}_{SU(N)} & \stackrel{N,\beta\gg 1}{=} q^{-Np-\frac{1}{2}(1-2p)+\frac{p^{2}}{2N}}
= q^{-\frac{1}{12} {\rm d}_{2[p]}}\,, \qquad \ \ \   {\rm d}_{2\, [p]}= Np(1+\tfrac{1}{2N})(1-\tfrac{p}{N})\,. 
\ea
These are also the same  expressions  as found in \cite{Mori:2014tca} using a saddle-point  evaluation of the matrix model integral. 
This proves that  in  the $N, \beta\gg 1$ limit  these expressions  are actually exact if one ignores the 
 exponentially small corrections.
 

\subsubsection*{Defect anomaly coefficients  ${\rm b}$ and ${\rm d}_{2}$ for  the $(n,m)$ representation of $SU(N)$}

For completeness, let us record   also  the derivation of the explicit form of ${\rm b}$ and ${\rm d}_{2}$
for  the more general  $(n,m)$ representation. 
For an $SU(N)$ representation with the Young tableaux 
 corresponding to a partition  $P=(\ell_{1}, \ell_{2}, \dots)$  one has  \cite{Estes:2018tnu}
\be
{\rm b} = 24(\rho, \l)+3(\l,\l)\,,\qquad\qquad 
{\rm d}_{2} = 24(\rho, \l)+6(\l,\l)\,, \la{H1}
\ee
where  $\l= (\l_1, \l_2, \dots)$, $\l_{n}=\ell_{n}-\ell_{n+1}$ are the Dynkin labels and  $\rho$ is the Weyl vector so that 
\ba
(\rho, \l) = \tfrac{1}{2}\sum_{q\geq1}(N-q)q\l_{q}\,, \qquad (\l,\l) = \tfrac{1}{N}\sum_{q\ge 1}(N-q)\l_{q}\big(-q\l_{q}+2\sum_{r=1}^{q}r\l_{r}\big).
\ea
For a representation with $n$ rows and $m$ columns we have 
$P=(m,m,\dots, m, 0, 0, \dots)$ with $n$ entries equal to $m$. Hence, 
$\l=(0, 0, \dots, 0, m, 0, 0, \dots)$, \ie\  we get  a single non-zero component $\l_{n}=m$. Then \rf{H1} implies that 
\ba
{\rm b} &= \te 12(N-n)nm+\frac{3}{N}(N-n)m(-mn+2mn) = 12N\, nm\, \big(1-\frac{n}{N}\big)\big(1+\frac{m}{4N}\big)\,, \la{H3} \\
{\rm d}_{2} &=\te  12 (N-n)nm+\frac{6}{N}(N-n)m(-mn+2mn) = 12N\, nm\, \big(1-\frac{n}{N}\big)\big(1+\frac{m}{2N}\big)\,. \la{H4}
\ea
In particular,  for the  special cases of $(n,m)$  representations  which are the symmetric  $(1,k)$  and  the antisymmetric $(k,1)$ 
ones we reproduce the expressions used in \rf{5},\rf{6} and  (\ref{1.8}), \ie\ 
\ba
\la{H.5}
 {\rm b}_{(k)} &\te = 12N\, k\,\big(1-\frac{1}{N}\big) \,  \big(1+\frac{k}{4N}\big)\,, \qquad\qquad 
{\rm d}_{2\, (k)} =12N\, k\, \big(1-\frac{1}{N}\big)\, \big(1+\frac{k}{2N}\big)\, , \\ 
\la{H.6}
 {\rm b}_{[k]} &= \te 12N\, k\, \big(1-\frac{k}{N}\big)\big(1+\frac{1}{4N}\big)\,, \qquad\qquad 
{\rm d}_{2\, [k]} =12N\, k\, \big(1-\frac{k}{N}\big)\big(1+\frac{1}{2N}\big)\,.
\ea

%

\small 
\bibliography{BT-Biblio}
\bibliographystyle{JHEP-v2.9}
\end{document}

\iffa 
\item 
one thing to stress  to avoid confusion:
we are *not* setting   $\kappa=0$  in computing one-loop correction; in
fact for $\kappa=0$ there is none --
look at  \eg\ 8.15, 8.16 --  \eg\ mixing  that depends on $H_{3}$  is
absent for $\kappa =0$ ;
and indeed for $\kappa =0$ when  probe   shape degenerates   result
should be trivial.
The point is that  one  absorbs $\kappa$ into rescaling of quantum fields
(plus extra
residual  dependence cancels).
\item It is useful also to keep in mind that 
having M5 on ads x s 
we may view this -- expanding in modes on S3--
as action of M2 on s3 coupled to infinite tower 
of massive modes . this is similar to how we treated M2 on ads2xs1 as a string on ads2 plus tower of KK modes on S1
The effect of these modes is to account for the fact that now we have not k =1 but any large. 
but thatvalso rescales M2 tension compared to
single M2 as now we have T5 in front of action 
ie get not $L^3$ but $L^6$ power in front 
This is just a rephrasing but may be useful 
to keep in mind
\end{itemize}
\fi